\documentclass[10pt,letterpaper]{article}
\usepackage[top=0.85in,left=2.0in,footskip=0.75in,marginparwidth=1.75in]{geometry}

\usepackage[utf8]{inputenc}

\usepackage[
backend=biber,
style=numeric,
sorting=none
]{biblatex}
\usepackage{nameref,hyperref}

\usepackage[right]{lineno}

\usepackage{microtype}
\DisableLigatures[f]{encoding = *, family = * }

\usepackage{setspace} 
\usepackage{changepage}
\usepackage{pdflscape}

\usepackage[aboveskip=1pt,labelfont=bf,labelsep=period,singlelinecheck=off]{caption}

\makeatletter
\renewcommand{\@biblabel}[1]{\quad#1.}
\makeatother

\usepackage{lastpage,fancyhdr,graphicx}
\usepackage{epstopdf}
\fancyheadoffset[L]{2.25in}
\fancyfootoffset[L]{2.25in}

\usepackage{color}

\definecolor{Gray}{gray}{.25}

\usepackage{graphicx}

\usepackage{sidecap}

\usepackage{wrapfig}
\usepackage[pscoord]{eso-pic}
\usepackage[fulladjust]{marginnote}
\reversemarginpar

\usepackage{amsmath}
\usepackage{booktabs}
\usepackage{longtable}
\usepackage{multirow}
\usepackage[capitalise]{cleveref}
\usepackage{tabularx}
\usepackage{threeparttable}
\usepackage{ragged2e}
\usepackage{enumitem}
\usepackage{wrapfig}
\usepackage{xfrac}
\usepackage{subcaption}
\usepackage[textsize=small,textwidth=1.5in]{todonotes}

\usepackage[acronym]{glossaries}
\usepackage{packages/glossary_table}

\newglossary[slg]{symbols}{not}{ntn}{Symbols}
\newglossary[slghm]{symbolshm}{nothm}{ntnhm}{Symbols Hierarchical Model}
\makeglossaries
\newacronym{ode}{ODE}{ordinary differential equation}  

\hypersetup{
    colorlinks,
    linkcolor={red!50!black},
    citecolor={blue!50!black},
    urlcolor={blue!80!black}
}

\crefformat{equation}{(Eq.~#2#1#3)}
\crefformat{figure}{(Fig.~#2#1#3)}
\crefformat{table}{(Tab.~#2#1#3)}
\crefname{equation}{Eq.}{equations}

\newcolumntype{b}{X}
\newcolumntype{s}{>{\hsize=.25\hsize}X}
\newcolumntype{v}{>{\hsize=.1\hsize}X}
\newcolumntype{u}{>{\hsize=.15\hsize}X}
\newcolumntype{i}{>{\hsize=.15\hsize}X}

\usepackage{xr-hyper}
\makeatletter
\newcommand*{\addFileDependency}[1]{
  \typeout{(#1)}
  \@addtofilelist{#1}
  \IfFileExists{#1}{}{\typeout{No file #1.}}
}
\makeatother

\newcommand*{\myexternaldocument}[1]{%
    \externaldocument{#1}%
    \addFileDependency{#1.tex}%
    \addFileDependency{#1.aux}%
}

\myexternaldocument{SI_main}

\usepackage{packages/si_setup}

\begin{document}
\vspace*{0.35in}

\begin{flushleft}
{\Large
\textbf\newline{From Lab to Landscape: Assessing the Impact of Pesticides on Pollinator Populations Based on Laboratory Data by Combining ALMaSS and BufferGUTS}
}

{
    \bf
    Florian Schunck\textsuperscript{1,*},
    Agnieszka Bednarska\textsuperscript{2},
    Leonhard Bürger\textsuperscript{1},
    Christopher John Topping\textsuperscript{3},
    Andreas Focks\textsuperscript{1},
    Xiaodong Duan\textsuperscript{3},
}
\\
\bigskip
\textsuperscript{1} Osnabrück University, Osnabrück, Germany
\\
\textsuperscript{2} Institute of Nature Conservation of the Polish Academy of Sciences, Krakow, Poland
\\
\textsuperscript{3} Aarhus University, Aarhus, Denmark
\\
\bigskip
* fschunck@uos.de

\end{flushleft}

\section*{Abstract}
\begin{wrapfigure}[12]{r}{75mm}
    \includegraphics[width=75mm]{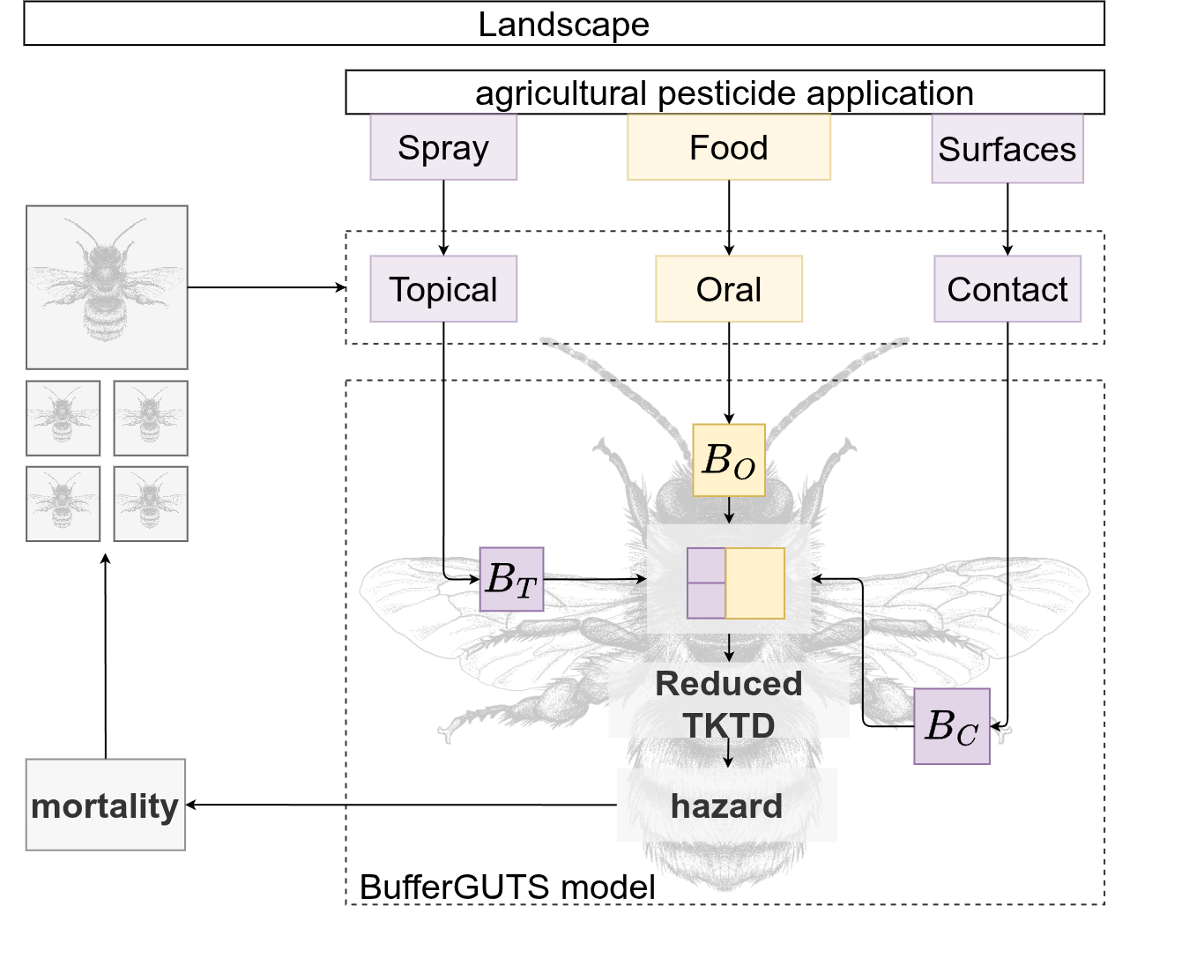}
\end{wrapfigure}

The application for pesticides is hazardous by design. Pesticides are designed to eradicate pests from crops, which fulfills an important role in the current agricultural system. However, nature conservation requires that pesticide applications are protective for non-target organisms, which provide ecosystem services on the other hand. Environmental risk assessment (ERA) is supposed to strike this balance, but the current use of laboratory derived toxicity thresholds in the landscape context, without consideration of population and landscape dynamics might be too coarse to achieve this task in cases.


Here, we propose to overcome this limitation by coupling the Animal, Landscape, and Man Simulation System (ALMaSS) with the recently developed BufferGUTS model for non-target arthropods. The ALMaSS system is a spatially explicit framework that models wildlife, habitats, and human activities. The BufferGUTS model uses a simplified toxicokinetic-toxicodynamic approach that estimates the effect of pesticide application via multiple exposure paths.

We investigated the potential of using systems based ERA for a case study of the solitary bee \textit{Osmia bicornis} exposed to the pesticide formulation 
Closer (a.i. sulfoxaflor). For this, laboratory survival data of topical and oral exposure to 
Closer were used to calibrate BufferGUTS models with Bayesian inference.
The resulting parameters from the posterior parameter distributions were then used to parametrise model organisms in ALMaSS simulations of multiple landscapes to extrapolate the effects of 
sulfoxaflor 
at different exposure levels on solitary bee population dynamics.

The integration of BufferGUTS into ALMaSS landscape simulation was achieved with high numerical precision, 
allowing for the calculation of exact daily survival probabilities for model organisms in the ALMaSS framework.

We found that an application rate 400 times above the regulatory acceptable concentration led only to negligible population effects in ALMaSS simulations, but an exploratory analysis of pesticide-driven larval mortality showed that effects might be more severe when additional pollinator life stages are considered. 

Despite serveral open research questions (mixture toxicity, larval toxicity), the work demonstrates how mechanistic modelling embedded into individual based modelling frameworks can support ERA by combining exposure and effect under the same umbrella. Therefore, the integration of the TKTD model BufferGUTS into the ALMaSS framework enables its use as a systems-based ERA tool, bridging the gap between controlled laboratory experiments and realistic landscape-scale risk assessments for next generation ERA.

\section{Introduction}
While environmental risk assessment (ERA) has previously focused on aquatic organisms, growing evidence of declines in terrestrial invertebrates \cite{Hallmann.2017,Wagner.2020,seibold.2019}, together with evidence linking pesticide exposure with pollinator risks \cite{rundlof2015seed,Potts.2016,Goulson.2015,Wood.2017} has increased the need for pollinator protection. However, not all species are under threat everywhere \cite{Crossley.2020}, because both winners and losers of the Anthropocene emerge at increasing rates \cite{Dornelas.2019}. This implied contextuality of insect and pollinator decline indicate that holistic methods are required to assess the risks associated with environmental contaminants for a multitude of species that may behave differently in a changing world. Systems-based ERA \cite{AXELMAN2024174526} is an advanced approach to address such challenges. However, it is necessary to develop a system that integrates 1) general species life-cycle and behaviour, 2) landscape structure, 3) environmental exposures to pesticides, and 4) species responses to these exposures. 

Various pollinator simulation frameworks exist. The BEEHAVE was developed to harmonize existing models of the colony-building western honey bee (\textit{Apis mellifera}), including foraging dynamics in real landscapes \cite{Becher.2013}. Bumble-BEEHAVE is a derived model for bumblebees \cite{Becher.2018}. For solitary bees, SolBeePop \cite{Schmolke.2023} has been developed. The Animal, Landscape, and Man Simulation System (ALMaSS) \cite{topping2013modelling} is a complex, spatially explicit \cite{topping2012spatial,Topping2024Landscape} agent- and subpopulation-based \cite{topping2024managing, Duan2024subpopulation} modelling framework that simulates the detailed life cycle and behaviour of 22 species.
For \textit{A. mellifera} \cite{duan2022apis} and the solitary red mason bee (\textit{Osmia bicornis}) \cite{ziolkowska2023formal}, ALMaSS employs individual-based models where each agent undergoes a complete life cycle, from egg to adult, with developmental rates and behaviours modulated by external conditions, including resource availability and pesticide exposure. Further formal model descriptions for additional pollinator species \textit{Noctua pronuba} \cite{Topping.2025}, \textit{Eristalis tenax} \cite{Ziolkowska.2025} and \textit{Pieris napi} \cite{Topping.2025a} are published and can be readily implemented in ALMaSS.

Recently ALMaSS was extended to simulate pesticide fate in landscapes in six environmental compartments (soil, in-plant, plant surface, pollen, nectar and seeds) \cite{Poulsen2023Pesticide}. This enables the estimation of exposure risks in wildlife habitats under different crop management practices and pesticide applications, e.g. floral resource exposure from pesticide seed coating \cite{bonmatin2015environmental}. To unfold the potential of ALMaSS for systems based risk assessment, the exposure to pesticides has to be connected to state-of-the-art effect modelling approaches.

For assessing species responses to chemical exposures, TKTD (toxicokinetic-toxicodynamic) models are continuous time \gls{ode} models that are frequently applied to predict pesticide effects for unknown exposure patterns \cite{Ashauer.2010}.
Particularly, the general unified threshold model for survival (GUTS) \cite{Jager.2011} is a commonly used TKTD framework to model time-resolved survival of chemical exposure \cite{Focks.2018,Singer.2023,Ashauer.2016,Dalhoff.2020} or chemical mixtures over time \cite{Cedergreen.2017,Bart.2022}. 
Recently, GUTS extensions were developed to model survival of terrestrial invertebrates under pesticide exposure. The BeeGUTS model \cite{Baas.2022} was developed specifically for the species \textit{Apis mellifera}, but has also proven useful for solitary bees \cite{Schmolke.2024}. The BufferGUTS model \cite{Burger.2025,Burger.2025b} takes a more generic, species agnostic approach to describe multi-route exposure as a buffered compartment in the insect body and on the insect surface. Thereby, the BufferGUTS model is ideally suited to serve as a generalised TKTD model for modelling toxic effects from multiple exposure paths in real landscapes for a wide range of Pollinator species.

In this work, we integrate the BufferGUTS model for multiple exposure pathways into the ALMaSS modelling framework to provide a computational method for calculating daily mortality risks from pesticide application for any model organism. We pay special attention to harmonizing large-scale, discrete-time simulations with small-scale, continuous-time effect models by calculating exact exposure-driven survival probabilities over arbitrary time intervals using a generalistic approach  transferable to any simulation framework. To lower the computational burden of the implementation, we provide an analytic solution for the BufferGUTS model for hourly constant exposure profiles. 
Finally, we use the ALMaSS--BufferGUTS integration to predict landscape scale effects from laboratory tests, by calibrating a BufferGUTS model on \textit{O. bicornis} datasets of sulfoxaflor
exposure using the obtained effect parameters in ALMaSS simulations 
to predict pesticide effects in different landscapes and management practices.

\section{Method}

\subsection{Study Approach}

To extrapolate toxicity information from the laboratory to the landscape scale, we followed a 2-step procedure. First, toxicity parameters of the BufferGUTS model \cite{Burger.2025b} were estimated from experimental survival data of \textit{O. bicornis} orally and topically exposed to 
Closer (a.i. sulfoxaflor) \cite{Misiewicz.2025,Zbrozek.2022}. Second, the toxicity parameters were used to parametrise modelled organisms in the ALMaSS framework, to calculate their survival chances following pesticide exposure.

\begin{figure}[htb]
    \centering
    \includegraphics[width=1.0\linewidth]{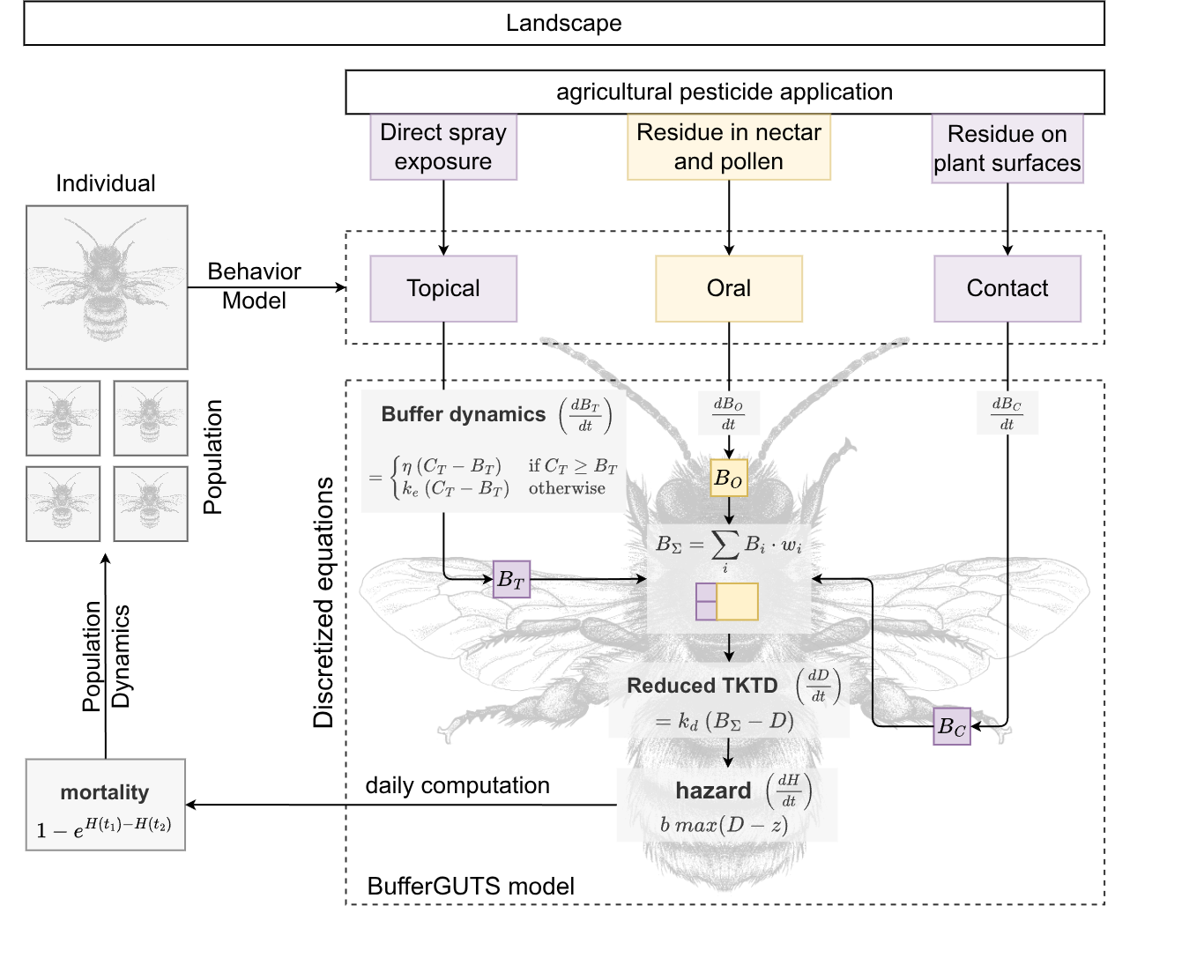}
    \caption{ALMaSS--BufferGUTS integration: Aggregation of exposure buffers to harmonize exposure pathways. Exposure to pesticides depends on the landscape, pesticide application scenarios and modelled behaviour of the individual pollinators. The exposure buffers can be considered as compartmentalized environments that the pollinator carries along and their dynamics follows simplifying assumptions (Sections~\ref{sssec_intake},~\ref{sec:exposure-surface-contact},~\ref{sec:exposure-topical-contact}). Buffers are weighted according to their toxicity and summed. The summed buffer concentration $B_\Sigma$ is coupled with the reduced GUTS model, which is used to calculate the daily mortality risk with the stochastic death (SD) model. The computed mortality risks feed back to the ALMaSS population model.}
    \label{fig:bufferguts-aggregation}
\end{figure}

For this, a robust integration of the BufferGUTS model in the ALMaSS framework (or any discrete modelling framework) is necessary, which guarantees equivalence between the continuous time solution and the discrete time solution and couples exposure information to the inputs of the BufferGUTS module (Fig.~\ref{fig:bufferguts-aggregation}).

In the following, the integration implementation is described before the calibration routine and scenarios for extrapolation are outlined. Due to the extent of the method, only brief descriptions of the methods are presented in the manuscript, while comprehensive model description and discretisation (Sec. \ref{si:sec:model-description}), source code (Sec.~\ref{si:sec:source-code}), data and results are made available in the supporting information.

\subsection{Landscape and Exposure simulation in ALMaSS}

The ALMaSS framework is structured into two core components: species modelling and landscape modelling, each contributing to a detailed representation of ecological systems. 
The species models simulate individual or group behaviours of wildlife species with high biological fidelity, capturing life-history traits, movement patterns, and responses to environmental variability. 
The landscape model \cite{Topping2024Landscape} in ALMaSS is a dynamic, high-resolution representation of the environment, incorporating temporally and spatially explicit agricultural practices. This includes crop rotations and pesticide applications \cite{Poulsen2023Pesticide}.  
A key feature of the landscape model is its detailed tracking of pesticide residues across six environmental compartments: soil, in-plant, plant surface, pollen and nectar, and seeds (the latter for crops treated with seed coatings). These residues are updated daily and mapped at a fine spatial resolution of $1~m^2$, allowing for precise estimation of exposure risks for foraging pollinators and other non-target organisms. 

Pollinators are exposed via multiple routes. Here we consider oral and contact exposure, the latter is subdivided into topical contact and surface contact (Fig.~\ref{fig:bufferguts-aggregation}). Oral exposure concentration $C_O$ acts via consumption of contaminated nectar or pollen ($mg/mg$). Surface contact exposure $C_S$ represents the absolute pesticide mass ($mg$) calculated as the pesticide surface concentration on the plant surface ($mg/m^2$) multiplied by the half body surface, which is integrated by the buffer. Topical exposure $C_T$ represents the absolute pesticide mass ($mg$) calculated as the pesticide application rate ($mg/m^2$) multiplied by the interception surface, approximated by the half body surface ($cm^2$), which is another input of the buffer. The duration of topical exposure for a given pollinator adult on a given day is set to be one hour. The one-hour exposure duration represents a conservative yet realistic assumption for topical exposure during pesticide application. It only happens when there is an application in the field where the adults are foraging. The chance of this occurrence is set to $1/24$. Applications may occur at any time during daylight hours (or occasionally at night for certain products), and this timing is not predictable within the daily time-step framework. The 1/24 probability acknowledges that whilst an application occurs on a given day, the specific hour is effectively random from the model's perspective. 


\subsection{The BufferGUTS model for multiple exposure pathways} \label{ssecbuffer}

The BufferGUTS model has been extensively described in previous work \cite{Burger.2025,Burger.2025b} and the comprehensive model description is available in Sec.~\ref{si:sec:buffer}. In Brief:
Conceptually, the BufferGUTS model introduces a separate buffer $B_i$ to track chemical concentrations $C_i$ of each exposure pathway $i$ (e.g. the stomach for the oral pathway and the exoskeleton or cuticle for the topical pathway) (Eq.~\ref{eq_ode_buffer}, Fig.~\ref{fig:bufferguts-aggregation}). These buffers can be seen as a portable environment that the agent carries along. In contrast to internal compartments such as organs, hemolymph, etc., buffers are not separated from the environment by membranes. Therefore physiological (feeding) or environmental (deposition) processes govern the input dynamics. Therefore, in discrete exposure events, the chemical concentration in a buffer for a given exposure pathway almost immediately approaches the environmental concentration with speed $\eta \gg 1$. When exposure stops or is reduced, the concentration of the buffer decays exponentially with a rate constant $k_e$. 

Mathematically the buffer concept is expressed as: 

\begin{align}
    \frac{dB_i}{dt} &= \begin{cases}\label{eq_ode_buffer}
        \eta \cdot (C_i - B_i) & B_i \leq C_i \\
        k_{e,i} \cdot (C_i - B_i) & B_i > C_i \\
    \end{cases}\\
    \frac{dD}{dt} &= k_d~\left(\sum_i w_i B_i - D(t)\right) \label{eq_ode_damage} \\
    \frac{dH}{dt} &= b ~ max(D(t)-z, 0) \label{eq_ode_hazard}
\end{align}

To keep parameters identifiable from only survival data, we make the simplification $k_{e,i} = k_d$, which assumes that buffer elimination and damage accrual have the same speed \cite{Burger.2025,Burger.2025b}. For simplicity, all buffers share the same elimination rate constant $k_d$ for the same chemical. These assumptions are discussed in Sec.~\ref{sec:limitations}. The outlined equations including the simplification corresponds to the BufferGUTS concentration addition (CA) variant, which was found to be the optimal variant for the majority of chemicals investigated in \cite{Burger.2025b}. Accordingly, the buffers are aggregated in the damage ODE (Eq.~\ref{eq_ode_damage}, compare RED CA mixture model~\citeauthor{Bart.2022}, \citeyear{Bart.2022}) with buffer specific weights $w_i$ that reflect the contribution to the overall toxicodynamic damage. For $n$ exposure pathways, $n-1$ weights need to be calibrated, because one buffer weight is always set equal to 1.

As a survival model, we choose the stochastic death (SD) model (Eq.~\ref{eq_ode_hazard}, compare \citeauthor{Jager.2011}, \citeyear{Jager.2011}), which is rooted in survival analysis \cite{Miller.1998,OQuigley.2021}. The mathematical framework of survival analysis allows for a simple and probabilistically consistent way of combining independent hazards $S(t)=exp(-\sum_i H_i(t))$, which is suitable for the ALMaSS framework. 
The background hazard $h_b$ is not explicitly listed in Eq.~\ref{eq_ode_hazard}, because it contributes \textbf{independently} to the hazard in the standard threshold model; it will only be used for calibration purposes but not in the ALMaSS framework, which has its own logic of calculating aging.

\subsection{Discretization}

To derive the expanded discretized equations (Sec.~\ref{si:sec:expanded-discretized-equations}), symbolic integration was used with the python package Sympy. From this, the condensed forms of equations were manually derived. Due to the extent of the equations, only the most important aspects are presented in the manuscript, while condensed equations, expanded forms and comprehensive explanations are presented in Sec. \ref{si:sec:discretization}. To demonstrate the correctness of the discretization, we provide graphical proof by comparing ODE solution, analytical solution and solutions from the implementation in ALMaSS in the BufferGUTS model (Fig.~\ref{fig:discretization_bufferguts}).

Although the standard time step of ALMaSS is one day, we have decided to increase the time resolution to hourly time steps to cover short topical exposure events, and consider the species day/night activity for correctly assessing the accumulation of a pesticide in the buffers. Thus $\Delta t = 1h$.

\paragraph{Buffer Dynamics} 

The buffer equations (Eq.~\ref{eq_ode_buffer}) can be exactly solved for one time step $\Delta t$ from $t-1$ to $t$ by using analytic solutions for first order kinetic processes $e^{-kt}$ and assuming that uptake is immediate if the environmental concentration is larger than the buffer. Leading to \ref{eq_buffer_guts_discrete}.

\paragraph{Buffer stacking} \label{sec:buffer-stacking} 

The assumptions of the standard BufferGUTS model (Eq.~\ref{eq_ode_buffer}) imply that the buffers cannot increase above the maximum environment concentration or dose. This is a realistic assumption for oral exposure, where it can be assumed that the oral buffer fills and empties so quickly that the dominant influence is the external concentration. For mass-based buffers, where theoretically no upper limit is available, it is unrealistic that a higher environmental dose replaces the mass that is left in the buffer. For this reason, a small modification is made for the discretization of mass-based buffers (topical, contact): 

\begin{equation}
C_i(t - \Delta t)= \begin{cases}
 C_i(t - \Delta t) + B_i(t - \Delta t) &  C_i(t - \Delta t) > 0 \\
 0 & \text{otherwise}
\end{cases}
\end{equation}

This can be understood as if the residual buffer mass is re-applied to the buffer in addition to the new environmental dose. This modification is a \textit{\textbf{temporary solution}} until a buffer model is available that explicitly models uptake and deposition processes in the buffer. In order not to overload buffers, the intercepted pesticide mass in $mg$ (pesticide application rate $mg/m^2$ $\times$ half body surface $cm^2$, Sec.~\ref{sec:exposure-topical-contact}), needs to be divided by the duration of the exposure, leading to an interception rate in $mg/h$. This results in a linear increase of the topical or contact buffer until the intercepted mass is reached. If, e.g., an exposure event happens immediately afterwards, the mass in the buffer continues to increase and does not settle on the new intercepted dose, as it happens with oral exposure buffers.  

\paragraph{Damage dynamics}

The discretisation of the damage equation (Eq.~\ref{eq_ode_damage}) involves two first-order kinetic processes in sequence. For multiple buffers there is a combinatorially increasing number of solutions with $2^n$ where $n$ is the number of buffers, resulting in a piecewise solution to the damage equation (\ref{eq_damage_guts_discrete}). 

\paragraph{Hazard}

Equation \ref{eq_ode_hazard} can be solved analytically up until the point where $D(t) = z$ (\ref{eq:hazard_discrete}). At this point, the integral changes and an intermediate step has to be taken. The time $t^*$ of this step can in some cases be solved analytically, but in the majority of cases it is much simpler and also computationally efficient to approximate the root of the equality numerically. 

In order to minimise the number of intermediate steps for ALMaSS, the analytical solution can be computed for each time step and conditionally, whether the sign of the equation $D(t) - z$ changes from the beginning to the end of the time step, we know that an intermediate step must be taken, and $H(t)$ can be re-computed with an intermediate step. Fortunately, only very volatile exposure profiles will lead to many hits of $D(t) = z$ and in most cases the numerical root finder will only be needed a few times in the lifetime of an individual.

\subsection{Mortality probability in time intervals}

Discrete time frameworks such as ALMaSS challenge the modelled organisms with daily tasks and trials. To evaluate survival over a given time interval, the probability of death within a given interval, conditional of being alive before the start of that period must be calculated. 

\begin{equation} \label{eq_conditional_lethality}
\Pr(t_1 < T \leq t_2 | T > t_1) = 1 - e^{- \int_{t_1}^{t_2} h(t')dt'} = 1 - e^{H(t_1) - H(t_2)}
\end{equation}

Where $H(t_1)$ is the cumulative hazard at the beginning of the interval, and $H(t_2)$ is the cumulative hazard at the end of the interval. Therefore the probability of death in a given interval is proportional to the accumulation of hazard throughout that interval. The survival probability is given by the complementary function $S(t_1, t_2) = e^{H(t_1) - H(t_2)}$. Because $h(t)$ is strictly non negative, $H(t_2) \geq H(t_1)$, which satisfies that $0 \leq S(t_1, t_2) \leq 1$. The detailed derivation of the survival probability over time intervals is given in the supporting information (Sec. \ref{si:sec:daily-mortality-chance}).

\subsection{Experimental Data}


To calibrate the TKTD module for running simulations, a dataset of \textit{O. bicornis} females' survival after exposure to the formulation Closer (active ingredient, a.i. sulfoxaflor) 
via topical \cite{Misiewicz.2025} and oral \cite{Zbrozek.2022} administration was used. For topical exposure, each individual bee was exposed to 1 $\mu$L of solution of a.i. in 0.01\% Triton X-100 (used to facilitate the adhesion of the solution to the bee body) at the following concentrations of sulfoxaflor as a.i.: 24, 120, 600, and 3000 ng/$\mu$L 
An additional control group was also used, in which the bees were exposed exclusively to 0.01\% Triton X-100, to assess the potential effects of this chemical. For oral exposure, the following concentrations of sulfoxaflor
as a.i. were used: 0.025, 0.1, 0.4, and 1.6 ng/mL. The doses were administered continuously in 33\% sucrose solutions (i.e., were allowed to feed ad libitum on the contaminated or control sucrose solution). Bees were checked daily for mortality. For the purpose of this study, 10-day survival data were used to focus on acute effects.


\subsection{Parameter Estimation}

Parameter estimation was conducted with the open source model-building platform pymob (\url{https://github.com/flo-schu/pymob}). The conditional binomial likelihood function used for the parameter estimation is given in (Sec. \ref{si:sec:likelhood-function}). We used hamiltonian Markov-Chain Monte-Carlo (MCMC) with the No-U-Turn-Sampler (NUTS) \cite{Hoffman.2011} with four parallel Markov chains to sample 50,000 parameter sets from the posterior distributions. These samples were used to estimate parameter uncertainty and model fits. For sulfoxaflor calibration, a simple 1-parameter background hazard model with a constant background hazard $dH_b/dt = h_b$ was chosen.

Software for parameter estimation and preparation of figures is publicly available: \url{https://gitlab.uni-osnabrueck.de/fschunck/bufferguts}, version 0.6.2 

\subsection{ALMaSS simulations of \textit{Osmia bicornis}} 

The ALMaSS implementation of \textit{O. bicornis} is an individual-based model that simulates the complete life cycle of this solitary bee species \cite{ziolkowska2023formal}. \textit{Osmia bicornis L.} is a univoltine, polylectic species that nests in pre-existing tube-shaped cavities. The model captures six main life stages: egg, larva, prepupa, pupa, cocooned adult, and adult. Development is temperature-driven using degree-day models with species-specific lower developmental thresholds (LDTs) and sum of effective temperatures (SET) for each stage.

Key behavioural and biological assumptions in the model include: (1) Females are monandrous while males are polygamous but males are not explicitly modelled due to their limited influence on reproduction beyond mating; (2) Sex ratio of offspring depends on maternal mass and reproductive age; (3) Nesting behaviour involves sequential provisioning of brood cells with pollen; (4) Foraging and dispersal patterns are based on homing distances characterised by $r_{50}$ and $r_{90}$ values; (5) Overwintering comprises three phases: pre-wintering, diapause, and post-diapause, each with distinct temperature requirements; (6) Mortality rates vary across life stages, with parasitism risk increasing with cell provisioning time; (7) Movement and nesting patterns follow beta distributions reflecting spatial heterogeneity in resource availability.

\subsection{Pesticide exposure scenarios}

The proposed BufferGUTS extension was implemented as a toxicity module for \textit{O. bicornis} females in the ALMaSS \textit{O. bicornis} \cite{ziolkowska2023formal} species model.

The adult life stage of \textit{O. bicornis} was exposed to pesticide in five scenarios, because laboratory data were only available to calibrate GUTS parameters for the adult life stage. One additional scenario considered the effects of pre-reproductive exposure with a simplified effect model. Two landscape windows, each measuring 10 km by 10 km, were utilised for testing purposes. Landscape 1 has more arable fields (ca. $73\%$) than Landscape 2 (ca. $52\%$ are covered by arable fields). ALMaSS implements farm type-specific rotations and crop area allocations, with individually tailored management plans for each crop \cite{Topping2024Landscape}. The pesticide applications follow the Good Agricultural Practices. Following pesticide application, the pesticide module models transfer amongst different environmental compartments and can track the fate \cite{Poulsen2023Pesticide}. Six scenarios were explored to test the new toxicity module with a dose of $1~g/m^2$, approximately 400-fold higher than the recommended sulfoxaflor application rate of $0.0024~g/m^2$ \cite{europeanfoodsafetyauthorityConclusionPeerReview2014} to test the model at extreme exposure conditions:

To study the effects of pesticides in a landscape context, we calculated ALMaSS simulations for the following scenarios:

\begin{enumerate}
    \item Baseline: No pesticide was sprayed in the landscape window, which means no pesticide exposure.
    \item Oral exposure pathway only: Pesticide exposure only occurs through oral intake of the contaminated floral resource.
    \item Topical exposure pathway only: Pesticide exposure only occurs through overspray.
    \item Contact exposure pathway only: Pesticide exposure only occurs through contact with the contaminated plant surface.
    \item All exposure pathways: All three pesticide exposure pathways are included.
    \item All-eggs: All three pesticide exposure pathways are included and eggs are tested once against a mortality threshold (details below).
\end{enumerate}

Since the implemented TKTD module does not include effects on eggs and larvae, we evaluate early-life mortality in an \textit{all-eggs} scenario. As a strong simplification, because laboratory data for larval mortality were unavailable, laid eggs were subjected to a single mortality test against a threshold ($0.01~mg$), which was set only for exploratory purpose, based on the per egg pesticide amount in the provision of pollen that females collect for nest building. If the pesticide amount in the provision is larger than $0.01~mg$, the egg will have a $50\%$ chance of dying from the pesticide. If the eggs survived this one-time survival test, the pesticide would be removed from the pollen, effectively implementing a threshold-based mortality for the early life stages. 

Each scenario has been replicated 30 times for each landscape, randomly varying in crop rotation, foraging activity, and BufferGUTS parameters drawn from the calibrated distributions for simulated \textit{O. bicornis} individuals. After initialisation, replicates were simulated for 30 years, where each individual passes a random mortality trial each day, based on the calculated daily survival probability, described above. The first twenty years were used to reach a steady state. The last 10 years' data were used for analysis. The number of new females, the number of dead females caused by pesticide exposure, and the population size were recorded daily.

For all scenarios except Baseline, we used two different parameter settings: \textit{single}, where a single set of the optimised parameters was used; and \textit{random}, where each simulated female used a random parameter set obtained from the Markov-Chain Monte-Carlo (MCMC) samples during parameter inference.
While the implemented BufferGUTS model uses the SD variant, using samples from parameter distributions in an individual based model mimics the behaviour also of the individual tolerance model \cite{Ashauer.2010a}

\section{Results}

\subsection{Corroboration of discretized solutions and ALMaSS implementation}

To demonstrate the correctness of the discretized solutions (analytical solutions) and their implementation in ALMaSS, random exposure profiles for oral, topical and contact exposure were generated over a duration of 180 days. Then, solutions of the respective implementations were compared.

\begin{figure}[htb]
    \centering
    \includegraphics[width=\linewidth]{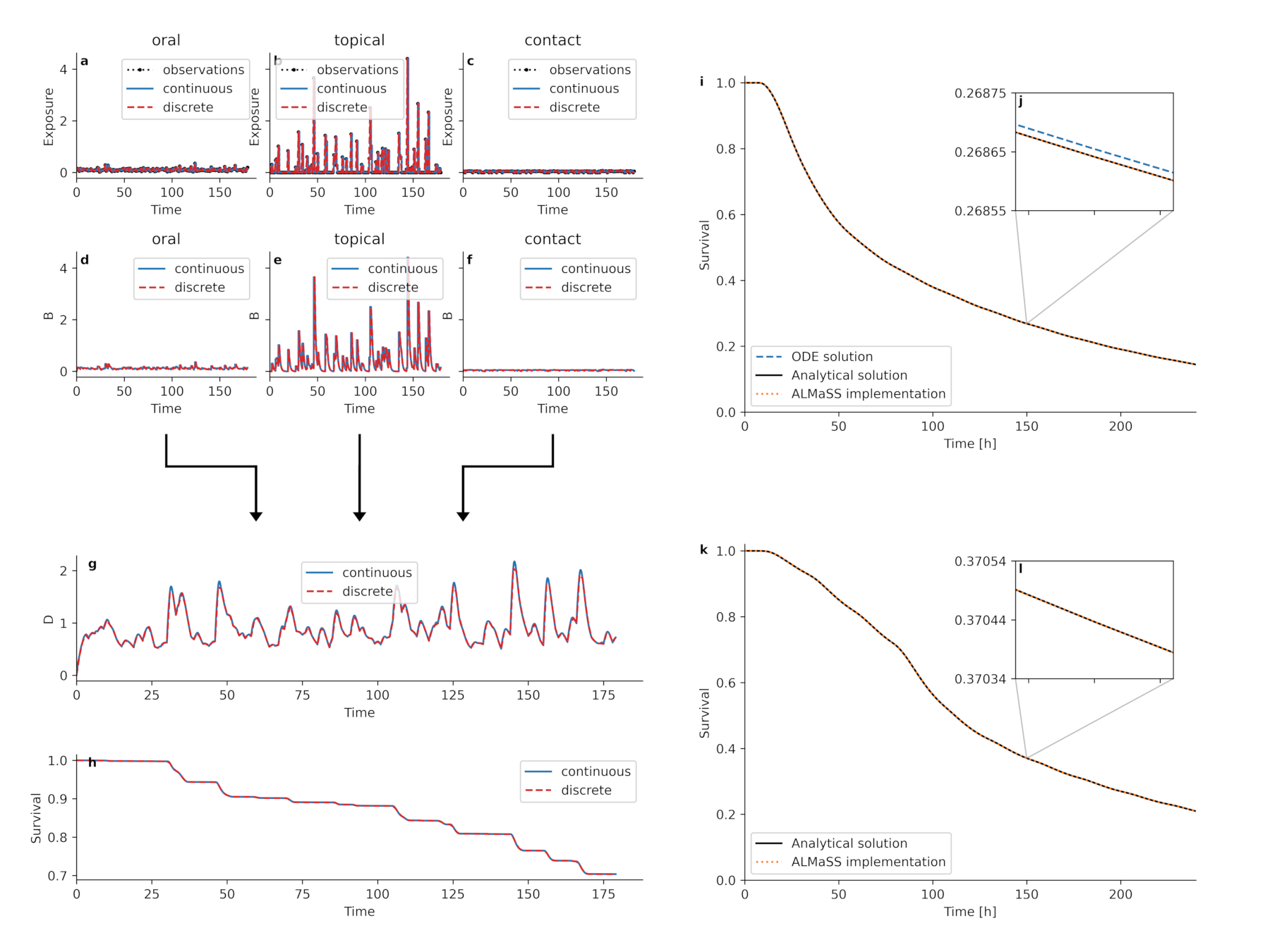}
    \caption{Synthetic exposure profiles of 180 days for oral ($\text{mg}_{a.i.}$ $\text{mg}_{nectar}^{-1}$), topical ($\text{mg}_{a.i.}$) and contact  ($\text{mg}_{a.i.}$) exposure \textbf{(a, b, c)} in and the continuous and discrete solutions of the buffer \textbf{(d, e, f)} damage \textbf{(g)} and survival equations \textbf{(h)}, for $\eta=1000$ and $k_d=0.1$. \textbf{i, j, k, l}: Alignment between discrete solutions and ALMaSS implementation for the parameter set used in the case study: $k_d=0.079625~\text{h}^{-1}$, $h_b=0.0000417 ~\text{h}^{-1}$, $b= 0.0000833 ~\text{h}^{-1} ~\text{ng}_{a.i.}^{-1}$, $m=0.091 ~\text{ng}_{a.i.}$, $w_o=1.75~10^9 ~\text{mg}_{nectar} ~\text{ng}_{a.i.} ~\text{mg}_{a.i.}^{-1} $ $w_t=w_c=10^6 ~\text{ng}_{a.i.}^{-1} ~\text{mg}_{a.i.}^{-1}$.}
    \label{fig:discretization_bufferguts}
\end{figure}

The resulting dynamic (Fig.~\ref{fig:discretization_bufferguts}a--f) clearly demonstrates that the discretized solution and ALMaSS implementation exactly match the continuous time solution up to a precision limit of the ODE solver.

This result provides a mathematically rigorous and general framework for integrating mechanistic survival models into discrete time simulations of terrestrial invertebrates. Where the BufferGUTS model resides at the interface between $n$ exposure paths and calculates survival hazards $H(t)$, which can be easily translated into a probability of death over an arbitrary time interval with $\Pr(t_1 < T \leq t_2 | T > t_1) = 1 - e^{H(t_1) - H(t_2)}$.

\subsection{BufferGUTS parameter estimation on \textit{Osmia bicornis} survival data from topical and oral exposure experiments}

\begin{figure}[htb]
    \centering
    \includegraphics[width=\linewidth]{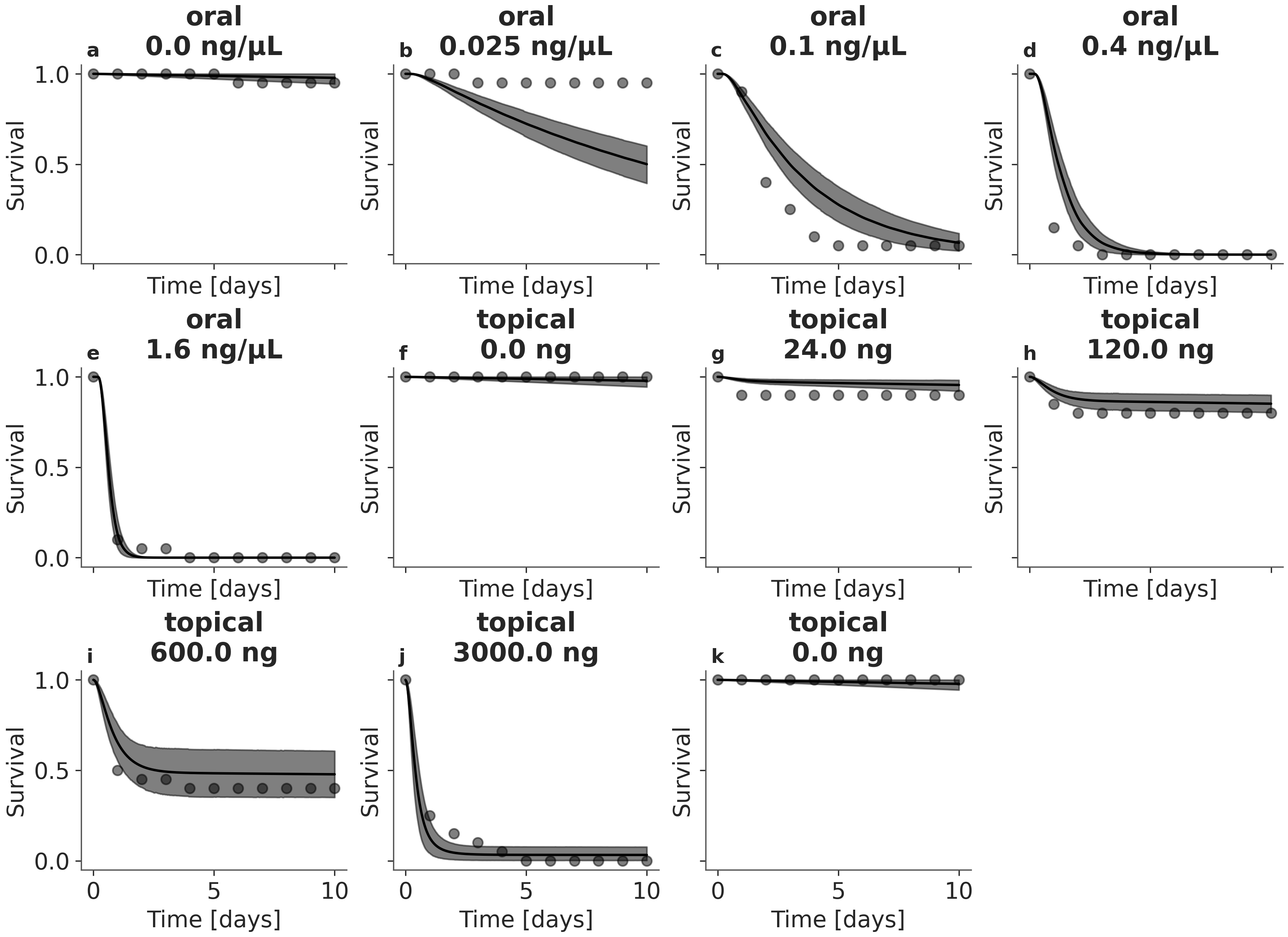}
    \caption{BufferGUTS joint model fits for \textit{O. bicornis} under sulfoxaflor exposure. Dots represent observations, solid lines represent the estimated survival probability and shaded areas indicate the uncertainty in the survival probability. \textbf{a--e}: Oral exposure. \textbf{f--k}: Topical exposure. Topical control treatments were set up with exposure to the surfactant Triton X-100 (\textbf{f}) and without exposure (\textbf{k}).}
    \label{fig:res-osmia-fits}
\end{figure}

The BufferGUTS model was fitted with high precision to the data (NRMSE 0.07) and estimated the dominant rate constant for both exposure routes (Fig.~\ref{fig:res-osmia-fits}). 
Notably the 0.025 ng/µL oral exposure treatment cannot be captured by the BufferGUTS model, indicating slight differences in the GUTS parameters between oral and topical exposure that cannot be compensated by the weight parameter. 
The parameters were estimated with high certainty, except for the threshold parameter $m$ (Tab.~\ref{tab:parameters--tktd-osmia--sulfoxaflor-ca-v01}).

\begin{table}[htb]
\centering
\caption{Maximum a posteriori (MAP) parameter estimates and 94\% highest density interval (HDI) of the posterior distributions of the BufferGUTS model calibrated to survival data after oral and topical exposure to sulfoxaflor. Parameters were converted from calibration units (oral: ng/$\mu$l, topical: ng, timescale: day) to work on the exposure units on which the ALMaSS toxicity module operates (oral: mg sulfoxaflor/mg nectar, topical: mg sulfoxaflor, timescale: hours). The ALMaSS units of the weights $w$, assert that the exposure unit is rescaled to the appropriate model compartment unit of ng, which is used during calibration (Explanation: Sec.~\ref{almass-unit-conversion}).}
\label{tab:parameters--tktd-osmia--sulfoxaflor-ca-v01}
\small

\begin{tabular}{llllll}
\toprule
Parameter & MAP orig. (94\% HDI) & Unit  & Conv. factor &   MAP conv. & ALMaSS unit \\
\midrule
$b$ & 0.002 (0.001--0.003) & 1/d/$\text{ng}_{a.i.}$ & 1/24 & 0.000083 & $1/\text{h}/\text{ng}_{a.i.}$ \\
$h_b$ & 0.001 (4.7e-5--0.005) & 1/d & 1/24 & 0.000042 & $1/\text{h}$ \\
$k_d$ & 1.91 (1.3--2.9) & 1/d & 1/24 & 0.0796 & $1/\text{h}$ \\
$m$ & 0.091 (0.0002--3.5) & $\text{ng}_{a.i.}$ & 1 & 0.091 & $\text{ng}_{a.i.}$ \\
$w_{oral}$ & 1560 (850--2300) & $\mu l_{nectar}$ & 1.12e+6 & 1.75e+9 & $\text{mg}_{nectar} \cdot \text{ng}_{a.i.}/\text{mg}_{a.i.}$ \\
$w_{topical}$ & 1 & - & 1e+6 & 1e+6 & $\text{ng}_{a.i.}^{-1} / \text{mg}_{a.i.}$ \\
\bottomrule
\end{tabular}
\end{table}

\subsection{Extrapolation of pollinator dynamics under pesticide stress with ALMaSS} \label{sec:casestudy}

\begin{figure}[htb] 
    \centering
    \begin{subfigure}{\textwidth}
         \centering
         \includegraphics[width=\linewidth]{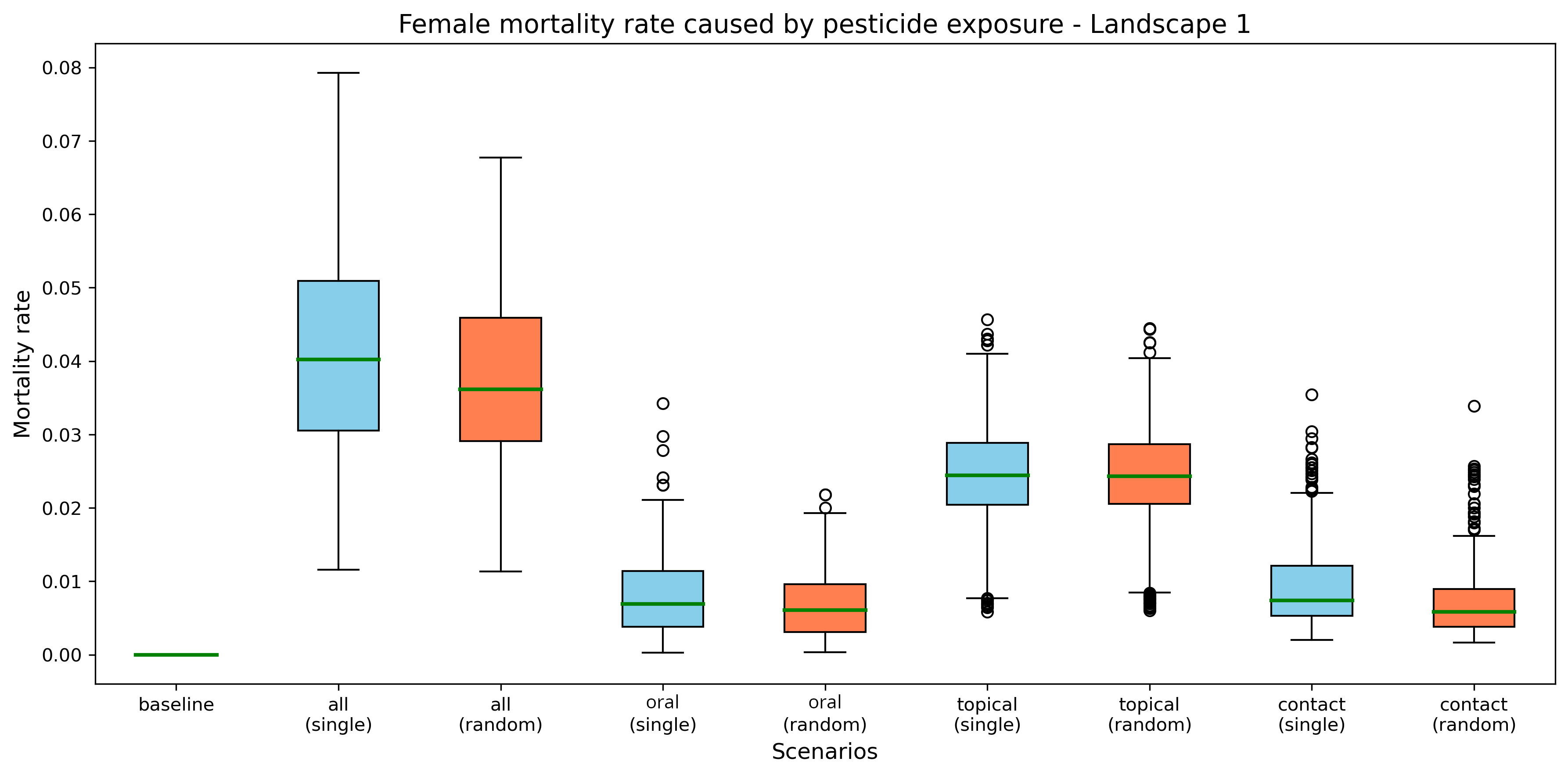}
    \end{subfigure}
    \begin{subfigure}{\textwidth}
         \centering
         \includegraphics[width=\linewidth]{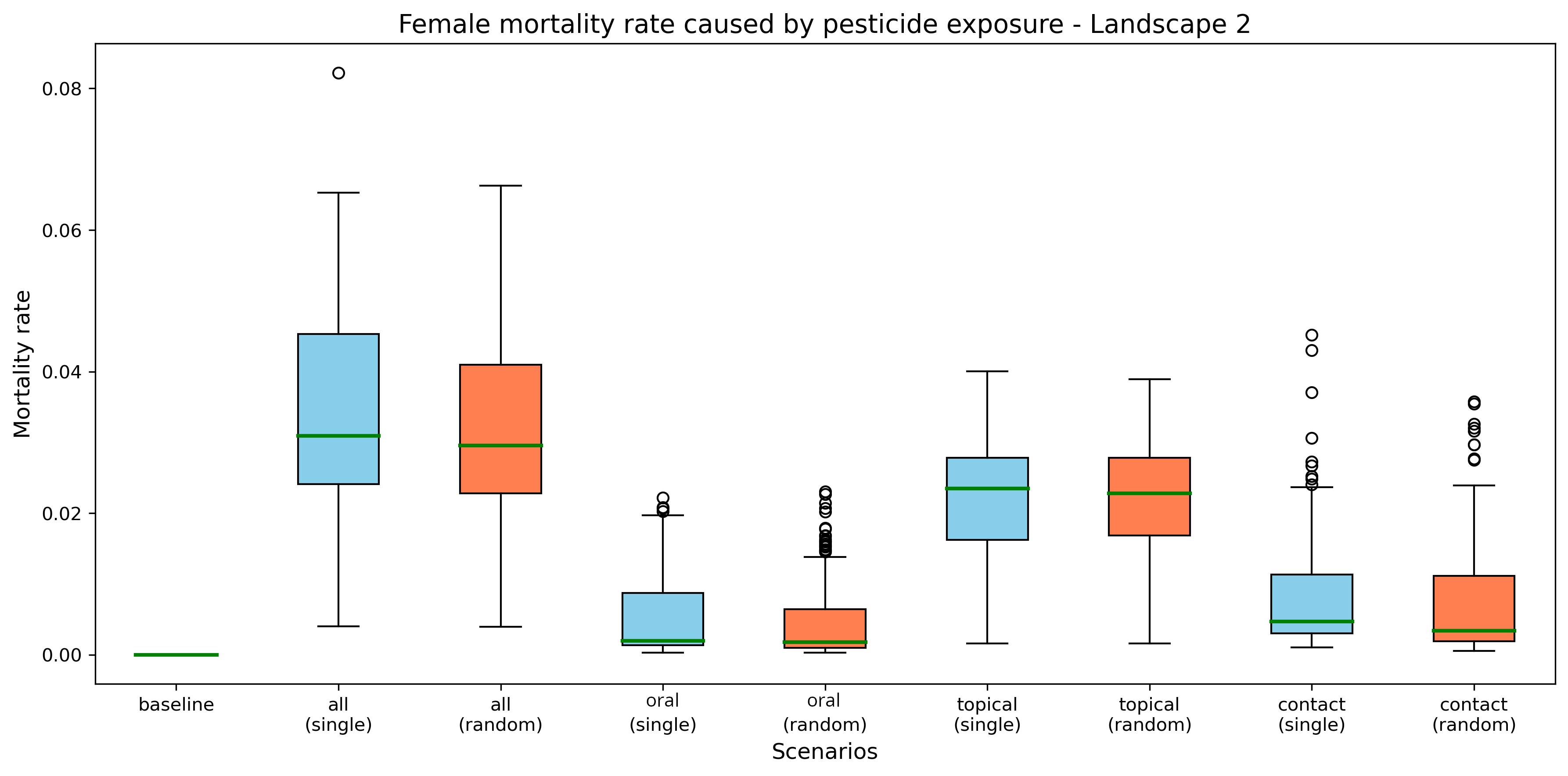}
    \end{subfigure}
    \caption{Yearly female mortality rate (female deaths per year caused by pesticide exposure divided by the total number of emerged females over the year) caused by pesticide exposure on Landscape 1 (73\% arable fields) and 2 (52\% arable fields) for the five adult exposure scenarios. Single: Only the maximum-a-posteriori GUTS parameter estimate was used to assess the population level impacts of pesticide exposure. Random: Random parameter estimates from the posterior distributions were used to assess the impacts of pesticide exposure. The baseline scenario shows zero mortality as no pesticides were applied. Box plots show the median (green line), interquartile range (box), range (whiskers), and outliers (circles) of mortality rates across 30 replications over the last 10 years of simulation.}
    \label{fig:mortality_pesticide}
\end{figure}

We calculated the mortality rate as female deaths per year caused by pesticide exposure divided by the total number of emerged females over the year, as shown in Fig.~\ref{fig:mortality_pesticide}. Mortality rates varied depending on the exposure pathway and the landscape. The "all exposure paths" scenario exhibited the highest mortality rates, with median values of approximately $0.04$ for both single and random parameter settings on both landscapes. Individual exposure pathways showed considerably lower mortality rates: oral intake scenarios had median mortality rates of approximately $0.007$--$0.02$, while topical and contact exposure scenarios showed median rates of approximately $0.02$--$0.03$. This result is surprising as the calibrated toxicity weights of oral exposure was 850-2300 times higher than topical exposure. A possible explanation for this is that effective exposure to sulfoxaflor in nectar is much lower than overspray exposure, leading to a relatively higher effect of the topical exposure paths. Regardless of these differences in mortality rates, the population-level impacts remained negligible, as evidenced by the stable population sizes across all scenarios.

The whisker plots of the accumulated yearly newborn female population for the two landscapes are presented in Fig.\ref{fig:accumulated_females}. The results demonstrate that female population sizes remained relatively stable across all scenarios, with no evident reduction compared to the baseline. Landscape 2, which has less arable land coverage ($52\%$), supported higher female populations (median $\sim$320,000) than Landscape 1 with $73\%$ arable coverage (median $\sim$210,000). Notably, neither the choice of exposure pathway nor the parameter setting (single vs. random) resulted in substantial population-level effects, suggesting that the tested pesticide exposure levels do not significantly impact landscape population size when the pesticide response is excluded for the other life stages.

Comparison between single and random parameter settings revealed minimal differences in both population sizes and mortality rates, indicating that individual variation in BufferGUTS parameters does not substantially affect population outcomes under the tested exposure levels. Some scenarios exhibited outliers with higher mortality rates, particularly in the contact and topical exposure pathways, suggesting occasional extreme exposure events that, while affecting individual survival, did not translate into population-level declines.

\begin{figure}[htb]
\centering
\includegraphics[width=.5\linewidth]{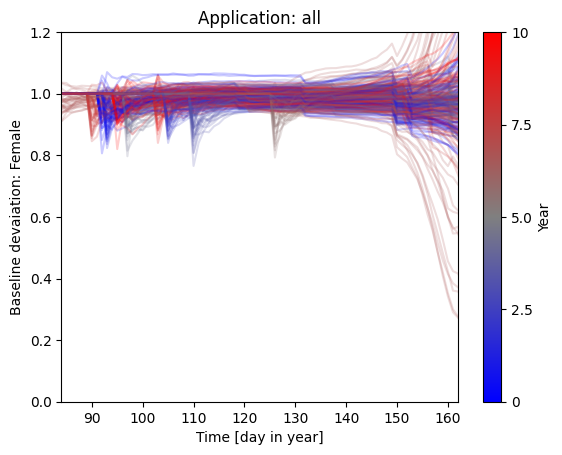}
\caption{Temporal dynamics of female population deviation from baseline on Landscape 2 for the "all exposure paths" scenario using random parameter sets. Each line represents one simulation year (coloured from blue to red for years 20--30), showing the ratio of female population to baseline across the active season (days 90--170). Values near 1.0 indicate no deviation from baseline, while lower values indicate population reduction.}
\label{fig:temporal_dynamics}
\end{figure}

To investigate the temporal dynamics underlying these annual aggregates, we examined the within-season population trajectories for Landscape 2 under the "all exposure paths" scenario. Fig.~\ref{fig:temporal_dynamics} compares the population dynamics between baseline and this test scenario across the active season of \textit{O. bicornis} (approximately April to June, corresponding to days 90--170). Most trajectories remained relatively smooth and close to 1.0 throughout the majority of the active season, indicating a minimal deviation from baseline during most years. However, a subset of trajectories exhibited population declines after day 150, with some dropping to approximately 0.4 of the baseline population. This decline is partly driven by overall low population numbers in late season. The frequency and severity of these declines showed a tendency to increase in later simulation years (indicated by redder colours), which could suggest either potential cumulative effects over time or the occurrence of occasional extreme exposure events during late-season pesticide applications.

\begin{figure}[htb]
    \centering
    \includegraphics[width=\linewidth]{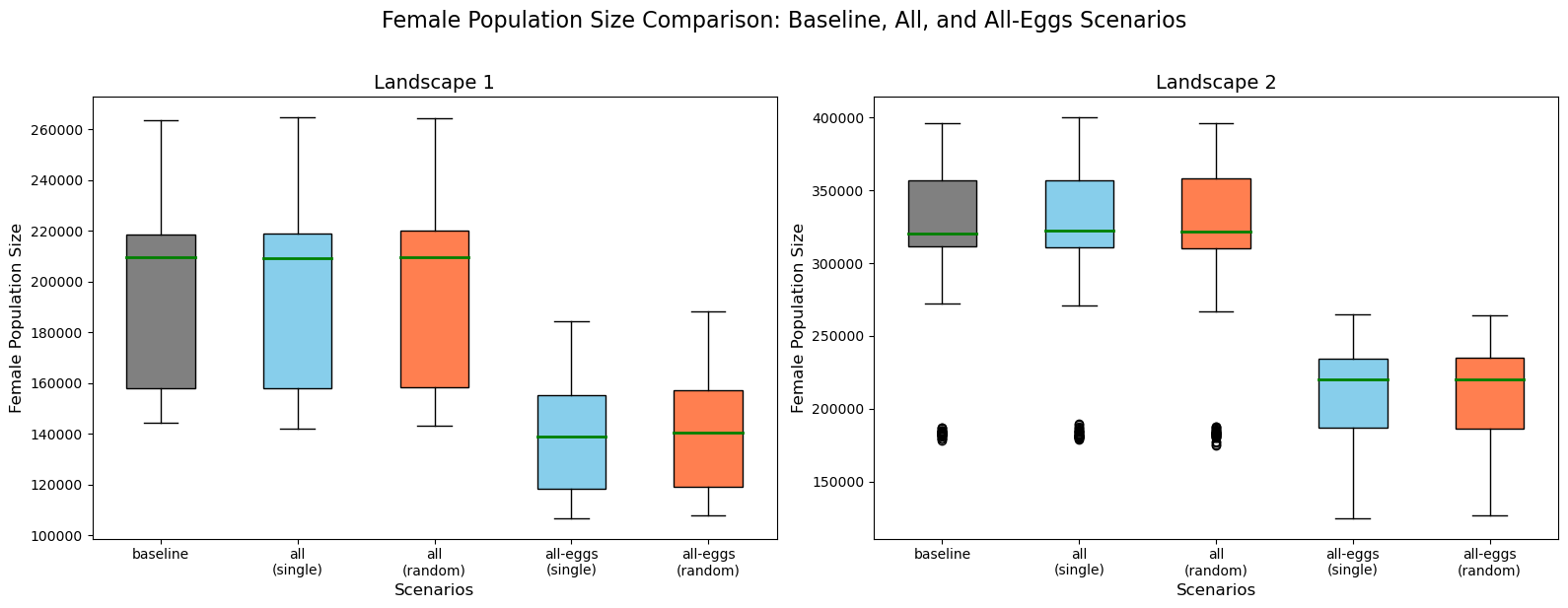}
    \caption{Female population size comparison between baseline, all exposure paths, and all-eggs scenarios on Landscape 1 and 2. The all-eggs scenarios include threshold-based egg mortality (mortality rate = 0.5) based on pesticide contamination in pollen provisions. Box plots show the median (green line), interquartile range (box), and range (whiskers) of female population across 30 replications over the last 10 years of simulation. Black dots indicate outliers.}
    \label{fig:all_eggs_comparison}
\end{figure}

The results of this all-eggs scenario are presented in Fig.~\ref{fig:all_eggs_comparison}, which compares the baseline, all exposure paths, and all-eggs scenarios. In stark contrast to the negligible population impacts observed when only adult females were exposed to pesticides, the inclusion of pre-reproductive mortality resulted in substantial population declines. On Landscape 1, the median female population size decreased from approximately 210,000 in the baseline and all scenarios to approximately 140,000 in the all-eggs scenarios, representing a reduction of approximately 33\%. Similarly, on Landscape 2, populations declined from approximately 320,000 to approximately 220,000, a reduction of approximately 31\%. These declines were consistent across both single and random parameter settings, with minimal difference between the two approaches. The all-eggs scenarios also exhibited occasional extreme outliers, particularly visible in Landscape 2, indicating rare events where populations crashed to exceptionally low levels. These results suggest that pre-reproductive life stages represent a critical vulnerability that can drive population-level impacts even when adult exposure alone does not.

\section{Discussion}

The TKTD framework for pollinator species models in ALMaSS, developed in this work, provides a consistent way to extrapolate chemical risks for pollinators from laboratory data to real landscapes, combining exposure and effect models. The BufferGUTS model handling the effect modelling of multiple pesticide exposure pathways was successfully integrated into the ALMaSS \textit{O. bicornis} model. This integration was used to derive daily survival probabilities for the modelled agents in a numerically efficient way, which is needed for large-scale individual-based simulations. While the TKTD modelling in the BufferGUTS model outlined in this report is limited to estimating only the dominant rate constant of the combined TK and TD processes, this limitation is bound to the lack of internal body residue data and can be  overcome if such data become available, because a general framework is in place to account for internal toxicant kinetics.
The case study's results demonstrate how risks of acute pesticide effects on non-target pollinators could be estimated in a landscape context. Although the application of $1~g/m^2$ sulfoxaflor in this case did not lead to a persistent reduction in population abundance, it showed that random variations in crop rotation, foraging activity, and mortality trials led to occasional population declines even if only acute lethality was considered. In addition to acute lethal effects, effects from chronic exposure and sublethal effects also contribute to the health of a population under chemical exposure. For example, repeated exposure to imidacloprid was found to reduce nesting probability, nesting rate, and proportion of females in the solitary bee \textit{Osmia lignaria} \cite{Stuligross.2021}.

The contrast between scenarios with and without pre-reproductive mortality suggests that pre-reproductive life stages can represent a critical vulnerability that may drive population-level impacts. When a threshold-based egg mortality was applied based on pesticide contamination in pollen provisions, populations declined by $31-33\%$ across both landscapes. This exploratory result could have profound implications for pollinator risk assessment, which has historically focused on adult exposure due to the availability of standardised testing protocols for adult bees.  A field study with artificial pollen dosing showed that exposure to pollen sulfoxaflor concentrations expected at a 400-fold field realistic application rate led to 20\% lethality of \textit{Osmia cornifrons} and an array of sublethal effects of four pesticides \cite{Phan.2024}, showing even stronger effects than this work. Also laboratory studies with \textit{O. bicornis} show that cypermethrin and chlorpyrifos affect, amongst sublethal effects, the survival probability for larvae \cite{Mokkapati.2021}, even at environmentally reaslistic concentrations.  Extending the TKTD framework in PollinERA to model explicit survival probabilities for larvae is feasible and should be considered in future work. To corroborate this early finding, case studies with environmental realistic pesticide applications need to be simulated and the calculated pollen loadings in nests need to be compared with residue measurements from field-collected provisions.

The disproportionate impact of larval mortality is probably due to several factors. First, developing larvae are continuously exposed to contaminated pollen provisions throughout their entire development period, whereas adults experience intermittent exposure during foraging. Second, larvae may have reduced detoxification capacity compared to adults \cite{Fine.2017,Maiwald.2023}. Third, and perhaps most importantly, mortality at the egg or larval stage prevents any reproductive output from affected individuals, whereas adult mortality may ocurr after successful reproduction and has therefore lower population consequences. This aligns with demographic theory showing that pre-reproductive survival has a stronger influence on population growth rate than survival in the reproductive period \cite{caswell2001matrix}.

\subsection{Limitations} \label{sec:limitations}

In this work, the BufferGUTS model was calibrated on exposure data from oral and topical exposure. Physiologically, this covers the internal uptake of the toxicant via the stomach and the external uptake via the cuticle. While these pathways probably comprise the most relevant (surface-related) uptake routes, there are a number of open questions.

The buffer-dynamics for pollinators is researched only to a low degree, yet it is the basis for the toxicity assessment in real-world exposure scenarios. This especially refers to accumulation dynamics in the buffers (i.e. gut and cuticle), which has not received sufficient attention. In contrast, degradation and distribution dynamics within the organisms have been investigated with radioactive $^{14}C$ labelling studies with \textit{Apis mellifera}, indicating that parent compounds quickly disappear from compartments associated with the ingestive system of the insect (head, thorax, abdomen). However, metabolites of the compounds are much longer, albeit at lower concentrations, present in the entire system \cite{Suchail.2004,Brunet.2005}. A similar fast removal of pesticides from the cuticle, associated with an increase in the whole organism concentration after topical exposure was observed for three tested neonicotinoids in a $^{14}C$ labelling study \cite{Zaworra.2019}, estimating a half-life on the buffer of approximately 24h. In all studies, conducted with \textit{A. mellifera}, oral and topical buffer decay dynamics follow exponential decay, which supports that buffer elimination is considered as a first-order kinetic process in the BufferGUTS approach. Although degradation and internal distribution dynamics have been studied, buffer accumulation dynamics from repeated exposures is not known. This may be related to the challenges of conducting repeated exposures (especially topical) and measuring residues in laboratory experiments. Recent work suggested improvements for the GUTS models for terrestrial organisms \cite{Jager.2026}. Future development of the BufferGUTS model should consider these improvements.   

While the pollen consumption of \textit{O. bicornis} is not well characterized, several pollinators consume pollen to complement their diet. Although there is a wealth of literature on quantifying pesticide residues in pollen and nectar \cite[e.g.]{Zioga.2020,Zioga.2023}, there is a pronounced lack of effect data for pollen exposure. Only one publication investigated co-exposure of nectar and pollen in \textit{O. bicornis} \cite{Azpiazu.2019} but did not investigate individual effects, which prohibits a comparison of the toxicities of food types. Within the current scope of the TKTD model implementation in the ALMaSS framework, we assume that pollen acts via the same exposure pathway and consequently has the same weight factor $w$ in the BufferGUTS model. The only differentiation is possible in the exposure. While we currently cannot assess whether the toxicity of pesticides in pollen is lower or higher than the toxicity of pesticides in nectar, we also ignore the beneficial effects of pollen consumption in the current framework.

In its current form, the BufferGUTS model fully depends on the availability of survival data from single-exposure experiments for each considered exposure path to predict effects of arbitrary multi-exposure-path scenarios. While this satisfies the needs for a pesticide risk assessment for arthropods, it depends on substantial animal testing for given species-substance combinations. Here, we propose ways forward to reduce animal testing and incorporate additional mechanistic information.

Currently neither the BufferGUTS model nor its integration into ALMaSS is fully validated. Additionally, calibration of the BufferGUTS model on sulfoxaflor laboratory data showed that the threshold parameter is unidentifiable despite the availability of a NOEC treatment indicating that toxicity thresholds between oral and topical exposure do not exactly converge. This highlights the need for validation studies and further development of the BufferGUTS model.

We strongly recommend the development of a TKTD/GUTS parameter database which could be used in the training of parameter estimators for toxicity predictions. Empirical toxicity estimators in the form of QSARs are rapidly gaining attention \cite{Benfenati.2019,Zubrod.2024}, but QSARs or general machine learning algorithms are notoriously data hungry. This conflicts with the scarcity of toxicity data for non-Apis pollinator species. We hypothesize that developing QSARs for TKTD parameters may profit from the implicit bio-physical relationships that are intrinsic to parameters of mechanistic models. 


\section{Conclusion}

This work demonstrates that toxicokinetic-toxicodynamic models can be successfully integrated into spatially explicit agent-based population models to bridge the gap between controlled laboratory experiments and realistic landscape-scale risk assessments. While in this case, TK processes could not be explicitly modelled, due to limitations in available data, the extension to full TKTD models is straightforward in the implemented framework. The BufferGUTS framework provides a computationally efficient and mathematically rigorous approach to aggregate multiple pesticide exposure pathways while capturing dominant biological processes through reduced mechanistic representations of uptake, elimination, and damage dynamics. 

However, several knowledge gaps require urgent attention. The disproportionate impact of larval mortality highlights the critical need to develop standardised toxicity testing protocols for immature life stages across multiple pollinator species. Current parametrisation relies entirely on adult survival data, yet our simulation results indicate that pre-reproductive survival may exert stronger effects on population growth rates. Additionally, direct measurements of buffer dynamics and internal concentration dynamics through laboratory studies would strengthen confidence in model predictions by differentiating between physiological/environmental processes, toxicokinetic processes and toxicodynamic responses, enabling more reliable extrapolation across species and compounds. Finally, sub-lethal and chronic effects should be considered for a rigorous assessment of chemical risks for pollinators.

In the end, this work demonstrates that protecting pollinator populations requires moving beyond adult mortality endpoints to a comprehensive assessment across all life stages, particularly those occurring before reproduction. The BufferGUTS-ALMaSS framework provides both the theoretical foundation and practical computational tools to achieve this goal, enabling science-based risk assessment that accounts for landscape complexity, temporal dynamics of exposure, and population-level consequences of individual-level effects. As regulatory frameworks evolve toward more mechanistic and ecologically relevant approaches, such integrated modelling platforms will become increasingly essential for making informed decisions that balance agricultural productivity with biodiversity conservation.

\section*{Acknowledgements}
We thank Ryszard Laskowski and Sergio Albacete for their insightful comments. 
The study was conducted in the PollinERA project funded by the European Union (Horizon Europe 101135005). 
Computations on the HPC of Osnabrück University were funded by the Deutsche Forschungsgemeinschaft (DFG, German Research Foundation) - 456666331.

\nolinenumbers



\printglossary[type=\acronymtype]
\printbibliography

@article{Ashauer.2016,
  author = {Ashauer, Roman and Albert, Carlo and Augustine, Starrlight
    and Cedergreen, Nina and Charles, Sandrine and Ducrot, Virginie and
    Focks, Andreas and Gabsi, Faten and Gergs, André and Goussen, Benoit
    and Jager, Tjalling and Kramer, Nynke I. and Nyman, Anna-Maija and
    Poulsen, Veronique and Reichenberger, Stefan and Schäfer, Ralf B.
    and Van Den Brink, Paul J. and Veltman, Karin and Vogel, Sören and
    Zimmer, Elke I. and Preuss, Thomas G.},
  title = {Modelling Survival: {Exposure} Pattern, Species Sensitivity
    and Uncertainty},
  journal = {Scientific Reports},
  volume = {6},
  number = {1},
  pages = {29178},
  date = {2016-07-06},
  year = {2016},
  urldate = {2024-02-28},
  url = {https://www.nature.com/articles/srep29178},
  doi = {10.1038/srep29178},
  issn = {2045-2322},
  langid = {en-US}
}

@article{Bart.2022,
  author = {Bart, Sylvain and Short, Stephen and Jager, Tjalling and
    Eagles, Emily J. and Robinson, Alex and Badder, Claire and Lahive,
    Elma and Spurgeon, David J. and Ashauer, Roman},
  title = {How to Analyse and Account for Interactions in Mixture
    Toxicity with Toxicokinetic-Toxicodynamic Models},
  journal = {The Science of the total environment},
  volume = {843},
  pages = {157048},
  date = {2022-01-01},
  year = {2022},
  doi = {10.1016/j.scitotenv.2022.157048},
  issn = {0048-9697}
}

@article{Cedergreen.2017,
  author = {Cedergreen, Nina and Dalhoff, Kristoffer and Li, Dan and
    Gottardi, Michele and Kretschmann, Andreas C.},
  title = {Can {{Toxicokinetic}} and {{Toxicodynamic} {Modeling} {Be}
    {Used}} to {{Understand}} and {{Predict} {Synergistic}
    {Interactions}} Between {{Chemicals}}?},
  journal = {Environmental science \& technology},
  volume = {51},
  number = {24},
  pages = {14379-14389},
  date = {2017-01-01},
  year = {2017},
  doi = {10.1021/acs.est.7b02723}
}

@article{Dalhoff.2020,
  author = {Dalhoff, Kristoffer and Hansen, Anna M. B. and Rasmussen,
    Jes J. and Focks, Andreas and Strobel, Bjarne W. and Cedergreen,
    Nina},
  title = {Linking {{Morphology}}, {{Toxicokinetic}}, and
    {{Toxicodynamic} {Traits}} of {{Aquatic} {Invertebrates}} to
    {{Pyrethroid} {Sensitivity}}},
  journal = {Environmental science \& technology},
  volume = {54},
  number = {9},
  pages = {5687-5699},
  date = {2020-01-01},
  year = {2020},
  doi = {10.1021/acs.est.0c00189}
}

@article{Jager.2011,
  author = {Jager, Tjalling and Albert, Carlo and Preuss, Thomas G. and
    Ashauer, Roman},
  title = {General Unified Threshold Model of Survival–a
    Toxicokinetic-Toxicodynamic Framework for Ecotoxicology},
  journal = {Environmental science \& technology},
  volume = {45},
  number = {7},
  pages = {2529-40},
  year = {2011},
  date = {2011-01-01},
  doi = {10.1021/es103092a}
}

@article{Singer.2023,
  author = {Singer, Alexander and Nickisch, Dirk and Gergs, André},
  title = {Joint Survival Modelling for Multiple Species Exposed to
    Toxicants},
  journal = {Science of The Total Environment},
  volume = {857},
  pages = {159266},
  date = {2023-01},
  urldate = {2024-03-27},
  year = {2023},
  url = {https://linkinghub.elsevier.com/retrieve/pii/S0048969722063653},
  doi = {10.1016/j.scitotenv.2022.159266},
  issn = {00489697},
  langid = {en-US}
}

@book{Jager.2018a,
  title = {Modelling {{Survival}} under {{Chemical Stress}}: {{A}} Comprehensive Guide to the {{GUTS}} Framework},
  author = {Jager, Tjalling and Ashauer, Roman},
  year = {2018},
  month = jan,
  edition = {2},
  publisher = {Leanpub},
  isbn = {978-1-9999705-1-2}
}

@article{topping2013modelling,
  title={Modelling skylarks (Alauda arvensis) to predict impacts of changes in land management and policy: development and testing of an agent-based model},
  author={Topping, Christopher J and Oddersk{\ae}r, Peter and Kahlert, Johnny},
  journal={PLoS One},
  volume={8},
  number={6},
  pages={e65803},
  year={2013},
  publisher={Public Library of Science San Francisco, USA}
}

@article{topping2024managing,
  title={Managing large and complex population operations with agent-based models: The ALMaSS Population\_Manager},
  author={Topping, Christopher John and Duan, Xiaodong},
  journal={Food and Ecological Systems Modelling Journal},
  volume={5},
  pages={e117593},
  year={2024},
  publisher={Pensoft Publishers}
}

@article{topping2012spatial,
  title={Spatial dynamic factors affecting population-level risk assessment for a terrestrial arthropod: an agent-based modeling approach},
  author={Topping, Chris J and Lagisz, Malgorzata},
  journal={Human and Ecological Risk Assessment: An International Journal},
  volume={18},
  number={1},
  pages={168--180},
  year={2012},
  publisher={Taylor \& Francis}
}

@article{ziolkowska2023formal,
  title={The Formal Model for the solitary bee Osmia bicornis L. agent-based model},
  author={Zi{\'o}{\l}kowska, El{\.z}bieta and Bednarska, Agnieszka and Laskowski, Ryszard and Topping, Christopher J},
  journal={Food and Ecological Systems Modelling Journal},
  volume={4},
  year={2023}
}

@article{duan2022apis,
  title={Apis RAM Formal Model Description},
  author={Duan, Xiaodong and Wallis, David and Hatjina, Fani and Simon-Delso, Noa and Bruun Jensen, Annette and Topping, Christopher John},
  journal={EFSA Supporting Publications},
  volume={19},
  number={2},
  pages={7184E},
  year={2022},
  publisher={Wiley Online Library}
}

@article{bonmatin2015environmental,
  title={Environmental fate and exposure; neonicotinoids and fipronil},
  author={Bonmatin, J-M and Giorio, Chiara and Girolami, Vincenzo and Goulson, D and Kreutzweiser, DP and Krupke, C and Liess, M and Long, E and Marzaro, Matteo and Mitchell, EAD and others},
  journal={Environmental science and pollution research},
  volume={22},
  pages={35--67},
  year={2015},
  publisher={Springer}
}

@article{rundlof2015seed,
  title={Seed coating with a neonicotinoid insecticide negatively affects wild bees},
  author={Rundl{\"o}f, Maj and Andersson, Georg KS and Bommarco, Riccardo and Fries, Ingemar and Hederstr{\"o}m, Veronica and Herbertsson, Lina and Jonsson, Ove and Klatt, Bj{\"o}rn K and Pedersen, Thorsten R and Yourstone, Johanna and others},
  journal={Nature},
  volume={521},
  number={7550},
  pages={77--80},
  year={2015},
  publisher={Nature Publishing Group UK London}
}

@article{Zaworra.2019,
  title = {Pharmacokinetics of {{Three Neonicotinoid Insecticides}} upon {{Contact Exposure}} in the {{Western Honey Bee}}, {{{\emph{Apis}}}}{\emph{ Mellifera}}},
  author = {Zaworra, Marion and Koehler, Harald and Schneider, Josef and Lagojda, Andreas and Nauen, Ralf},
  year = {2019},
  month = jan,
  journal = {Chemical Research in Toxicology},
  volume = {32},
  number = {1},
  pages = {35--37},
  issn = {0893-228X, 1520-5010},
  doi = {10.1021/acs.chemrestox.8b00315},
  urldate = {2024-06-24}
}

@book{Miller.1998,
  title = {Survival Analysis},
  author = {Miller, Rupert G. and Gong, Gail and Mu{\~n}oz, Alvaro},
  year = {1998},
  series = {Wiley Series in Probability and Mathematical Statistics},
  edition = {Wiley classics library ed (Online-Ausg.)},
  publisher = {Wiley-Interscience},
  address = {New York},
  isbn = {978-1-118-03106-3 978-0-471-25548-2 978-0-471-09434-0}
}

@book{OQuigley.2021,
  title = {Survival {{Analysis}}: {{Proportional}} and {{Non-Proportional Hazards Regression}}},
  shorttitle = {Survival {{Analysis}}},
  author = {O'Quigley, John},
  year = {2021},
  publisher = {Springer International Publishing},
  address = {Cham},
  doi = {10.1007/978-3-030-33439-0},
  urldate = {2024-11-27},
  copyright = {https://www.springer.com/tdm},
  isbn = {978-3-030-33438-3 978-3-030-33439-0}
}

@article{Wood.2017,
  title = {The Environmental Risks of Neonicotinoid Pesticides: A Review of the Evidence Post 2013},
  author = {Wood, Thomas James and Goulson, Dave},
  year = {2017},
  month = jan,
  journal = {Environmental science and pollution research international},
  volume = {24},
  number = {21},
  pages = {17285--17325},
  doi = {10.1007/s11356-017-9240-x}
}

@article{Hallmann.2017,
  title = {More than 75 Percent Decline over 27 Years in Total Flying Insect Biomass in Protected Areas},
  author = {Hallmann, Caspar A. and Sorg, Martin and Jongejans, Eelke and Siepel, Henk and Hofland, Nick and Schwan, Heinz and Stenmans, Werner and M{\"u}ller, Andreas and Sumser, Hubert and H{\"o}rren, Thomas and Goulson, Dave and Kroon, Hans},
  year = {2017},
  month = jan,
  journal = {PloS one},
  volume = {12},
  number = {10},
  pages = {0185809},
  doi = {10.1371/journal.pone.0185809}
}

@article{Wagner.2020,
  title = {Insect {{Declines}} in the {{Anthropocene}}},
  author = {Wagner, David L.},
  year = {2020},
  month = jan,
  journal = {Annual review of entomology},
  volume = {65},
  pages = {457--480},
  issn = {0066-4170},
  doi = {10.1146/annurev-ento-011019-025151}
}

@article{Crossley.2020,
  title = {No Net Insect Abundance and Diversity Declines across {{US Long Term Ecological Research}} Sites},
  author = {Crossley, Michael S. and Meier, Amanda R. and Baldwin, Emily M. and Berry, Lauren L. and Crenshaw, Leah C. and Hartman, Glen L. and {Lagos-Kutz}, Doris and Nichols, David H. and Patel, Krishna and Varriano, Sofia and Snyder, William E. and Moran, Matthew D.},
  year = {2020},
  month = jan,
  journal = {Nature ecology \& evolution},
  volume = {4},
  number = {10},
  pages = {1368--1376},
  doi = {10.1038/s41559-020-1269-4}
}

@article{Dornelas.2019,
  title = {A Balance of Winners and Losers in the {{Anthropocene}}},
  author = {Dornelas, Maria and Gotelli, Nicholas J. and Shimadzu, Hideyasu and Moyes, Faye and Magurran, Anne E. and McGill, Brian J.},
  year = {2019},
  month = jan,
  journal = {Ecology letters},
  volume = {22},
  number = {5},
  pages = {847--854},
  doi = {10.1111/ele.13242}
}

@article{AXELMAN2024174526,
title = {A systems-based analysis to rethink the European environmental risk assessment of regulated chemicals using pesticides as a pilot case},
journal = {Science of The Total Environment},
volume = {948},
pages = {174526},
year = {2024},
issn = {0048-9697},
doi = {https://doi.org/10.1016/j.scitotenv.2024.174526},
url = {https://www.sciencedirect.com/science/article/pii/S0048969724046746},
author = {Johan Axelman and Annette Aldrich and Sabine Duquesne and Thomas Backhaus and Stephan Brendel and Andreas Focks and Sheila Holz and Saskia Knillmann and Silvia Pieper and Emilia Silva and Maria Schmied-Tobies and Christopher John Topping and Louise Wipfler and James Williams and José Paulo Sousa}
}

@article{Topping2024Landscape,
	author = {Christopher John Topping and Xiaodong Duan},
	title = {ALMaSS Landscape and Farming Simulation: software classes and methods},
	volume = {5},
	number = {},
	year = {2024},
	doi = {10.3897/fmj.5.121215},
	publisher = {Pensoft Publishers},
	issn = {},
	pages = {e121215},
	URL = {https://doi.org/10.3897/fmj.5.121215},
	eprint = {https://doi.org/10.3897/fmj.5.121215},
	journal = {Food and Ecological Systems Modelling Journal}
}

@article{Duan2024subpopulation,
	author = {Xiaodong Duan and Christopher J. Topping},
	title = {A general subpopulation model for the Animal Landscape and Man Simulation System (ALMaSS)},
	volume = {5},
	number = {},
	year = {2024},
	doi = {10.3897/fmj.5.122467},
	publisher = {Pensoft Publishers},
	issn = {},
	pages = {e122467},
	URL = {https://doi.org/10.3897/fmj.5.122467},
	eprint = {https://doi.org/10.3897/fmj.5.122467},
	journal = {Food and Ecological Systems Modelling Journal}
}

@article{Poulsen2023Pesticide,
	author = {Trine Poulsen and Xiaodong Duan and Christopher J. Topping},
	title = {Modelling dynamic pesticide amounts in multiple environmental compartments at landscape scales in ALMaSS},
	volume = {4},
	number = {},
	year = {2023},
	doi = {10.3897/fmj.4.107849},
	publisher = {Pensoft Publishers},
	issn = {},
	pages = {e107849},
	URL = {https://doi.org/10.3897/fmj.4.107849},
	eprint = {https://doi.org/10.3897/fmj.4.107849},
	journal = {Food and Ecological Systems Modelling Journal}
}

@misc{Hoffman.2011,
    title = {The {No}-{U}-{Turn} {Sampler}: {Adaptively} {Setting} {Path} {Lengths} in {Hamiltonian} {Monte} {Carlo}},
    shorttitle = {The {No}-{U}-{Turn} {Sampler}},
    url = {http://arxiv.org/abs/1111.4246},
    language = {en},
    urldate = {2023-12-14},
    publisher = {arXiv},
    author = {Hoffman, Matthew D. and Gelman, Andrew},
    month = nov,
    year = {2011},
    note = {arXiv:1111.4246 [cs, stat]}
}

@article{Azpiazu.2019,
  title = {Chronic Oral Exposure to Field-Realistic Pesticide Combinations via Pollen and Nectar: Effects on Feeding and Thermal Performance in a Solitary Bee},
  shorttitle = {Chronic Oral Exposure to Field-Realistic Pesticide Combinations via Pollen and Nectar},
  author = {Azpiazu, Celeste and Bosch, Jordi and Vi{\~n}uela, Elisa and Medrzycki, Piotr and Teper, Dariusz and Sgolastra, Fabio},
  year = {2019},
  month = sep,
  journal = {Scientific Reports},
  volume = {9},
  number = {1},
  pages = {13770},
  issn = {2045-2322},
  doi = {10.1038/s41598-019-50255-4},
  urldate = {2025-06-06}
}

@article{Zioga.2020,
  title = {Plant Protection Product Residues in Plant Pollen and Nectar: {{A}} Review of Current Knowledge},
  shorttitle = {Plant Protection Product Residues in Plant Pollen and Nectar},
  author = {Zioga, Elena and Kelly, Ruth and White, Bl{\'a}naid and Stout, Jane C.},
  year = {2020},
  month = oct,
  journal = {Environmental Research},
  volume = {189},
  pages = {109873},
  issn = {00139351},
  doi = {10.1016/j.envres.2020.109873},
  urldate = {2025-06-06}
}

@article{Zioga.2023,
  title = {Pesticide Mixtures Detected in Crop and Non-Target Wild Plant Pollen and Nectar},
  author = {Zioga, Elena and White, Bl{\'a}naid and Stout, Jane C.},
  year = {2023},
  month = jun,
  journal = {Science of The Total Environment},
  volume = {879},
  pages = {162971},
  issn = {00489697},
  doi = {10.1016/j.scitotenv.2023.162971},
  urldate = {2025-06-06}
}

@article{Burger.2025,
    title = {From water to land—{Usage} of {Generalized} {Unified} {Threshold} models of {Survival} ({GUTS}) in an above-ground terrestrial context exemplified by honeybee survival data},
    volume = {44},
    copyright = {https://creativecommons.org/licenses/by/4.0/},
    issn = {0730-7268, 1552-8618},
    url = {https://academic.oup.com/etc/article/44/2/589/7943034},
    doi = {10.1093/etojnl/vgae058},
    language = {en},
    number = {2},
    urldate = {2025-04-11},
    journal = {Environmental Toxicology and Chemistry},
    author = {Bürger, Leonhard Urs and Focks, Andreas},
    month = feb,
    year = {2025},
    pages = {589--598}
}

@article{Benfenati.2019,
    title = {Integrating in silico models and read-across methods for predicting toxicity of chemicals: {A} step-wise strategy},
    volume = {131},
    doi = {10.1016/j.envint.2019.105060},
    journal = {Environment international},
    author = {Benfenati, Emilio and Chaudhry, Qasim and Gini, Giuseppina and Dorne, Jean Lou},
    month = jan,
    year = {2019},
    pmid = {31377600},
    note = {Num Pages: 15},
    pages = {105060}
}

@article{Zubrod.2024,
    title = {Bio-{QSARs} 2.0: {Unlocking} a new level of predictive power for machine learning-based ecotoxicity predictions by exploiting chemical and biological information},
    volume = {186},
    issn = {01604120},
    shorttitle = {Bio-{QSARs} 2.0},
    url = {https://linkinghub.elsevier.com/retrieve/pii/S0160412024001934},
    doi = {10.1016/j.envint.2024.108607},
    language = {en},
    urldate = {2024-05-05},
    journal = {Environment International},
    author = {Zubrod, Jochen P. and Galic, Nika and Vaugeois, Maxime and Dreier, David A.},
    month = apr,
    year = {2024},
    pages = {108607}
}

@article{Suchail.2004,
  title = {{\emph{In Vivo}} Distribution and Metabolisation Of{\textsuperscript{14}} {{C}}-imidacloprid in Different Compartments of {{{\emph{Apis}}}}{\emph{ Mellifera}} {{L}}},
  author = {Suchail, S{\'e}verine and De Sousa, Georges and Rahmani, Roger and Belzunces, Luc P},
  year = {2004},
  month = nov,
  journal = {Pest Management Science},
  volume = {60},
  number = {11},
  pages = {1056--1062},
  issn = {1526-498X, 1526-4998},
  doi = {10.1002/ps.895},
  urldate = {2025-06-20}
}

@article{Brunet.2005,
  title = {{\emph{In Vivo}} Metabolic Fate of [{\textsuperscript{14}} {{C}}]-acetamiprid in Six Biological Compartments of the Honeybee, {{{\emph{Apis}}}}{\emph{ Mellifera}} {{L}}},
  author = {Brunet, Jean-Luc and Badiou, Alexandra and Belzunces, Luc P},
  year = {2005},
  month = aug,
  journal = {Pest Management Science},
  volume = {61},
  number = {8},
  pages = {742--748},
  issn = {1526-498X, 1526-4998},
  doi = {10.1002/ps.1046},
  urldate = {2025-06-20}
}

@article{Stuligross.2021,
  title = {Past Insecticide Exposure Reduces Bee Reproduction and Population Growth Rate},
  author = {Stuligross, Clara and Williams, Neal M.},
  year = 2021,
  month = jan,
  journal = {Proceedings of the National Academy of Sciences of the United States of America},
  volume = {118},
  number = {48},
  issn = {0027-8424},
  doi = {10.1073/pnas.2109909118}
}

@book{caswell2001matrix,
  title={Matrix population models},
  author={Caswell, Hal},
  year={2001},
  publisher={Sinauer Associates Sunderland, MA}
}

@article{europeanfoodsafetyauthorityConclusionPeerReview2014,
  title = {Conclusion on the Peer Review of the Pesticide Risk Assessment of the Active Substance Sulfoxaflor},
  author = {{European Food Safety Authority}},
  year = 2014,
  month = may,
  journal = {EFSA Journal},
  volume = {12},
  number = {5},
  issn = {18314732, 18314732},
  doi = {10.2903/j.efsa.2014.3692},
  urldate = {2025-11-04}
}

@article{seibold.2019,
  title = {Arthropod Decline in Grasslands and Forests Is Associated with Landscape-Level Drivers},
  author = {Seibold, Sebastian and Gossner, Martin M. and Simons, Nadja K. and Bl{\"u}thgen, Nico and M{\"u}ller, J{\"o}rg and Ambarl{\i}, Didem and Ammer, Christian and Bauhus, J{\"u}rgen and Fischer, Markus and Habel, Jan C. and Linsenmair, Karl Eduard and Nauss, Thomas and Penone, Caterina and Prati, Daniel and Schall, Peter and Schulze, Ernst-Detlef and Vogt, Juliane and W{\"o}llauer, Stephan and Weisser, Wolfgang W.},
  year = 2019,
  month = oct,
  journal = {Nature},
  volume = {574},
  number = {7780},
  pages = {671--674},
  issn = {0028-0836, 1476-4687},
  doi = {10.1038/s41586-019-1684-3},
  urldate = {2025-11-27},
  langid = {english}
}

@article{Burger.2025b,
  title = {Combined {{Modeling}} of {{Multiple Exposure Routes}} for {{Terrestrial Arthropods Using}} the {{Toxicokinetic}}--{{Toxicodynamic BufferGUTS Model}}},
  author = {B{\"u}rger, Leonhard U. and Schunck, Florian and Focks, Andreas},
  year = 2025,
  month = jul,
  journal = {Environmental Science \& Technology},
  pages = {acs.est.5c03925},
  issn = {0013-936X, 1520-5851},
  doi = {10.1021/acs.est.5c03925},
  urldate = {2025-07-31},
  copyright = {https://creativecommons.org/licenses/by/4.0/}
}

@thesis{Zbrozek.2022,
  title = {Interactive Effects of Two Commonly Used Insecticides on Survival of the Red Mason Bee (\textit{{{Osmia}} bicornis})},
  author = {Zbrozek, Maryellen},
  year = 2022,
  address = {Krak\'ow, Poland},
  school = {Institute of Environmental Sciences, Biology Faculty, Jagiellonian University}
}

@article{Misiewicz.2025,
  title = {Combined Effects of Commercial Insecticides on Survival of the Red Mason Bee {{Osmia}} Bicornis},
  author = {Misiewicz, Anna and Zbrozek, Maryellen and Laskowski, Ryszard and Bednarska, Agnieszka J.},
  year = 2025,
  month = oct,
  journal = {Ecotoxicology and Environmental Safety},
  volume = {304},
  pages = {119023},
  issn = {01476513},
  doi = {10.1016/j.ecoenv.2025.119023},
  urldate = {2025-11-27}
}

@article{Phan.2024,
  title = {Systemic Pesticides in a Solitary Bee Pollen Food Store Affect Larval Development and Increase Pupal Mortality},
  author = {Phan, Ngoc T. and Joshi, Neelendra K. and Rajotte, Edwin G. and Zhu, Fang and Peter, Kari A. and {L{\'o}pez-Uribe}, Margarita M. and Biddinger, David J.},
  year = 2024,
  month = mar,
  journal = {Science of The Total Environment},
  volume = {915},
  pages = {170048},
  issn = {00489697},
  doi = {10.1016/j.scitotenv.2024.170048},
  urldate = {2025-11-28}
}

@article{Mokkapati.2021,
  title = {The Development of the Solitary Bee {{Osmia}} Bicornis Is Affected by Some Insecticide Agrochemicals at Environmentally Relevant Concentrations},
  author = {Mokkapati, Jaya Sravanthi and Bednarska, Agnieszka J. and Laskowski, Ryszard},
  year = 2021,
  month = jun,
  journal = {Science of The Total Environment},
  volume = {775},
  pages = {145588},
  issn = {00489697},
  doi = {10.1016/j.scitotenv.2021.145588},
  urldate = {2025-11-27}
}

@article{Potts.2016,
  title = {Safeguarding Pollinators and Their Values to Human Well-Being},
  author = {Potts, Simon G. and {Imperatriz-Fonseca}, Vera and Ngo, Hien T. and Aizen, Marcelo A. and Biesmeijer, Jacobus C. and Breeze, Thomas D. and Dicks, Lynn V. and Garibaldi, Lucas A. and Hill, Rosemary and Settele, Josef and Vanbergen, Adam J.},
  year = 2016,
  month = dec,
  journal = {Nature},
  volume = {540},
  number = {7632},
  pages = {220--229},
  issn = {0028-0836, 1476-4687},
  doi = {10.1038/nature20588},
  urldate = {2026-02-17}
}

@article{Ashauer.2010a,
  title = {Toxicokinetic--Toxicodynamic Modelling in an Individual Based Context---{{Consequences}} of Parameter Variability},
  author = {Ashauer, Roman},
  year = 2010,
  month = may,
  journal = {Ecological Modelling},
  volume = {221},
  number = {9},
  pages = {1325--1328},
  issn = {03043800},
  doi = {10.1016/j.ecolmodel.2010.01.015},
  urldate = {2026-02-17}
}

@article{Goulson.2015,
  title = {Bee Declines Driven by Combined Stress from Parasites, Pesticides, and Lack of Flowers},
  author = {Goulson, Dave and Nicholls, Elizabeth and Bot{\'i}as, Cristina and Rotheray, Ellen L.},
  year = 2015,
  month = mar,
  journal = {Science},
  volume = {347},
  number = {6229},
  pages = {1255957},
  issn = {0036-8075, 1095-9203},
  doi = {10.1126/science.1255957},
  urldate = {2026-02-17}
}

@article{Baas.2022,
  title = {{{BeeGUTS}}---{{A Toxicokinetic}}--{{Toxicodynamic Model}} for the {{Interpretation}} and {{Integration}} of {{Acute}} and {{Chronic Honey Bee Tests}}},
  author = {Baas, Jan and Goussen, Benoit and Miles, Mark and Preuss, Thomas G. and Roessink, Ivo},
  year = 2022,
  month = sep,
  journal = {Environmental Toxicology and Chemistry},
  volume = {41},
  number = {9},
  pages = {2193--2201},
  issn = {0730-7268, 1552-8618},
  doi = {10.1002/etc.5423},
  urldate = {2024-03-19}
}

@article{Schmolke.2023,
  title = {{{SolBeePop}}: {{A}} Model of Solitary Bee Populations in Agricultural Landscapes},
  shorttitle = {{{SolBeePop}}},
  author = {Schmolke, Amelie and Galic, Nika and Hinarejos, Silvia},
  year = 2023,
  month = dec,
  journal = {Journal of Applied Ecology},
  volume = {60},
  number = {12},
  pages = {2573--2585},
  issn = {0021-8901, 1365-2664},
  doi = {10.1111/1365-2664.14541},
  urldate = {2026-02-23}
}

@article{Schmolke.2024,
  title = {{{SolBeePop}} {\emph{Ecotox}} : {{A Population Model}} for {{Pesticide Risk Assessments}} of {{Solitary Bees}}},
  shorttitle = {{{SolBeePop}} {\emph{Ecotox}}},
  author = {Schmolke, Amelie and Galic, Nika and Roeben, Vanessa and Preuss, Thomas G. and Miles, Mark and Hinarejos, Silvia},
  year = 2024,
  month = dec,
  journal = {Environmental Toxicology and Chemistry},
  volume = {43},
  number = {12},
  pages = {2645--2661},
  issn = {0730-7268, 1552-8618},
  doi = {10.1002/etc.5990},
  urldate = {2026-02-23},
  copyright = {https://academic.oup.com/pages/standard-publication-reuse-rights}
}

@article{Focks.2018,
  title = {Calibration and Validation of Toxicokinetic-Toxicodynamic Models for Three Neonicotinoids and Some Aquatic Macroinvertebrates},
  author = {Focks, Andreas and Belgers, Dick and Boerwinkel, Marie-Claire and Buijse, Laura and Roessink, Ivo and {van den Brink}, Paul J.},
  year = 2018,
  month = jan,
  journal = {Ecotoxicology (London, England)},
  volume = {27},
  number = {7},
  pages = {992--1007},
  doi = {10.1007/s10646-018-1940-6}
}

@article{Ashauer.2010,
  title = {Advantages of Toxicokinetic and Toxicodynamic Modelling in Aquatic Ecotoxicology and Risk Assessment},
  author = {Ashauer, Roman and Escher, Beate I.},
  year = 2010,
  month = jan,
  journal = {Journal of environmental monitoring : JEM},
  volume = {12},
  number = {11},
  pages = {2056--61},
  issn = {1464-0325},
  doi = {10.1039/c0em00234h}
}

@article{Becher.2013,
  title = {{{REVIEW}}: {{Towards}} a Systems Approach for Understanding Honeybee Decline: A Stocktaking and Synthesis of Existing Models},
  shorttitle = {{{REVIEW}}},
  author = {Becher, Matthias A. and Osborne, Juliet L. and Thorbek, Pernille and Kennedy, Peter J. and Grimm, Volker},
  editor = {Steffan-Dewenter, Ingolf},
  year = 2013,
  month = aug,
  journal = {Journal of Applied Ecology},
  volume = {50},
  number = {4},
  pages = {868--880},
  issn = {0021-8901, 1365-2664},
  doi = {10.1111/1365-2664.12112},
  urldate = {2026-02-24},
  copyright = {http://creativecommons.org/licenses/by/3.0/}
}

@article{Becher.2018,
  title = {{\emph{Bumble}} - {{{\textsc{BEEHAVE}}}} : {{A}} Systems Model for Exploring Multifactorial Causes of Bumblebee Decline at Individual, Colony, Population and Community Level},
  shorttitle = {{\emph{Bumble}} -},
  author = {Becher, Matthias A. and Twiston-Davies, Grace and Penny, Tim D. and Goulson, Dave and Rotheray, Ellen L. and Osborne, Juliet L.},
  editor = {Beggs, Jacqueline},
  year = 2018,
  month = nov,
  journal = {Journal of Applied Ecology},
  volume = {55},
  number = {6},
  pages = {2790--2801},
  issn = {0021-8901, 1365-2664},
  doi = {10.1111/1365-2664.13165},
  urldate = {2026-02-24}
}

@article{Topping.2025a,
  title = {The {{Formal Model}} for the Butterfly {{Pieris}} Napi ({{Lepidoptera}}, {{Pieridae}}) Agent-Based Model in the {{Animal Landscape}} and {{Man Simulation System}} ({{ALMaSS}})},
  author = {Topping, Christopher John and Duan, Xiaodong},
  year = 2025,
  month = feb,
  journal = {Food and Ecological Systems Modelling Journal},
  volume = {6},
  pages = {e142802},
  issn = {2815-3197},
  doi = {10.3897/fmj.6.142802},
  urldate = {2026-02-25},
  copyright = {http://creativecommons.org/licenses/by/4.0/}
}

@article{Topping.2025,
  title = {The {{Formal Model}} for {{Noctua}} Pronuba ({{Lepidoptera}}, {{Noctuidae}}) Representing a Typical Agricultural-Landscape Night-Flying Moth Species for Pesticide Risk Assessment},
  author = {Topping, Christopher John and Pettersson, Lars B.},
  year = 2025,
  month = dec,
  journal = {Food and Ecological Systems Modelling Journal},
  volume = {6},
  pages = {e165232},
  issn = {2815-3197},
  doi = {10.3897/fmj.6.165232},
  urldate = {2026-02-25},
  copyright = {http://creativecommons.org/licenses/by/4.0/}
}

@article{Ziolkowska.2025,
  title = {The {{Formal Model}} for the Hoverfly {{Eristalis}} Tenax ({{Diptera}}, {{Syrphidae}}) Agent-Based Model in the {{Animal Landscape}} and {{Man Simulation System}} ({{ALMaSS}})},
  author = {Zi{\'o}{\l}kowska, El{\.z}bieta and Walczy{\'n}ska, Aleksandra and Topping, Christopher John},
  year = 2025,
  month = jul,
  journal = {Food and Ecological Systems Modelling Journal},
  volume = {6},
  pages = {e152847},
  issn = {2815-3197},
  doi = {10.3897/fmj.6.152847},
  urldate = {2026-02-25},
  copyright = {http://creativecommons.org/licenses/by/4.0/}
}

@article{Jager.2026,
  title = {Extending {{GUTS}} to Account for Oral Exposure in Bees: {{Preconditions}}, Model Options, and Evaluation of Published Approaches},
  shorttitle = {Extending {{GUTS}} to Account for Oral Exposure in Bees},
  author = {Jager, Tjalling},
  year = 2026,
  month = jul,
  journal = {Ecological Modelling},
  volume = {517},
  pages = {111614},
  issn = {03043800},
  doi = {10.1016/j.ecolmodel.2026.111614},
  urldate = {2026-05-22}
}

@article{Fine.2017,
  title = {Metabolism of {{{\emph{N}}}} -{{Methyl-2-Pyrrolidone}} in {{Honey Bee Adults}} and {{Larvae}}: {{Exploring Age Related Differences}} in {{Toxic Effects}}},
  shorttitle = {Metabolism of {{{\emph{N}}}} -{{Methyl-2-Pyrrolidone}} in {{Honey Bee Adults}} and {{Larvae}}},
  author = {Fine, Julia D. and Mullin, Christopher A.},
  year = 2017,
  month = oct,
  journal = {Environmental Science \& Technology},
  volume = {51},
  number = {19},
  pages = {11412--11422},
  issn = {0013-936X, 1520-5851},
  doi = {10.1021/acs.est.7b03291},
  urldate = {2026-06-22}
}

@article{Maiwald.2023,
  title = {Expression Profile of the Entire Detoxification Gene Inventory of the Western Honeybee, {{Apis}} Mellifera across Life Stages},
  author = {Maiwald, Frank and Haas, Julian and Hertlein, Gillian and Lueke, Bettina and Roesner, Janin and Nauen, Ralf},
  year = 2023,
  month = may,
  journal = {Pesticide Biochemistry and Physiology},
  volume = {192},
  pages = {105410},
  issn = {00483575},
  doi = {10.1016/j.pestbp.2023.105410},
  urldate = {2026-06-22}
}

\pagebreak

\appendix

\section*{Supporting information}\label{supporting-information}

\setcounter{section}{0}
\renewcommand{\thesection}{S\arabic{section}}
\setcounter{subsection}{0}
\renewcommand{\thesubsection}{S\arabic{section}.\arabic{subsection}}
\setcounter{subsubsection}{0}
\renewcommand{\thesubsubsection}{S\arabic{section}.\arabic{subsection}.\arabic{subsubsection}}
\setcounter{equation}{0}
\renewcommand{\theequation}{Eq. S\arabic{equation}}
\setcounter{figure}{0}
\setcounter{table}{0}
\renewcommand{\thefigure}{S\arabic{figure}}
\renewcommand{\thetable}{S\arabic{table}}

\section{Model Description---A Framework for integrating TKTD models into ALMaSS} \label{si:sec:model-description}

The buffers in the BufferGUTS models are the interfaces through which a given organism exchanges molecules with the physical environment. Although it would theoretically be possible to develop elaborate uptake and removal processes from the exposed surfaces of the modelled agents, reducing the operational complexity of such models requires some simplifying assumptions for calculating the effective exposure concentrations and doses.

\subsection{Modelling multiple exposure paths}

In the following, the exposure pathways that are modelled in ALMaSS and are subsequently fed into the BufferGUTS model (Eq.~\ref{eq_buffer_guts}) are briefly outlined. Note that environmental exposures are denoted as $C_i$, where $C$ may be the concentration or dose (depending on the path), and $i$ is the pathway. 
Although the standard time step of ALMaSS is one day, we have decided to increase the time resolution to hourly time steps to cover short topical exposure events, and consider the species day/night activity for correctly assessing the accumulation of a pesticide in the buffers. Thus $\Delta t = 1h$.

\subsubsection{Oral exposure pathway} \label{sssec_intake}
Oral exposure occurs when a model pollinator adult consumes a specific mass of resource ($R$), such as nectar or pollen, with a pesticide concentration of $C_{O}$ ($mg/mg$). In ALMaSS, pesticide concentrations ($C_{O}$) in pollen and nectar are expressed as the mass of the pesticide residue divided by the mass of the corresponding flower resource. This exposure concentration will be the oral input quantity for the buffer. 

\subsubsection{Contact exposure pathway} \label{sec:exposure-surface-contact}
Contact exposure is defined exclusively as the interaction between a foraging adult pollinator and a contaminated plant surface. Other contaminated surfaces, e.g., soil, are not included yet but can be added if it is important for the model species. In ALMaSS, the pesticide surface concentration on the plant surface ($mg/m^2$) multiplied by the half body surface is the pesticide mass ($mg$) represented by $C_S$ which will be integrated by the buffer, which means that for contact exposure the buffer takes the absolute pesticide mass as input. If stacking of buffer residues should be considered, the intercepted pesticide mass needs to be divided by the duration of the exposure (Sec.~\ref{sec:buffer-stacking}).

\subsubsection{Topical exposure pathway} \label{sec:exposure-topical-contact} 

Topical exposure occurs exclusively when a model pollinator adult is actively spending time in a field (e.g., collecting nectar and pollen, searching for a nesting site) where simultaneous pesticide spraying is taking place and/or where drift is occurring in the adjacent area. Consequently, topical exposure is treated as a mass transfer, which occurs at a constant rate during the exposure spray application. The duration of exposure for a given pollinator adult on a given day is set to be one hour. The one-hour exposure duration represents a conservative yet realistic assumption for topical exposure during pesticide application. It only happens when there is an application in the field where the adults are foraging. The chance of occurrence is set to $1/24$. Applications may occur at any time during daylight hours (or occasionally at night for certain products), and this timing is not predictable within the daily time-step framework. The 1/24 probability acknowledges that whilst an application occurs on a given day, the specific hour is effectively random from the model's perspective. In ALMaSS, the pesticide application rate ($mg/m^2$) is multiplied by the interception surface, approximated by the half body surface ($cm^2$), and the intercepted pesticide mass $C_T$ ($mg$) is another input of the buffer. If stacking of buffer residues should be considered, the intercepted pesticide mass needs to be divided by the duration of the exposure (Sec.~\ref{sec:buffer-stacking}), which has no effect for the (applied) special case of 1h exposure durations.

\subsection{The BufferGUTS model for multiple exposure pathways} \label{si:sec:buffer}

\subsubsection{Buffer dynamics}

The BufferGUTS model has been extensively described in previous work \cite{Burger.2025,Burger.2025b}. In Brief:
Conceptually, the BufferGUTS model introduces a separate buffer for each exposure pathway (e.g. the stomach for the oral pathway and the exoskeleton or cuticle for the topical pathway). These buffers can be seen as a portable environment that the agent carries along. In contrast to internal compartments such as organs, hemolymph, etc., buffers are not separated from the environment by membranes. Therefore physiological (feeding) or environmental (deposition) processes govern the input dynamics. Therefore, in discrete exposure events, the chemical concentration in a buffer for a given exposure pathway almost immediately approaches the environmental concentration. When exposure stops or is reduced, the concentration of the buffer decays exponentially with a rate constant $k_e$, which in turn depends on the physicochemical properties.

Mathematically the buffer concept is expressed as: 

\begin{align}\label{eq_buffer_guts}
    \frac{dB}{dt} = \begin{cases}
        \eta \cdot (C - B) & B \leq C \\
        k_e \cdot (C - B) & B > C
    \end{cases}
\end{align}

$\eta$ expresses the buffer-filling speed, and is set to very large values, so the buffer is almost instantly filled. This constant could be informed by prior knowledge about the exposure pathway, which depends on exposure and species. $k_e$ expresses the elimination of the substance from the buffer. The BufferGUTS model assumes that elimination from the buffer ($k_e$) and the dominant rate constant for damage and toxicant dynamics ($k_d$) are directly connected, that means $k_e = k_d$ \cite{Burger.2025}. This assumption of proportionality between the elimination dynamics of the substance in the buffer and the accrual dynamics of the scaled damage and concentration is supported to some extent by the observation of inverse correlations between the outer chemical residues and the inner chemical concentrations of three chemicals, using $^{14}C$ labelling experiments \cite{Zaworra.2019}. The main benefit of this simplification is the reduction of parameters to a number that can be estimated by survival data alone. This limitation is further discussed in Sec.~\ref{sec:limitations}. From here on, $k_d$ is used to denote the buffer decay rate constant. For each exposure pathway, there is a separate buffer that is filled and depleted independently. For simplicity, all buffers share the same elimination rate constant $k_e=k_d$ for the same chemical. This corresponds to the BufferGUTS concentration addition (CA) variant.

\subsubsection{Aggregation of exposure pathways}

Although exposure to pesticides in the terrestrial environment occurs via multiple pathways, it is assumed that the mode of action will be equivalent for all pathways \cite{Burger.2025}. If this assumption is correct, the remaining uncertainty lies in the contribution of the different pathways to the effective concentration at the target sites. Such differences arise from different diffusion coefficients through exposed surfaces such as the stomach, guts, different parts of the exoskeleton, and openings for glands or gas exchange to the target sites.

In order to account for these physiological and behavioural differences between exposure pathways, the contribution to the aggregated buffer $B_\Sigma$ is the weighted sum of all buffers (intake $B_O$, contact $B_c$, and overspray $B_T$). 

\begin{equation} \label{eq_pesticide_new}
    B_\Sigma(t) = \sum_i w_i B_i = w_I~B_O(t) + w_C~B_C(t) + w_T~B_T(t)
\end{equation}

The weights $w$ need to be fitted from experimental data. They are determined by fixing the weight of a selected buffer at 1 and calibrating the other two from the data. This procedure is necessary to avoid parameter identifiability issues. While it does not matter in principle which buffer is selected, two considerations guide its choice: 

\begin{enumerate}
    \item The unit of the fixed buffer is also the unit of the aggregated buffer.
    \item Because the buffer will enter the sum unmodified, it should be the buffer that is most accurately described and matches the reality best. 
\end{enumerate}

The approach requires experimental data for oral, contact and topical exposure paths. If such data are unavailable, either further assumptions to rescale the buffers before summing have to be taken, which heavily rely on physiological traits. 

Once the different exposure pathways have been aggregated into a single buffer $B_\Sigma$, this quantity can be considered as the effective dose metric for TKTD models. Starting from this point, the TKTD and GUTS models can be applied in a manner similar to the methods applied in aquatic risk assessment.

\subsubsection{Damage dynamics}

Because toxicity data for terrestrial organisms are often limited, the damage dynamics of the reduced GUTS model is used \cite{Jager.2018a}.

\begin{align} \label{eq_tktd_reduced_guts}
    \frac{dD}{dt} &= k_d~(B_\Sigma(t) - D(t)) ~,
\end{align}

where $B_\Sigma$ is the summed buffer concentration and $D$ is the toxicodynamic damage scaled to have the same unit as the buffer $B_\Sigma$. Eq.~\ref{eq_tktd_reduced_guts} is typically referred to as the reduced GUTS (TKTD) model, and it can estimate the combined dynamics of residue and damage with a single dominant rate constant $k_d$, which describes the combined speed of toxicokinetics and damage dynamics \cite{Jager.2018a}. 

\subsubsection{Hazard dynamics: Threshold model of pesticide risk}

As a survival model, we choose the stochastic death model \cite{Jager.2011}, which is rooted in survival analysis \cite{Miller.1998,OQuigley.2021}. The mathematical framework of survival analysis allows for an extremely simple and probabilistically logical way of combining independent hazards $S(t)=exp(-\sum_i H_i(t))$, which is suitable for the ALMaSS framework. 
It assumes that we have incomplete knowledge of the exact processes that lead to the binary outcome \textit{dead or alive}, and that this uncertainty is appropriately expressed by modelling lethality as a random process. Nevertheless, individual differences in pesticide tolerance are accounted for by parametrising modelled agents from the posterior threshold parameter distribution, as outlined in Sec.~\ref{sec:casestudy}.

In survival analysis, the hazard rate from pesticides represents the instantaneous probability of death for a given organism. In addition to the pesticide hazard rate, the pollinators can also experience a background hazard related to the pollinator's age. The differential equation for hazard from a stressor for a stochastic death process within the GUTS framework is:

\begin{equation}\label{eq_guts_hazard}
    \frac{dH}{dt} = b ~ max(D(t)-z, 0)
\end{equation}

Note that the background hazard $h_b$ is not explicitly listed in this equation, because it contributes \textbf{independently} to the hazard in the standard threshold model, and can therefore be added later on. The same goes for other hazard rates, which can be simply added if they are assumed to contribute independently to the survival probability.

\subsection{Discretization} \label{si:sec:discretization}

\subsubsection{Buffer dynamics}

The buffer equations can be exactly solved for one time step $\Delta t$ from $t-1$ to $t$ by using analytic solutions for first order kinetic processes $e^{-kt}$. Because the buffer guts equation considers kinetics between a gradient of two concentrations, this results in:

\begin{align}\label{eq_buffer_guts_discrete}
    B_i(t) = \begin{cases}
        C_{i}(t-\Delta t) & B_{i}(t-\Delta t) \leq C_{i}(t-\Delta t) \\
        C_{i}(t-\Delta t) + (B_{i}(t-\Delta t) - C_{i}(t-\Delta t))~e^{-k_e\Delta t} & B_{i}(t-\Delta t) > C_{i}(t-\Delta t)
    \end{cases}
\end{align}

where $i$ is the respective environmental concentration or buffer. Eq.~\ref{eq_buffer_guts_discrete}, assumes that $\eta$ is very large and therefore assumes that uptake is immediate if the concentration is larger than the buffer, it is also assumed that degradation can be neglected during buffer filling. 
An expanded form of the equation is available in Annex~\ref{si:buffer_solution}.

\textbf{It should be noted that} $C_i(t-\Delta t)$ denotes the constant concentration over the entire time step $[t-\Delta t, t]$. If $C_i$ is not considered constant in one time step, Eq.~\ref{eq_buffer_guts_discrete} is not valid and needs to be adapted for the change in $C_i$ within one time step.

\subsubsection{Damage dynamics}

The discretisation of the damage equation is more complex, because it involves 2 first-order kinetic processes in sequence. Although it is again assumed that $C$ remains constant during the time step, $B$ changes throughout the time step if the buffer concentration is smaller than the external concentration. Therefore, the second term of Eq.~\ref{eq_damage_guts_discrete_change_b} adds a second term to Eq.~\ref{eq_damage_guts_discrete} that accounts for the change in B if B changes throughout the time step, i.e. $B > C$. 

For a single buffer, the exact solution can be easily written down because we only have 2 conditional branches due to the buffer conditions $C \geq B$ and $C < B$, but for multiple buffers there is a combinatorially increasing number of solutions with $2^n$ where $n$ is the number of buffers. In addition, for buffers where $C_i < B_i$, i.e., decay is assumed to be the dominant process, an additional second-order term is necessary to correct for the reduction in the buffer during the time step $\Delta t$.

\begin{align}\label{eq_damage_guts_discrete}
    D(t) &= C_\Sigma(t-\Delta t) + (D(t-\Delta t) - C_{\Sigma}(t-\Delta t))~e^{-k_e~\Delta t} \\ 
    &+ \sum_i{
    \begin{cases}
        0 & B_i(t-\Delta t) \leq C_i(t-\Delta t) \\
        k_e~\Delta t~(w_i B_i(t-\Delta t) - w_i C_i(t-\Delta t))~e^{-k_e~\Delta t} & B_i(t-\Delta t) > C_i(t-\Delta t)
    \end{cases}
    }\label{eq_damage_guts_discrete_change_b}
\end{align}

where $C_{\Sigma,~t-1}$ is the weighted sum of external concentrations, where the same weights are applied as in $B_\Sigma$. Again, $C_{t-1}$ refers to the external concentration that is assumed to be constant throughout the entire time step. The resulting solution is an exact match of the continuous time equations (Fig.~\ref{fig:discretization_bufferguts}g). An expanded set of equations is available in Annex~\ref{si:damage_solution}.


\subsubsection{Hazard dynamics: Threshold model of pesticide risk}

Equation \ref{eq_guts_hazard} can be solved analytically up until the point where $D(t) = z$. At this point, the integral changes and an intermediate step has to be taken, resulting in:

\begin{align}\label{eq:hazard_discrete}
    H(t, y, \theta) &= H(t-\Delta t) + h_b \Delta t + \begin{cases}
        0 & \text{if } D(t-\Delta t) - z \leq 0~\And~D(t) - z \leq 0\\
        \bar{H}(t, y(t-\Delta t, \theta)) & \text{if } D(t-\Delta t) - z > 0~\And~D(t) - z > 0\\
        \bar{H}(t^*, y(t-\Delta t), \theta)& \text{if } D(t-\Delta t) - z > 0~\And~D(t) - z \leq 0\\
        \bar{H}(t-t^*, y(t^*), \theta) &\text{if } D(t-\Delta t) - z \leq 0~\And~D(t) - z > 0\\
    \end{cases}\\
    \label{eq:hazard_discrete_part}
    \bar{H}(t,y,\theta) &= b \Delta t \left(\sum_i w_i C_i(t-\Delta t) - z \right) + \frac{b}{k_d} \left(D(t-\Delta t)-\sum_i w_i C_i (t-\Delta t) \right)\left(1- e^{-k_d \Delta t}\right) \\
             &+ \sum_i{
            \begin{cases}
                0 & B_i(t-\Delta t) \leq C_i(t-\Delta t) \\
                \frac{w_i~b}{k_d}(B_i(t-\Delta t) - C_i(t-\Delta t))\left(1 - e^{-k_d \Delta t}\right)  & B_i(t-\Delta t) > C_i(t-\Delta t)
            \end{cases}
            }\\
             &+ \sum_i{
            \begin{cases}
                0 & B_i(t-\Delta t) \leq C_i(t-\Delta t) \\
                -w_i b \Delta t (B_i(t-\Delta t) - C_i(t-\Delta t)) e^{-k_d \Delta t}  & B_i(t-\Delta t) > C_i(t-\Delta t)
            \end{cases}
        }
\end{align}
\normalsize

With the state variables $y: (C_i,~B_i,~D,~H)$ and the parameters $\theta: (w_i,~k_d,~b,~z,~h_b)$. The time $t^*$ of this step can in some cases be solved analytically, but in the majority of cases it is much simpler and also computationally efficient to approximate the root of the equality numerically. The expanded form is provided in Annex~\ref{si:root_finding_d_eq_z}. After finding $t^*$, Eq.~\ref{eq:hazard_discrete_part} is solved until $t^*$ to estimate the hazard if $D$ falls below the threshold $z$ at $t^*$. Eq.~\ref{eq:hazard_discrete_part} is solved from $t^*$ to $t$ if $D$ rises above the threshold $z$ at $t^*$ with updated state variables $y_{t^*}$. Multiple transitions in one time step are not possible as long as $C$ is constant within the time step.

\subsection{Daily mortality chance} \label{si:sec:daily-mortality-chance}

\subsubsection{Relationship between survival, hazard and lethality density}

By definition, the probability of survival $S$ past a time $t$ of an organism with a lifetime $T$ is defined as follows:

\begin{equation} \label{eq_definition_survival}
S(t) = \Pr(T > t) = \int_t^{\infty}~f(t')dt' = 1 - F(t)
\end{equation}

\begin{figure}[htb]
    \centering
    \includegraphics[width=0.6\linewidth]{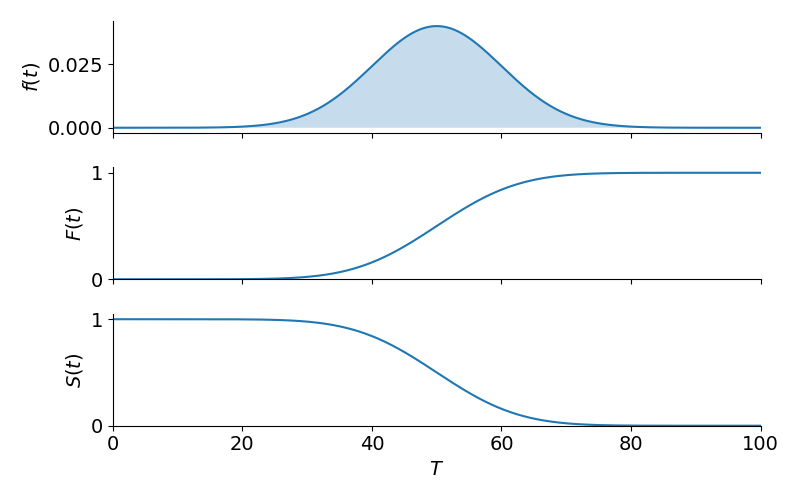}
    \caption{Relationship between the probability density function of lethality $f(t)$, the cumulative density function of lethality $F(t)$ and the complementary survival function $S(t) = 1- F(t).$}
    \label{fig:survival-relationship}
\end{figure}

$f(t)$ is the lethality density function. So, the probability density function of the random variable $T$ gives the probability of living until the time $t$ (or dying at the time $t$). This can be, of course, any distribution; in the example below, we assume $f(t)$ is a normal distribution.
$f(t)$ is an empirical distribution that has been learnt from the data. It integrates to the cumulative density function $L = F(t)$. It's inverse is the survival function $S(t) = 1 - F(t)$ (Fig.~\ref{fig:survival-relationship}).

It is important to note that $f(t)$ is not the hazard rate $h(t)$; these two terms are often confused. There is also no mechanistic information in the processes described up until this point. The hazard rate is also often written as $\lambda(t)$, however this work follows the $h(t)$ notation to refer to the hazard rate.

The hazard rate is defined as the probability to die in a very short interval after the time $t$ conditional on the organisms having survived until the time $t$ , and can be rewritten as a conditional probability.

\begin{equation}
h(t) = \lim_{\Delta t \to 0} \frac{\Pr(t < T \leq t + \Delta t~|~T > t)}{\Delta t} = \lim_{\Delta t \to 0} \frac{\Pr(t < T \leq t + \Delta t) \cap \Pr(T > t)}{\Pr(T > t) ~\Delta t}
\end{equation}

\begin{figure}[htb]
    \centering
    \includegraphics[width=0.6\linewidth]{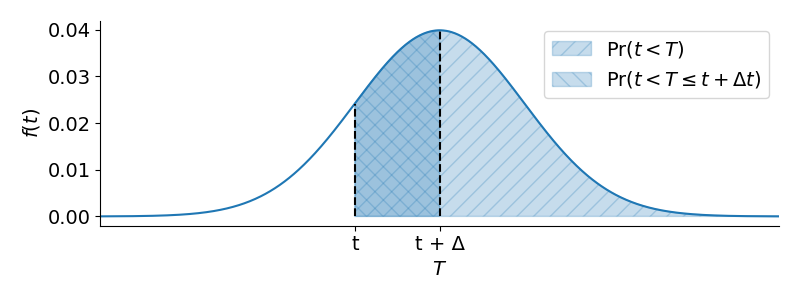}
    \caption{Graphical proof that the intersection between $\Pr(t < T \leq t + \Delta t)$ and $\Pr(t < T)$ is $\Pr(t < T \leq t + \Delta t)$}
    \label{fig:interval-survival-proof}
\end{figure}

The term $\Pr(t < T \leq t + \Delta t) \cap \Pr(T > t)$ is the intersection between both probabilities and by logical reasoning (Fig.~\ref{fig:interval-survival-proof}) the intersection between $\Pr(t < T \leq t + \Delta t)$ and $\Pr(T > t)$ is $\Pr(t < T \leq t + \Delta t)$.

Substituting $P(T > t) = S(t)$ in the denominator and rearranging the probability term in the numerator: $\Pr(t < T \leq t + \Delta t) = \Pr(T \leq t + \Delta t) - \Pr(t \leq T) = F(t + \Delta t) - F(t)$ leads to 

\begin{equation} \label{eq_hazard_rate}
 h(t) = \lim_{\Delta t \to 0} \frac{\Pr(t < T \leq t + \Delta t)}{\Pr(T > t)~\Delta t} =\lim_{\Delta t \to 0}~\frac{F(t + \Delta t) - F(t)}{S(t) ~\Delta t}  = \frac{f(t)}{S(t)} 
\end{equation}

This leads to the well known equality

\begin{equation} \label{eq_survival_cumhaz}
    S(t) = e^{-H(t)}
\end{equation}

\subsubsection{Deriving lethality probabilities for time intervals}

In order to accommodate time intervals, we simply omit the limit notation and division by the time interval from \ref{eq_hazard_rate} and assume that the conditional probability of dying in the interval $[t_1, t_2]$ is 

\begin{equation}
    p(t_1, t_2) = \Pr(t_1 < T \leq t_2 | T > t_1).
\end{equation}

$p(t_1, t_2)$ can again be rewritten as a conditional probability in the terms $F$ and $S$, where $\Delta = t_2 - t_1$.

\begin{equation} \label{eq_prob_death_in_interval}
    p(t_1, t_2) = \frac{F(t_2) - F(t_1)}{S(t_1)}
\end{equation}

Due to Equation~\ref{eq_definition_survival}, we can substitute $F(t) = 1 - S(t)$, and due to equation~\ref{eq_survival_cumhaz}, we can substitute $S(t) = e^{-H(t)}$. Given that we know the hazard function $h(t)$, we have the probability of death in the interval $[t_1, t_2]$:

\begin{equation}
p(t_0, t_1) = \frac{(1- S(t_2)) - (1 - S(t_1))}{S(t_1)} = 1 - \frac{S(t_2)}{S(t_1)} = 1 - \frac{e^{-H(t_2)}}{e^{-H(t_1)}} = 1 - e^{H(t_1)-H(t_2)} 
\end{equation}

Because $H(t)$ is a monotonically increasing function of $t$, we know that $H(t_2) \geq H(t_1)$; this means that we can rewrite the exponent in the numerator.

\begin{equation}
H(t_1) - H(t_2) = - H(t_1, t_2) = - \int_{t_1}^{t_2} h(t')dt'
\end{equation}

Leading to 

\begin{equation} \label{si:eq_conditional_lethality}
\Pr(t_1 < T \leq t_2 | T > t_1) = p(t_1, t_2) = 1 - e^{-H(t_1, t_2)}
\end{equation}

which defines the probability of dying in an interval $[t_1, t_2]$, conditional on having lived until $t_1$. Equation~\ref{eq_conditional_lethality} can be used to calculate probabilities from cumulative hazards (integrated hazard rates over a time interval). These probabilities can be used to generate random draws from a Bernoulli, binomial or multinomial distribution to simulate survival in an agent-based framework. Eq.~\ref{eq_conditional_lethality} can also be applied directly in the conditional binomial probability model to estimate parameters from survival data generated with repeated observations, which is usually the case in ecotoxicological testing.

Finally Eq.~\ref{eq_conditional_lethality} perfectly integrates with the vitality concept implemented in the ALMaSS model to combine various stressors. For this, interval hazards simply need to be summed before inserting them into the survival function $\Pr(t_1 < T \leq t_2 | T > t_1) = \exp\left(- \sum H_i(t_1, t_2)\right)$

\subsubsection{Probability model and likelihood function for survival data} \label{si:sec:likelhood-function}

The likelihood function used in the TKTD/GUTS methods is applied to estimate the model parameters. For the survival test, it is assumed that the daily number of deaths follows a conditional binomial distribution that is equivalent to the multinomial distribution \cite{Jager.2018a}. The conditional binomial distribution was chosen because it integrates better with the computational framework pymob used in parameter estimation.
In addition, conditional binomial survival or lethality probabilities can be directly estimated from the integrated hazard state, which is numerically more stable than calculating it from survival probabilities.

The probability model is described as follows:

\begin{align}
    k_i &\sim \text{Binomial}(n=n_i, p=p_i) \\
    k_i &= S^{obs}_i \\
    n_i &= S^{obs}_{i-1} \\
    p_i &= \Pr(t < T~|~t_{i-1} < T) = \frac{S_i}{S_{i-1}} = e^{-(H_i - H_{i-1})} \\
    H_i &= f(t_i, X, \theta)
\end{align}

where $p_{S_i|S_{i-1}}$ is the conditional probability to be alive at the $i$-th observation, when survival was observed at the previous observation, $H_i$ is the solution of the ODE model $f$ for the observation time $t$ of the $i$-th observation, and $\theta$ is the vector of parameters that defines the ODE model.

Finally, the likelihood function of the binomial probability distribution is 

\begin{align} \label{likelihood}
    \mathcal{L}(\theta|S^{obs}) &= \prod_i p_i^{k_i}~(1-p_i)^{n_i-k_i} \\
    &= \prod_i \left(e^{-(H_i - H_{i-1})}\right)^{S^{obs}_i}  \left(1 - e^{-(H_i - H_{i-1})}\right)^{S^{obs}_{i-1} - S^{obs}_i}\\
    \ell(\theta|S^{obs}) &= \sum_i k_i~\ln(p_i) + (n_i-k_i)~\ln(1-p_i) \\
    &= \sum_i S^{obs}_i \ln\left(e^{-(H_i - H_{i-1})}\right) + (S^{obs}_{i-1} - S^{obs}_i) \ln\left(1 - e^{-(H_i - H_{i-1})}\right)
\end{align}

with $k_i$ denoting the number of organisms alive at the $i$-th observation and $n_i$ denoting the number of organisms alive at the $i-1$-th observation. If censored data exist, and are not excluded from the dataset, censoring could be applied (Section~\ref{si:censoring}).

\section{Censoring}\label{si:censoring}

By default, the binomial model accounts for right hand censored data. Right hand censored data are the organisms that are alive at the end of the experiment and where death is no longer observed. However, other processes can contribute to an earlier need for censoring.
Thereby, we can account for organisms that \emph{left} the experiment. These include escaped, damaged or otherwise removed organisms. 
Escaped or removed organisms are accounted for in the quantity $k^*_i$, which is the number of organisms escaped after $i$. 

\textbf{Calculation example}: At time $t_{i-1}$, $k_{i-1} = 5$, after 1 day at time $t$, $k_{i} = 2$, but of the 3 potentially died organisms, one organism escaped in the interval; therefore $k^*_{i} = 1$. This leads to $5 - 2 - 1 = 2$; two organisms died, one escaped in the interval and is censored, the remaining number of organisms is two.

Censored observations are considered using the survival function at time $t_i$, where the individuals where missing. This is straightforward, as outlined in Eq.~\ref{eq_definition_survival}, the survival function $p^*_i = \exp(-H_{i-1})$ gives the probability of living longer than time $t_{i-1}$. This is exactly the information contained in censored information. It is known that the individual did not die until $t_{i-1}$. At this time it was last observed alive and escaped or was removed from the experiment afterwards. Therefore, the joint likelihood of recorded dead and censored organisms at any observation time $t_i$ is:

\begin{align} \label{eq_censored_likelihood}
    \ell(\theta|S^{obs}) &= \sum_i k_i \ln(p_i) + k^* \ln(p^*_i) + (n_i-k_i - k^*_i) \ln(1-p_i) \\
\end{align}

Coming back to the calculation example, the interpretation is as follows: We have $k_i=2$ conditional survivors at observation $i$ that were also alive at $i-1$. We have $k^*_i = 1$ organism that survived until $i-1$ and then we don't know any more about it (censored). And finally, we have $n_i = 5$ and therefore $5-2-1 = 2$ conditional deaths that were alive at $i-1$.

\section{Piecewise approach to discretising the Buffer GUTS model} \label{si:sec:expanded-discretized-equations}
\subsection{ODE system of BufferGUTS model}

The complete Bufferguts ODE system is defined as follows:\\

$\frac{d}{d t} B_{O}{\left(t \right)} = k_{b O} \left(C_{O} - B_{O}{\left(t \right)}\right)$

$\frac{d}{d t} B_{C}{\left(t \right)} = k_{b C} \left(C_{C} - B_{C}{\left(t \right)}\right)$

$\frac{d}{d t} B_{T}{\left(t \right)} = k_{b T} \left(C_{T} - B_{T}{\left(t \right)}\right)$

$\frac{d}{d t} D{\left(t \right)} = k_{d} \left(w_{C} B_{C}{\left(t \right)} + w_{O} B_{O}{\left(t \right)} + w_{T} B_{T}{\left(t \right)} - D{\left(t \right)}\right)$

$\frac{d}{d t} H{\left(t \right)} = b \max\left(0, - z + D{\left(t \right)}\right) + h_{b}$

$\frac{d}{d t} S{\left(t \right)} = \left(- b \max\left(0, - z + D{\left(t \right)}\right) - h_{b}\right) S{\left(t \right)}$
\\\\
B and D Terms cannot be solved symbolically together with different coefficients this is an approach with a  the below is a general function, valid for e.g. calculating internal  concentrations, but also for a buffer guts model, where the $k_d$ coefficients are inserted before the computation. 

\subsection{Approach to solving the model}

\subsubsection{Exact solution to the buffer equations}\label{si:buffer_solution}

Assuming $k_b = k_d$\\

$\frac{d}{d t} B_{O}{\left(t \right)} = k_{d} \left(C_{O,0} - B_{O}{\left(t \right)}\right)$

$\frac{d}{d t} B_{C}{\left(t \right)} = k_{d} \left(C_{C,0} - B_{C}{\left(t \right)}\right)$

$\frac{d}{d t} B_{T}{\left(t \right)} = k_{d} \left(C_{T,0} - B_{T}{\left(t \right)}\right)$
\\\\
then the exact solution for the buffers are\\

$B_{O}{\left(t \right)} = C_{O,0} + \left(B_{O,0} - C_{O,0}\right) e^{- k_{d} t}$

$B_{C}{\left(t \right)} = C_{C,0} + \left(B_{C,0} - C_{C,0}\right) e^{- k_{d} t}$

$B_{T}{\left(t \right)} = C_{T,0} + \left(B_{T,0} - C_{T,0}\right) e^{- k_{d} t}$
\\\\
The bufferGUTS model assumes that if the Concentration $C \geq B$, the buffer is filled very fast $\eta >> 1$. This leads to the simplification that for $C \geq B$, B(t) = C

$B(t) = \left[\begin{smallmatrix}\begin{cases} C_{O,0} & \text{for}\: B_{O,0} < C_{O,0} \\C_{O,0} + \left(B_{O,0} - C_{O,0}\right) e^{- k_{d} t} & \text{otherwise} \end{cases}\\\begin{cases} C_{C,0} & \text{for}\: B_{C,0} < C_{C,0} \\C_{C,0} + \left(B_{C,0} - C_{C,0}\right) e^{- k_{d} t} & \text{otherwise} \end{cases}\\\begin{cases} C_{T,0} & \text{for}\: B_{T,0} < C_{T,0} \\C_{T,0} + \left(B_{T,0} - C_{T,0}\right) e^{- k_{d} t} & \text{otherwise} \end{cases}\end{smallmatrix}\right]$

\subsubsection{Damage solutions} \label{si:damage_solution}

\begin{align}
\frac{d}{d t} D{\left(t \right)} &= k_{d} w_{C} \left(\begin{cases} C_{C,0} & \text{for}\: B_{C,0} < C_{C,0} \\C_{C,0} + \left(B_{C,0} - C_{C,0}\right) e^{- k_{d} t} & \text{otherwise} \end{cases}\right) \\
&+ k_{d}  w_{O} \left(\begin{cases} C_{O,0} & \text{for}\: B_{O,0} < C_{O,0} \\C_{O,0} + \left(B_{O,0} - C_{O,0}\right) e^{- k_{d} t} & \text{otherwise} \end{cases}\right) \\
&+ k_{d} w_{T} \left(\begin{cases} C_{T,0} & \text{for}\: B_{T,0} < C_{T,0} \\C_{T,0} + \left(B_{T,0} - C_{T,0}\right) e^{- k_{d} t} & \text{otherwise} \end{cases}\right) \\
&- k_dD{\left(t \right)}
\end{align}

leading to the exact solutions:
{
\footnotesize
\begin{align}    
D{\left(t \right)} &= C_{C,0} w_{C} + C_{O,0} w_{O} + C_{T,0} w_{T} + \left(- C_{C,0} w_{C} - C_{O,0} w_{O} - C_{T,0} w_{T} + D_{0}\right) e^{- k_{d} t} \\
&+ \begin{cases} 0 & \text{for}\: B_{C,0} < C_{C,0} \wedge B_{O,0} < C_{O,0} \wedge \\ & B_{T,0} < C_{T,0} \\k_{d} t \left(B_{T,0} w_{T} - C_{T,0} w_{T}\right) e^{- k_{d} t} & \text{for}\: B_{C,0} < C_{C,0} \wedge B_{O,0} < C_{O,0} \\k_{d} t \left(B_{O,0} w_{O} - C_{O,0} w_{O}\right) e^{- k_{d} t} & \text{for}\: B_{C,0} < C_{C,0} \wedge B_{T,0} < C_{T,0} \\k_{d} t \left(B_{O,0} w_{O} + B_{T,0} w_{T} - C_{O,0} w_{O} - C_{T,0} w_{T}\right) e^{- k_{d} t} & \text{for}\: B_{C,0} < C_{C,0} \\k_{d} t \left(B_{C,0} w_{C} - C_{C,0} w_{C}\right) e^{- k_{d} t} & \text{for}\: B_{O,0} < C_{O,0} \wedge B_{T,0} < C_{T,0} \\k_{d} t \left(B_{C,0} w_{C} + B_{T,0} w_{T} - C_{C,0} w_{C} - C_{T,0} w_{T}\right) e^{- k_{d} t} & \text{for}\: B_{O,0} < C_{O,0} \\k_{d} t \left(B_{C,0} w_{C} + B_{O,0} w_{O} - C_{C,0} w_{C} - C_{O,0} w_{O}\right) e^{- k_{d} t} & \text{for}\: B_{T,0} < C_{T,0} \\k_{d} t \left(B_{C,0} w_{C} + B_{O,0} w_{O} + B_{T,0} w_{T} - C_{C,0} w_{C} - C_{O,0} w_{O} - C_{T,0} w_{T}\right) e^{- k_{d} t} & \text{otherwise} 
\end{cases}
\end{align}}

By defining the different branches of the solution, the system can be solved one by one for each branch.

\pagebreak
\begin{landscape}
\subsubsection{Finding the time $t$ of where $D(t) = z$} \label{si:root_finding_d_eq_z}
\small
An exact solution to the Damage function is helpful. Because it allows to find The time at which the damage reaches the threshold value $z$.If this value can be found, the system can be solved exactly by using an intermediate step. Note, such a solution gets significantly more difficult, when $k_d \neq k_b$, i.e. when there are different rate constants in the buffer term and the damage term.In such cases, only approximate solutions are feasible for the Damage $D$
In some cases the equality $D(t) = z$ can be solved exactly, but the equations become easily so large that numerical solutions for finding the location of the intermediate step are much simpler and are only a minor performance issue. For this the equalities $D(t) = z$ are needed and if possible the derivatives.
\textbf{Note} that the equations in the conditional branches of the piecewise function $t_{jump}\left(t,Y_{0},\theta \right)$ are simply the different conditional functions of $D(t) - z$.
\\\\
\scalebox{.6}{
$t_{jump}{\left(t,Y_{0},\theta \right)} \\= \begin{cases} \left(- C_{C,0} w_{C} - C_{O,0} w_{O} - C_{T,0} w_{T} + D_{0} - z e^{k_{d} t} + \left(C_{C,0} w_{C} + C_{O,0} w_{O} + C_{T,0} w_{T}\right) e^{k_{d} t}\right) e^{- k_{d} t} & \text{for}\: B_{C,0} < C_{C,0} \wedge B_{O,0} < C_{O,0} \wedge B_{T,0} < C_{T,0} \\\left(B_{T,0} k_{d} t w_{T} + C_{C,0} w_{C} e^{k_{d} t} - C_{C,0} w_{C} + C_{O,0} w_{O} e^{k_{d} t} - C_{O,0} w_{O} - C_{T,0} k_{d} t w_{T} + C_{T,0} w_{T} e^{k_{d} t} - C_{T,0} w_{T} + D_{0} - z e^{k_{d} t}\right) e^{- k_{d} t} & \text{for}\: B_{C,0} < C_{C,0} \wedge B_{O,0} < C_{O,0} \\\left(B_{O,0} k_{d} t w_{O} + C_{C,0} w_{C} e^{k_{d} t} - C_{C,0} w_{C} - C_{O,0} k_{d} t w_{O} + C_{O,0} w_{O} e^{k_{d} t} - C_{O,0} w_{O} + C_{T,0} w_{T} e^{k_{d} t} - C_{T,0} w_{T} + D_{0} - z e^{k_{d} t}\right) e^{- k_{d} t} & \text{for}\: B_{C,0} < C_{C,0} \wedge B_{T,0} < C_{T,0} \\\left(B_{O,0} k_{d} t w_{O} + B_{T,0} k_{d} t w_{T} + C_{C,0} w_{C} e^{k_{d} t} - C_{C,0} w_{C} - C_{O,0} k_{d} t w_{O} + C_{O,0} w_{O} e^{k_{d} t} - C_{O,0} w_{O} - C_{T,0} k_{d} t w_{T} + C_{T,0} w_{T} e^{k_{d} t} - C_{T,0} w_{T} + D_{0} - z e^{k_{d} t}\right) e^{- k_{d} t} & \text{for}\: B_{C,0} < C_{C,0} \\\left(B_{C,0} k_{d} t w_{C} - C_{C,0} k_{d} t w_{C} + C_{C,0} w_{C} e^{k_{d} t} - C_{C,0} w_{C} + C_{O,0} w_{O} e^{k_{d} t} - C_{O,0} w_{O} + C_{T,0} w_{T} e^{k_{d} t} - C_{T,0} w_{T} + D_{0} - z e^{k_{d} t}\right) e^{- k_{d} t} & \text{for}\: B_{O,0} < C_{O,0} \wedge B_{T,0} < C_{T,0} \\\left(B_{C,0} k_{d} t w_{C} + B_{T,0} k_{d} t w_{T} - C_{C,0} k_{d} t w_{C} + C_{C,0} w_{C} e^{k_{d} t} - C_{C,0} w_{C} + C_{O,0} w_{O} e^{k_{d} t} - C_{O,0} w_{O} - C_{T,0} k_{d} t w_{T} + C_{T,0} w_{T} e^{k_{d} t} - C_{T,0} w_{T} + D_{0} - z e^{k_{d} t}\right) e^{- k_{d} t} & \text{for}\: B_{O,0} < C_{O,0} \\\left(B_{C,0} k_{d} t w_{C} + B_{O,0} k_{d} t w_{O} - C_{C,0} k_{d} t w_{C} + C_{C,0} w_{C} e^{k_{d} t} - C_{C,0} w_{C} - C_{O,0} k_{d} t w_{O} + C_{O,0} w_{O} e^{k_{d} t} - C_{O,0} w_{O} + C_{T,0} w_{T} e^{k_{d} t} - C_{T,0} w_{T} + D_{0} - z e^{k_{d} t}\right) e^{- k_{d} t} & \text{for}\: B_{T,0} < C_{T,0} \\\left(B_{C,0} k_{d} t w_{C} + B_{O,0} k_{d} t w_{O} + B_{T,0} k_{d} t w_{T} - C_{C,0} k_{d} t w_{C} + C_{C,0} w_{C} e^{k_{d} t} - C_{C,0} w_{C} - C_{O,0} k_{d} t w_{O} + C_{O,0} w_{O} e^{k_{d} t} - C_{O,0} w_{O} - C_{T,0} k_{d} t w_{T} + C_{T,0} w_{T} e^{k_{d} t} - C_{T,0} w_{T} + D_{0} - z e^{k_{d} t}\right) e^{- k_{d} t} & \text{otherwise} \end{cases}$
}
\\\\\\
\scalebox{.6}{
$\frac{\partial}{\partial t} t_{jump}{\left(t,Y_{0},\theta \right)} \\= \begin{cases} k_{d} \left(C_{C,0} w_{C} + C_{O,0} w_{O} + C_{T,0} w_{T} - D_{0}\right) e^{- k_{d} t} & \text{for}\: B_{C,0} < C_{C,0} \wedge B_{O,0} < C_{O,0} \wedge B_{T,0} < C_{T,0} \\k_{d} \left(- B_{T,0} k_{d} t w_{T} + B_{T,0} w_{T} + C_{C,0} w_{C} + C_{O,0} w_{O} + C_{T,0} k_{d} t w_{T} - D_{0}\right) e^{- k_{d} t} & \text{for}\: B_{C,0} < C_{C,0} \wedge B_{O,0} < C_{O,0} \\k_{d} \left(- B_{O,0} k_{d} t w_{O} + B_{O,0} w_{O} + C_{C,0} w_{C} + C_{O,0} k_{d} t w_{O} + C_{T,0} w_{T} - D_{0}\right) e^{- k_{d} t} & \text{for}\: B_{C,0} < C_{C,0} \wedge B_{T,0} < C_{T,0} \\k_{d} \left(- B_{O,0} k_{d} t w_{O} + B_{O,0} w_{O} - B_{T,0} k_{d} t w_{T} + B_{T,0} w_{T} + C_{C,0} w_{C} + C_{O,0} k_{d} t w_{O} + C_{T,0} k_{d} t w_{T} - D_{0}\right) e^{- k_{d} t} & \text{for}\: B_{C,0} < C_{C,0} \\k_{d} \left(- B_{C,0} k_{d} t w_{C} + B_{C,0} w_{C} + C_{C,0} k_{d} t w_{C} + C_{O,0} w_{O} + C_{T,0} w_{T} - D_{0}\right) e^{- k_{d} t} & \text{for}\: B_{O,0} < C_{O,0} \wedge B_{T,0} < C_{T,0} \\k_{d} \left(- B_{C,0} k_{d} t w_{C} + B_{C,0} w_{C} - B_{T,0} k_{d} t w_{T} + B_{T,0} w_{T} + C_{C,0} k_{d} t w_{C} + C_{O,0} w_{O} + C_{T,0} k_{d} t w_{T} - D_{0}\right) e^{- k_{d} t} & \text{for}\: B_{O,0} < C_{O,0} \\k_{d} \left(- B_{C,0} k_{d} t w_{C} + B_{C,0} w_{C} - B_{O,0} k_{d} t w_{O} + B_{O,0} w_{O} + C_{C,0} k_{d} t w_{C} + C_{O,0} k_{d} t w_{O} + C_{T,0} w_{T} - D_{0}\right) e^{- k_{d} t} & \text{for}\: B_{T,0} < C_{T,0} \\k_{d} \left(- B_{C,0} k_{d} t w_{C} + B_{C,0} w_{C} - B_{O,0} k_{d} t w_{O} + B_{O,0} w_{O} - B_{T,0} k_{d} t w_{T} + B_{T,0} w_{T} + C_{C,0} k_{d} t w_{C} + C_{O,0} k_{d} t w_{O} + C_{T,0} k_{d} t w_{T} - D_{0}\right) e^{- k_{d} t} & \text{otherwise} \end{cases}$
}
\\\\\\
\scalebox{0.6}{
$\frac{\partial^{2}}{\partial t^{2}} t_{jump}{\left(t,Y_{0},\theta \right)} \\= \begin{cases} k_{d}^{2} \left(- C_{C,0} w_{C} - C_{O,0} w_{O} - C_{T,0} w_{T} + D_{0}\right) e^{- k_{d} t} & \text{for}\: B_{C,0} < C_{C,0} \wedge B_{O,0} < C_{O,0} \wedge B_{T,0} < C_{T,0} \\k_{d}^{2} \left(B_{T,0} k_{d} t w_{T} - B_{T,0} w_{T} - C_{C,0} w_{C} - C_{O,0} w_{O} - C_{T,0} k_{d} t w_{T} + D_{0} - w_{T} \left(B_{T,0} - C_{T,0}\right)\right) e^{- k_{d} t} & \text{for}\: B_{C,0} < C_{C,0} \wedge B_{O,0} < C_{O,0} \\k_{d}^{2} \left(B_{O,0} k_{d} t w_{O} - B_{O,0} w_{O} - C_{C,0} w_{C} - C_{O,0} k_{d} t w_{O} - C_{T,0} w_{T} + D_{0} - w_{O} \left(B_{O,0} - C_{O,0}\right)\right) e^{- k_{d} t} & \text{for}\: B_{C,0} < C_{C,0} \wedge B_{T,0} < C_{T,0} \\k_{d}^{2} \left(B_{O,0} k_{d} t w_{O} - 2 B_{O,0} w_{O} + B_{T,0} k_{d} t w_{T} - 2 B_{T,0} w_{T} - C_{C,0} w_{C} - C_{O,0} k_{d} t w_{O} + C_{O,0} w_{O} - C_{T,0} k_{d} t w_{T} + C_{T,0} w_{T} + D_{0}\right) e^{- k_{d} t} & \text{for}\: B_{C,0} < C_{C,0} \\k_{d}^{2} \left(B_{C,0} k_{d} t w_{C} - B_{C,0} w_{C} - C_{C,0} k_{d} t w_{C} - C_{O,0} w_{O} - C_{T,0} w_{T} + D_{0} - w_{C} \left(B_{C,0} - C_{C,0}\right)\right) e^{- k_{d} t} & \text{for}\: B_{O,0} < C_{O,0} \wedge B_{T,0} < C_{T,0} \\k_{d}^{2} \left(B_{C,0} k_{d} t w_{C} - 2 B_{C,0} w_{C} + B_{T,0} k_{d} t w_{T} - 2 B_{T,0} w_{T} - C_{C,0} k_{d} t w_{C} + C_{C,0} w_{C} - C_{O,0} w_{O} - C_{T,0} k_{d} t w_{T} + C_{T,0} w_{T} + D_{0}\right) e^{- k_{d} t} & \text{for}\: B_{O,0} < C_{O,0} \\k_{d}^{2} \left(B_{C,0} k_{d} t w_{C} - 2 B_{C,0} w_{C} + B_{O,0} k_{d} t w_{O} - 2 B_{O,0} w_{O} - C_{C,0} k_{d} t w_{C} + C_{C,0} w_{C} - C_{O,0} k_{d} t w_{O} + C_{O,0} w_{O} - C_{T,0} w_{T} + D_{0}\right) e^{- k_{d} t} & \text{for}\: B_{T,0} < C_{T,0} \\k_{d}^{2} \left(B_{C,0} k_{d} t w_{C} - 2 B_{C,0} w_{C} + B_{O,0} k_{d} t w_{O} - 2 B_{O,0} w_{O} + B_{T,0} k_{d} t w_{T} - 2 B_{T,0} w_{T} - C_{C,0} k_{d} t w_{C} + C_{C,0} w_{C} - C_{O,0} k_{d} t w_{O} + C_{O,0} w_{O} - C_{T,0} k_{d} t w_{T} + C_{T,0} w_{T} + D_{0}\right) e^{- k_{d} t} & \text{otherwise} \end{cases}$
}
\\\\\\

\small
This system is extensible for any number of buffers, but of course the Combinatorial complexity of the buffer increases

\end{landscape}

\begin{landscape}
        
\subsubsection{Hazard solutions}\label{si:hazard_solution}

From the different branches of the damage $D$ solution, ODEs for the hazard can be formulated
\footnotesize
\\\\
\scalebox{0.6}{
$\frac{d}{d t} H{\left(t \right)} = \begin{cases} b \max\left(0, - z + \left(- C_{C,0} w_{C} - C_{O,0} w_{O} - C_{T,0} w_{T} + D_{0} + \left(C_{C,0} w_{C} + C_{O,0} w_{O} + C_{T,0} w_{T}\right) e^{k_{d} t}\right) e^{- k_{d} t}\right) + h_{b} & \text{for}\: B_{C,0} < C_{C,0} \wedge B_{O,0} < C_{O,0} \wedge B_{T,0} < C_{T,0} \\b \max\left(0, - z + \left(B_{T,0} k_{d} t w_{T} + C_{C,0} w_{C} e^{k_{d} t} - C_{C,0} w_{C} + C_{O,0} w_{O} e^{k_{d} t} - C_{O,0} w_{O} - C_{T,0} k_{d} t w_{T} + C_{T,0} w_{T} e^{k_{d} t} - C_{T,0} w_{T} + D_{0}\right) e^{- k_{d} t}\right) + h_{b} & \text{for}\: B_{C,0} < C_{C,0} \wedge B_{O,0} < C_{O,0} \\b \max\left(0, - z + \left(B_{O,0} k_{d} t w_{O} + C_{C,0} w_{C} e^{k_{d} t} - C_{C,0} w_{C} - C_{O,0} k_{d} t w_{O} + C_{O,0} w_{O} e^{k_{d} t} - C_{O,0} w_{O} + C_{T,0} w_{T} e^{k_{d} t} - C_{T,0} w_{T} + D_{0}\right) e^{- k_{d} t}\right) + h_{b} & \text{for}\: B_{C,0} < C_{C,0} \wedge B_{T,0} < C_{T,0} \\b \max\left(0, - z + \left(B_{O,0} k_{d} t w_{O} + B_{T,0} k_{d} t w_{T} + C_{C,0} w_{C} e^{k_{d} t} - C_{C,0} w_{C} - C_{O,0} k_{d} t w_{O} + C_{O,0} w_{O} e^{k_{d} t} - C_{O,0} w_{O} - C_{T,0} k_{d} t w_{T} + C_{T,0} w_{T} e^{k_{d} t} - C_{T,0} w_{T} + D_{0}\right) e^{- k_{d} t}\right) + h_{b} & \text{for}\: B_{C,0} < C_{C,0} \\b \max\left(0, - z + \left(B_{C,0} k_{d} t w_{C} - C_{C,0} k_{d} t w_{C} + C_{C,0} w_{C} e^{k_{d} t} - C_{C,0} w_{C} + C_{O,0} w_{O} e^{k_{d} t} - C_{O,0} w_{O} + C_{T,0} w_{T} e^{k_{d} t} - C_{T,0} w_{T} + D_{0}\right) e^{- k_{d} t}\right) + h_{b} & \text{for}\: B_{O,0} < C_{O,0} \wedge B_{T,0} < C_{T,0} \\b \max\left(0, - z + \left(B_{C,0} k_{d} t w_{C} + B_{T,0} k_{d} t w_{T} - C_{C,0} k_{d} t w_{C} + C_{C,0} w_{C} e^{k_{d} t} - C_{C,0} w_{C} + C_{O,0} w_{O} e^{k_{d} t} - C_{O,0} w_{O} - C_{T,0} k_{d} t w_{T} + C_{T,0} w_{T} e^{k_{d} t} - C_{T,0} w_{T} + D_{0}\right) e^{- k_{d} t}\right) + h_{b} & \text{for}\: B_{O,0} < C_{O,0} \\b \max\left(0, - z + \left(B_{C,0} k_{d} t w_{C} + B_{O,0} k_{d} t w_{O} - C_{C,0} k_{d} t w_{C} + C_{C,0} w_{C} e^{k_{d} t} - C_{C,0} w_{C} - C_{O,0} k_{d} t w_{O} + C_{O,0} w_{O} e^{k_{d} t} - C_{O,0} w_{O} + C_{T,0} w_{T} e^{k_{d} t} - C_{T,0} w_{T} + D_{0}\right) e^{- k_{d} t}\right) + h_{b} & \text{for}\: B_{T,0} < C_{T,0} \\b \max\left(0, - z + \left(B_{C,0} k_{d} t w_{C} + B_{O,0} k_{d} t w_{O} + B_{T,0} k_{d} t w_{T} - C_{C,0} k_{d} t w_{C} + C_{C,0} w_{C} e^{k_{d} t} - C_{C,0} w_{C} - C_{O,0} k_{d} t w_{O} + C_{O,0} w_{O} e^{k_{d} t} - C_{O,0} w_{O} - C_{T,0} k_{d} t w_{T} + C_{T,0} w_{T} e^{k_{d} t} - C_{T,0} w_{T} + D_{0}\right) e^{- k_{d} t}\right) + h_{b} & \text{otherwise} \end{cases}$
}
\\\\\\
leading to the exact solutions for each respective branch

\scalebox{.3}{
$H{\left(t \right)} = \begin{cases} \begin{cases} H_{0} + h_{b} t & \text{for}\: \left(- C_{C,0} w_{C} - C_{O,0} w_{O} - C_{T,0} w_{T} + D_{0} + \left(C_{C,0} w_{C} + C_{O,0} w_{O} + C_{T,0} w_{T} - z\right) e^{k_{d} t}\right) e^{- k_{d} t} \leq 0 \\\left(b \left(C_{C,0} w_{C} + C_{O,0} w_{O} + C_{T,0} w_{T} - D_{0}\right) + k_{d} t \left(b \left(C_{C,0} w_{C} + C_{O,0} w_{O} + C_{T,0} w_{T} - z\right) + h_{b}\right) e^{k_{d} t} + \left(- C_{C,0} b w_{C} - C_{O,0} b w_{O} - C_{T,0} b w_{T} + D_{0} b + H_{0} k_{d}\right) e^{k_{d} t}\right) e^{- k_{d} t} / k_{d} & \text{otherwise} \end{cases} & \text{for}\: B_{C,0} < C_{C,0} \wedge B_{O,0} < C_{O,0} \wedge B_{T,0} < C_{T,0} \\\begin{cases} H_{0} + h_{b} t & \text{for}\: \left(B_{T,0} k_{d} t w_{T} - C_{C,0} w_{C} - C_{O,0} w_{O} - C_{T,0} k_{d} t w_{T} - C_{T,0} w_{T} + D_{0} + \left(C_{C,0} w_{C} + C_{O,0} w_{O} + C_{T,0} w_{T} - z\right) e^{k_{d} t}\right) e^{- k_{d} t} \leq 0 \\\left(b \left(- B_{T,0} k_{d} t w_{T} - B_{T,0} w_{T} + C_{C,0} w_{C} + C_{O,0} w_{O} + C_{T,0} k_{d} t w_{T} + 2 C_{T,0} w_{T} - D_{0}\right) + k_{d} t \left(b \left(C_{C,0} w_{C} + C_{O,0} w_{O} + C_{T,0} w_{T} - z\right) + h_{b}\right) e^{k_{d} t} + \left(B_{T,0} b w_{T} - C_{C,0} b w_{C} - C_{O,0} b w_{O} - 2 C_{T,0} b w_{T} + D_{0} b + H_{0} k_{d}\right) e^{k_{d} t}\right) e^{- k_{d} t} / k_{d} & \text{otherwise} \end{cases} & \text{for}\: B_{C,0} < C_{C,0} \wedge B_{O,0} < C_{O,0} \\\begin{cases} H_{0} + h_{b} t & \text{for}\: \left(B_{O,0} k_{d} t w_{O} - C_{C,0} w_{C} - C_{O,0} k_{d} t w_{O} - C_{O,0} w_{O} - C_{T,0} w_{T} + D_{0} + \left(C_{C,0} w_{C} + C_{O,0} w_{O} + C_{T,0} w_{T} - z\right) e^{k_{d} t}\right) e^{- k_{d} t} \leq 0 \\\left(b \left(- B_{O,0} k_{d} t w_{O} - B_{O,0} w_{O} + C_{C,0} w_{C} + C_{O,0} k_{d} t w_{O} + 2 C_{O,0} w_{O} + C_{T,0} w_{T} - D_{0}\right) + k_{d} t \left(b \left(C_{C,0} w_{C} + C_{O,0} w_{O} + C_{T,0} w_{T} - z\right) + h_{b}\right) e^{k_{d} t} + \left(B_{O,0} b w_{O} - C_{C,0} b w_{C} - 2 C_{O,0} b w_{O} - C_{T,0} b w_{T} + D_{0} b + H_{0} k_{d}\right) e^{k_{d} t}\right) e^{- k_{d} t} / k_{d} & \text{otherwise} \end{cases} & \text{for}\: B_{C,0} < C_{C,0} \wedge B_{T,0} < C_{T,0} \\\begin{cases} H_{0} + h_{b} t & \text{for}\: \left(B_{O,0} k_{d} t w_{O} + B_{T,0} k_{d} t w_{T} - C_{C,0} w_{C} - C_{O,0} k_{d} t w_{O} - C_{O,0} w_{O} - C_{T,0} k_{d} t w_{T} - C_{T,0} w_{T} + D_{0} + \left(C_{C,0} w_{C} + C_{O,0} w_{O} + C_{T,0} w_{T} - z\right) e^{k_{d} t}\right) e^{- k_{d} t} \leq 0 \\\left(b \left(- B_{O,0} k_{d} t w_{O} - B_{O,0} w_{O} - B_{T,0} k_{d} t w_{T} - B_{T,0} w_{T} + C_{C,0} w_{C} + C_{O,0} k_{d} t w_{O} + 2 C_{O,0} w_{O} + C_{T,0} k_{d} t w_{T} + 2 C_{T,0} w_{T} - D_{0}\right) + k_{d} t \left(b \left(C_{C,0} w_{C} + C_{O,0} w_{O} + C_{T,0} w_{T} - z\right) + h_{b}\right) e^{k_{d} t} + \left(B_{O,0} b w_{O} + B_{T,0} b w_{T} - C_{C,0} b w_{C} - 2 C_{O,0} b w_{O} - 2 C_{T,0} b w_{T} + D_{0} b + H_{0} k_{d}\right) e^{k_{d} t}\right) e^{- k_{d} t} / k_{d} & \text{otherwise} \end{cases} & \text{for}\: B_{C,0} < C_{C,0} \\\begin{cases} H_{0} + h_{b} t & \text{for}\: \left(B_{C,0} k_{d} t w_{C} - C_{C,0} k_{d} t w_{C} - C_{C,0} w_{C} - C_{O,0} w_{O} - C_{T,0} w_{T} + D_{0} + \left(C_{C,0} w_{C} + C_{O,0} w_{O} + C_{T,0} w_{T} - z\right) e^{k_{d} t}\right) e^{- k_{d} t} \leq 0 \\\left(b \left(- B_{C,0} k_{d} t w_{C} - B_{C,0} w_{C} + C_{C,0} k_{d} t w_{C} + 2 C_{C,0} w_{C} + C_{O,0} w_{O} + C_{T,0} w_{T} - D_{0}\right) + k_{d} t \left(b \left(C_{C,0} w_{C} + C_{O,0} w_{O} + C_{T,0} w_{T} - z\right) + h_{b}\right) e^{k_{d} t} + \left(B_{C,0} b w_{C} - 2 C_{C,0} b w_{C} - C_{O,0} b w_{O} - C_{T,0} b w_{T} + D_{0} b + H_{0} k_{d}\right) e^{k_{d} t}\right) e^{- k_{d} t} / k_{d} & \text{otherwise} \end{cases} & \text{for}\: B_{O,0} < C_{O,0} \wedge B_{T,0} < C_{T,0} \\\begin{cases} H_{0} + h_{b} t & \text{for}\: \left(B_{C,0} k_{d} t w_{C} + B_{T,0} k_{d} t w_{T} - C_{C,0} k_{d} t w_{C} - C_{C,0} w_{C} - C_{O,0} w_{O} - C_{T,0} k_{d} t w_{T} - C_{T,0} w_{T} + D_{0} + \left(C_{C,0} w_{C} + C_{O,0} w_{O} + C_{T,0} w_{T} - z\right) e^{k_{d} t}\right) e^{- k_{d} t} \leq 0 \\\left(b \left(- B_{C,0} k_{d} t w_{C} - B_{C,0} w_{C} - B_{T,0} k_{d} t w_{T} - B_{T,0} w_{T} + C_{C,0} k_{d} t w_{C} + 2 C_{C,0} w_{C} + C_{O,0} w_{O} + C_{T,0} k_{d} t w_{T} + 2 C_{T,0} w_{T} - D_{0}\right) + k_{d} t \left(b \left(C_{C,0} w_{C} + C_{O,0} w_{O} + C_{T,0} w_{T} - z\right) + h_{b}\right) e^{k_{d} t} + \left(B_{C,0} b w_{C} + B_{T,0} b w_{T} - 2 C_{C,0} b w_{C} - C_{O,0} b w_{O} - 2 C_{T,0} b w_{T} + D_{0} b + H_{0} k_{d}\right) e^{k_{d} t}\right) e^{- k_{d} t} / k_{d} & \text{otherwise} \end{cases} & \text{for}\: B_{O,0} < C_{O,0} \\\begin{cases} H_{0} + h_{b} t & \text{for}\: \left(B_{C,0} k_{d} t w_{C} + B_{O,0} k_{d} t w_{O} - C_{C,0} k_{d} t w_{C} - C_{C,0} w_{C} - C_{O,0} k_{d} t w_{O} - C_{O,0} w_{O} - C_{T,0} w_{T} + D_{0} + \left(C_{C,0} w_{C} + C_{O,0} w_{O} + C_{T,0} w_{T} - z\right) e^{k_{d} t}\right) e^{- k_{d} t} \leq 0 \\\left(b \left(- B_{C,0} k_{d} t w_{C} - B_{C,0} w_{C} - B_{O,0} k_{d} t w_{O} - B_{O,0} w_{O} + C_{C,0} k_{d} t w_{C} + 2 C_{C,0} w_{C} + C_{O,0} k_{d} t w_{O} + 2 C_{O,0} w_{O} + C_{T,0} w_{T} - D_{0}\right) + k_{d} t \left(b \left(C_{C,0} w_{C} + C_{O,0} w_{O} + C_{T,0} w_{T} - z\right) + h_{b}\right) e^{k_{d} t} + \left(B_{C,0} b w_{C} + B_{O,0} b w_{O} - 2 C_{C,0} b w_{C} - 2 C_{O,0} b w_{O} - C_{T,0} b w_{T} + D_{0} b + H_{0} k_{d}\right) e^{k_{d} t}\right) e^{- k_{d} t} / k_{d} & \text{otherwise} \end{cases} & \text{for}\: B_{T,0} < C_{T,0} \\\begin{cases} H_{0} + h_{b} t & \text{for}\: \left(B_{C,0} k_{d} t w_{C} + B_{O,0} k_{d} t w_{O} + B_{T,0} k_{d} t w_{T} - C_{C,0} k_{d} t w_{C} - C_{C,0} w_{C} - C_{O,0} k_{d} t w_{O} - C_{O,0} w_{O} - C_{T,0} k_{d} t w_{T} - C_{T,0} w_{T} + D_{0} + \left(C_{C,0} w_{C} + C_{O,0} w_{O} + C_{T,0} w_{T} - z\right) e^{k_{d} t}\right) e^{- k_{d} t} \leq 0 \\\left(b \left(- B_{C,0} k_{d} t w_{C} - B_{C,0} w_{C} - B_{O,0} k_{d} t w_{O} - B_{O,0} w_{O} - B_{T,0} k_{d} t w_{T} - B_{T,0} w_{T} + C_{C,0} k_{d} t w_{C} + 2 C_{C,0} w_{C} + C_{O,0} k_{d} t w_{O} + 2 C_{O,0} w_{O} + C_{T,0} k_{d} t w_{T} + 2 C_{T,0} w_{T} - D_{0}\right) + k_{d} t \left(b \left(C_{C,0} w_{C} + C_{O,0} w_{O} + C_{T,0} w_{T} - z\right) + h_{b}\right) e^{k_{d} t} + \left(B_{C,0} b w_{C} + B_{O,0} b w_{O} + B_{T,0} b w_{T} - 2 C_{C,0} b w_{C} - 2 C_{O,0} b w_{O} - 2 C_{T,0} b w_{T} + D_{0} b + H_{0} k_{d}\right) e^{k_{d} t}\right) e^{- k_{d} t} / k_{d} & \text{otherwise} \end{cases} & \text{otherwise} \end{cases}$
}
\end{landscape}

\section{Discretizing the Buffer GUTS model}

Let's start with the ODEs: please note that these are the piecewise functions for the range of $(t-\Delta t, t]$

$\frac{d}{d t} D{\left(t \right)} = k_{d}\left( W_CB_{C}{\left(t \right)} +  W_OB_{O}{\left(t \right)} +  W_TB_{T}{\left(t \right)} - D{\left(t \right)}\right)$

where,

$B(t) = \begin{cases}
        C(t-\Delta t) & B(t-\Delta t) \leq C(t-\Delta t) \\
        C(t-\Delta t) + \left(B(t-\Delta t\right) - C(t-\Delta t))~e^{-k_\Delta t} & B(t-\Delta t) > C(t-\Delta t)
    \end{cases} $

please note that $C(t-\Delta t)$ and $B(t-\Delta t)$ are constant in the range of $(t-\Delta t, t]$.

Then we can calculate $D(t)$ as below.

\begin{align} 
D{\left(t \right)} &= W_CC_{C}(t-\Delta t) + W_OC_{O}(t-\Delta t) + W_TC_{T}(t-\Delta t) \\
&+ \left(-  W_CC_{C}(t-\Delta t) - W_OC_{O}(t-\Delta t) - W_TC_{T}(t-\Delta t) + D(t-\Delta t)\right) e^{- k_{d} t}\\
&+ f_O(t) + f_C(t) + f_T(t)
\end{align}

where

$f_O(t) = \begin{cases}
        0 & B_O(t-\Delta t) \leq C_O(t-\Delta t) \\
        k_{d} t W_O \left(B_{O}(t-\Delta t) - C_{O}(t-\Delta t)\right) e^{- k_{d} t} & B_O(t-\Delta t) > C_O(t-\Delta t)
    \end{cases} $

$f_C(t) = \begin{cases}
        0 & B_C(t-\Delta t) \leq C_C(t-\Delta t) \\
        k_{d} t W_C \left(B_{C}(t-\Delta t) - C_{C}(t-\Delta t)\right) e^{- k_{d} t} & B_C(t-\Delta t) > C_C(t-\Delta t)
    \end{cases} $

$f_T(t) = \begin{cases}
        0 & B_T(t-\Delta t) \leq C_T(t-\Delta t) \\
        k_{d} t W_T \left(B_{T}(t-\Delta t) - C_{T}(t-\Delta t)\right) e^{- k_{d} t} & B_T(t-\Delta t) > C_T(t-\Delta t)
    \end{cases} $

For the hazard

$\frac{d}{d t} H{\left(t \right)} = b \max\left(0, - z + D{\left(t \right)}\right) + h_{b}$

we can get $H(t)$

\begin{align} 
H{\left(t \right)} &= H{\left(t - \Delta t\right)} + h_{b}t \\ 
&+ b\left(W_CC_{C}(t-\Delta t) + W_OC_{O}(t-\Delta t) + W_TC_{T}(t-\Delta t) \right)t \\
& + \frac{b\left(-  W_CC_{C}(t-\Delta t) - W_OC_{O}(t-\Delta t) - W_TC_{T}(t-\Delta t) + D(t-\Delta t)\right) \left(1-e^{- k_{d} t}\right)}{k}\\
&+ b\left(g_O(t) + g_C(t) + g_T(t) \right) -bzt
\end{align}

where

$g_O(t) = \begin{cases}
        0 & B_O(t-\Delta t) \leq C_O(t-\Delta t) \\
         W_O \left(B_{O}(t-\Delta t) - C_{O}(t-\Delta t)\right) \left(\frac{1- (k_{d}t+1)e^{- k_{d} t}}{k_{d}} \right) & B_O(t-\Delta t) > C_O(t-\Delta t)
    \end{cases} $

$g_C(t) = \begin{cases}
        0 & B_C(t-\Delta t) \leq C_C(t-\Delta t) \\
         W_C \left(B_{C}(t-\Delta t) - C_{C}(t-\Delta t)\right) \left(\frac{1- (k_{d}t+1)e^{- k_{d} t}}{k_{d}} \right) & B_C(t-\Delta t) > C_C(t-\Delta t)
    \end{cases} $

$g_T(t) = \begin{cases}
        0 & B_T(t-\Delta t) \leq C_T(t-\Delta t) \\
         W_T \left(B_{T}(t-\Delta t) - C_{T}(t-\Delta t)\right) \left(\frac{1- (k_{d}t+1)e^{- k_{d} t}}{k_{d}} \right) & B_T(t-\Delta t) > C_T(t-\Delta t)
    \end{cases} $

when the it is not always larger than the threshold, 5 points are used to estimate the hazards.

\section{Source code and Data} \label{si:sec:source-code}

\begin{itemize}
    \item Bufferguts model: \url{https://gitlab.uni-osnabrueck.de/fschunck/bufferguts}
    \item ALMaSS model: \url{https://gitlab.com/ALMaSS}
    \item Sulfoxaflor data and calibration:\\ \url{https://gin.g-node.org/pollinera/osmia-bicornis-sulfoxaflor}
\end{itemize}

\section{Report(case\_study=bufferguts,
\\ scenario=osmia\_sulfoxaflor\_fluctuating\_exposure\_16\_8)}\label{reportcase_studybufferguts-scenarioosmia_sulfoxaflor_fluctuating_exposure_16_8}

\begin{itemize}
\tightlist
\item
  Using \texttt{bufferguts==0.5.10}
\item
  Using \texttt{pymob==0.5.18}
\item
  Using backend: \texttt{NumpyroBackend}
\item
  Using settings:
  \texttt{../bufferguts/scenarios/\\ osmia\_sulfoxaflor\_fluctuating\_exposure\_16\_8/settings.cfg}
\end{itemize}

\subsection{Report: Model}\label{report-model}

\subsubsection{Model}\label{model}

\begin{Shaded}
\begin{Highlighting}[]
\KeywordTok{def}\NormalTok{ buffer\_guts\_v2(t, y, x\_in, eta, k\_d, b, z, h\_b, w):}
    \CommentTok{"""Bufferguts model used for visualizations and method development"""}

\NormalTok{    B }\OperatorTok{=}\NormalTok{ y[}\DecValTok{0}\NormalTok{]}
\NormalTok{    D }\OperatorTok{=}\NormalTok{ y[}\DecValTok{1}\NormalTok{]}
\NormalTok{    H }\OperatorTok{=}\NormalTok{ y[}\DecValTok{2}\NormalTok{]}

\NormalTok{    C }\OperatorTok{=}\NormalTok{ x\_in.evaluate(t)}

\NormalTok{    buffer\_filling\_rate }\OperatorTok{=}\NormalTok{ eta }\OperatorTok{*}\NormalTok{ (C }\OperatorTok{{-}}\NormalTok{ B)}
\NormalTok{    buffer\_depletion\_rate }\OperatorTok{=}\NormalTok{ k\_d }\OperatorTok{*}\NormalTok{ (C }\OperatorTok{{-}}\NormalTok{ B)}

    \CommentTok{\# \# \# 1 if B \textgreater{} C, 0 otherwise, with smooth transition at B\textasciitilde{}=C}
    \CommentTok{\# switch = 0.5 + (1 / jnp.pi) * jnp.arctan(1e5 * (B / bc\_scaler {-} C / bc\_scaler))}

    \CommentTok{\# dB\_dt = buffer\_filling\_rate {-} (buffer\_filling\_rate {-}}
    \CommentTok{\#     buffer\_depletion\_rate) * switch}

\NormalTok{    dB\_dt }\OperatorTok{=}\NormalTok{ jnp.where(C }\OperatorTok{\textgreater{}=}\NormalTok{ B, buffer\_filling\_rate, buffer\_depletion\_rate)}

\NormalTok{    dD\_dt }\OperatorTok{=}\NormalTok{ k\_d }\OperatorTok{*}\NormalTok{ (jnp.}\BuiltInTok{sum}\NormalTok{(B }\OperatorTok{*}\NormalTok{ w) }\OperatorTok{{-}}\NormalTok{ D)}

    \CommentTok{\# the numerical problem lies in the calculation of the buffer.}
    \CommentTok{\# switchDS = 0.5 + (1 / jnp.pi) * jnp.arctan(2e6 * (D {-} z))}
    \CommentTok{\# dH\_dt = (b * switchDS * (D {-} z) + h\_b)}
    
\NormalTok{    dH\_dt }\OperatorTok{=}\NormalTok{ (b }\OperatorTok{*}\NormalTok{ jnp.maximum((D }\OperatorTok{{-}}\NormalTok{ z),jnp.array([}\FloatTok{0.0}\NormalTok{], dtype}\OperatorTok{=}\BuiltInTok{float}\NormalTok{)) }\OperatorTok{+}\NormalTok{ h\_b)}
    
    \ControlFlowTok{return}\NormalTok{ dB\_dt, dD\_dt, dH\_dt}
\end{Highlighting}
\end{Shaded}

\subsubsection{Solver post processing}\label{solver-post-processing}

\begin{Shaded}
\begin{Highlighting}[]
    \AttributeTok{@staticmethod}
    \KeywordTok{def}\NormalTok{ \_solver\_post\_processing(results, t, interpolation):}
\NormalTok{        results[}\StringTok{"survival"}\NormalTok{] }\OperatorTok{=}\NormalTok{ jnp.exp(}\OperatorTok{{-}}\NormalTok{results[}\StringTok{"H"}\NormalTok{])}
\NormalTok{        results[}\StringTok{"exposure"}\NormalTok{] }\OperatorTok{=}\NormalTok{ jax.vmap(interpolation.evaluate)(t)}
        \ControlFlowTok{return}\NormalTok{ results}
\end{Highlighting}
\end{Shaded}

\subsubsection{Probability model}\label{probability-model}

\begin{figure}
\centering
\pandocbounded{\includegraphics[keepaspectratio,alt={Directed acyclic graph (DAG) of the probability model.}]{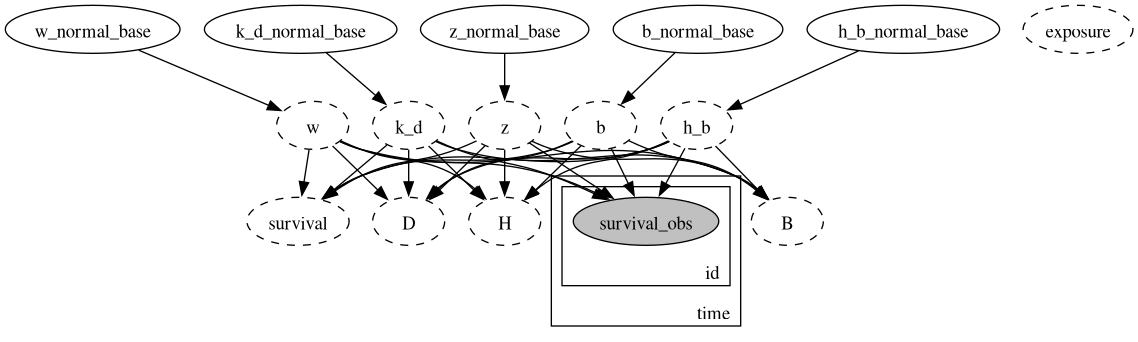}}
\caption{Directed acyclic graph (DAG) of the probability model.}
\end{figure}

\subsection{Report: Parameters}\label{report-parameters}

\subsubsection{Free parameters}\label{free-parameters}

\begin{itemize}
\tightlist
\item
  w \(\sim\)
  lognorm(scale={[}1000,1.0{]},s={[}2,1e-05{]},high={[}10000,1.01{]},low={[}1e-05,0.99{]},\\ dims=(`exposure\_path',))
\item
  k\_d \(\sim\) lognorm(scale=0.25,s=2,low=1e-07,high=25,dims=())
\item
  z \(\sim\) lognorm(scale=1.0,s=2,low=1e-08,high=100,dims=())
\item
  b \(\sim\) lognorm(scale=0.25,s=2,low=1e-07,high=250,dims=())
\item
  h\_b \(\sim\) lognorm(scale=0.01,s=2,low=1e-07,high=2.5,dims=())
\end{itemize}

\subsubsection{Fixed parameters}\label{fixed-parameters}

\begin{itemize}
\tightlist
\item
  eta \(=\) 1000.0, dims=()
\end{itemize}

\subsection{Report: Table parameter estimates
}\label{report-table-parameter-estimates}

\begin{longtable}[]{@{}rlrrrl@{}}
\toprule\noalign{}
& index & mean & hdi 3\% & hdi 97\% & unit \\
\midrule\noalign{}
\endhead
\bottomrule\noalign{}
\endlastfoot
0 & k\_d & 2.03 & 1.33 & 2.87 & 1/d \\
1 & z & 1.13 & 0.000196 & 3.49 & ng \\
2 & b & 0.00225 & 0.00145 & 0.00309 & 1/d/ng \\
3 & h\_b & 0.00226 & 4.77e-05 & 0.00544 & 1/d \\
4 & w\hyperref[oral]{oral} & 1530 & 850 & 2290 & $\mu$l \\
5 & w\hyperref[topical]{topical} & 1 & 1 & 1 & \\
\end{longtable}

\subsection{Report: Goodness of fit}\label{report-goodness-of-fit}

\begin{longtable}[]{@{}lrr@{}}
\toprule\noalign{}
& survival & model \\
\midrule\noalign{}
\endhead
\bottomrule\noalign{}
\endlastfoot
NRMSE & 0.0658973 & nan \\
NRMSE (95\%-hdi{[}lower{]}) & 0.0495851 & nan \\
NRMSE (95\%-hdi{[}upper{]}) & 0.0820703 & nan \\
Log-Likelihood & -80.4668 & -80.4668 \\
Log-Likelihood (95\%-hdi{[}lower{]}) & -83.3823 & -83.3823 \\
Log-Likelihood (95\%-hdi{[}upper{]}) & -78.2988 & -78.2988 \\
n (data) & 121 & 121 \\
k (parameters) & nan & 6 \\
BIC & nan & 189.708 \\
BIC (95\%-hdi{[}lower{]}) & nan & 185.372 \\
BIC (95\%-hdi{[}upper{]}) & nan & 195.539 \\
\end{longtable}

Report `goodness\_of\_fit' was successfully generated and saved in
`../bufferguts/results/osmia\_sulfoxaflor\_fluctuating\_exposure\_16\_8/goodness\_of\_fit.csv'

\subsection{Report: Diagnostics}\label{report-diagnostics}

\begin{figure}
\centering
\pandocbounded{\includegraphics[keepaspectratio,alt={Paired parameter estimates}]{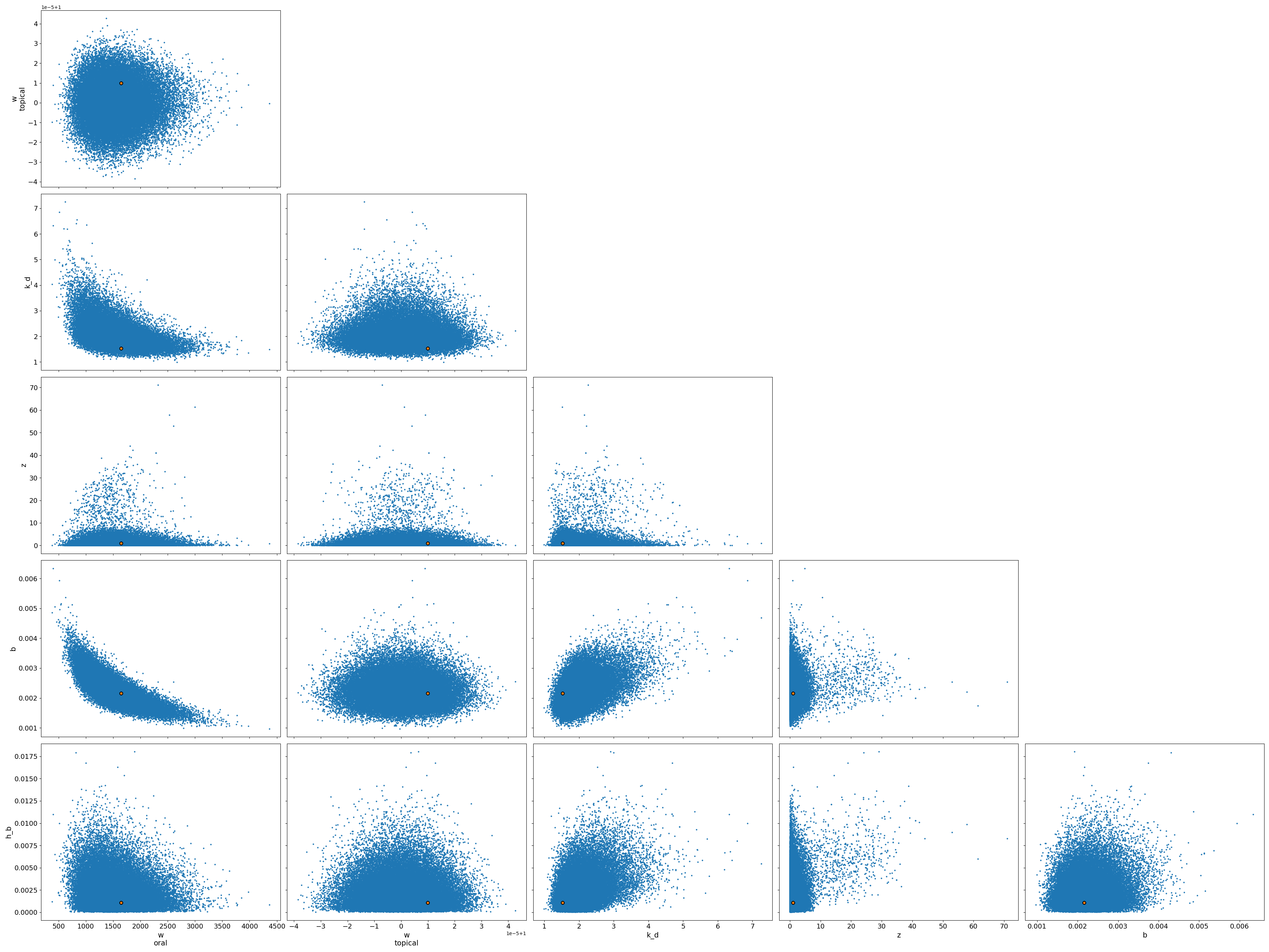}}
\caption{Paired parameter estimates}
\end{figure}

\begin{figure}
\centering
\pandocbounded{\includegraphics[keepaspectratio,alt={Psuedo trace, generated for draws from the optimized SVI distribution}]{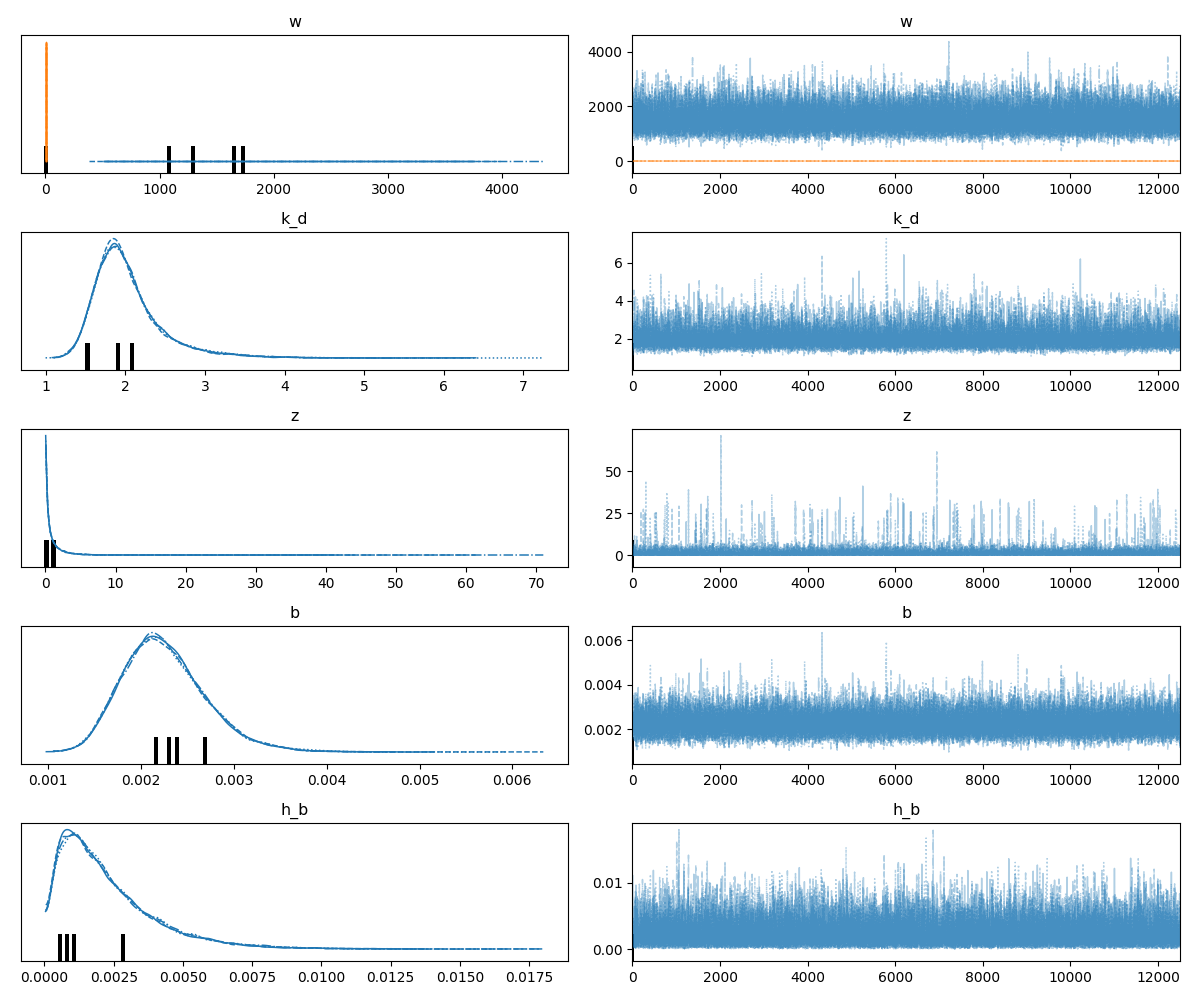}}
\caption{MCMC Trace, generated for draws from the posterior
distribution}
\end{figure}

Report `diagnostics' was successfully generated and saved in
'../bufferguts/results/\\osmia\_sulfoxaflor\_fluctuating\_exposure\_16\_8/posterior\_pairs.png',\\
'../bufferguts/results/osmia\_sulfoxaflor\_fluctuating\_exposure\_16\_8/posterior\_trace.png'

\subsection{Report: Model input}\label{report-model-input}

\subsubsection{Exposure conditions}\label{exposure-conditions}

These are the exposure conditions that were assumed for parameter
inference. Double check if they are aligned with your expectations.
Especially short exposure durations may not be perceivable in this view.
In this case it is recommended to have a look at the exposure conditions
in the numerical tables provided below.

\begin{figure}
\centering
\pandocbounded{\includegraphics[keepaspectratio,alt={Exposure model fits.}]{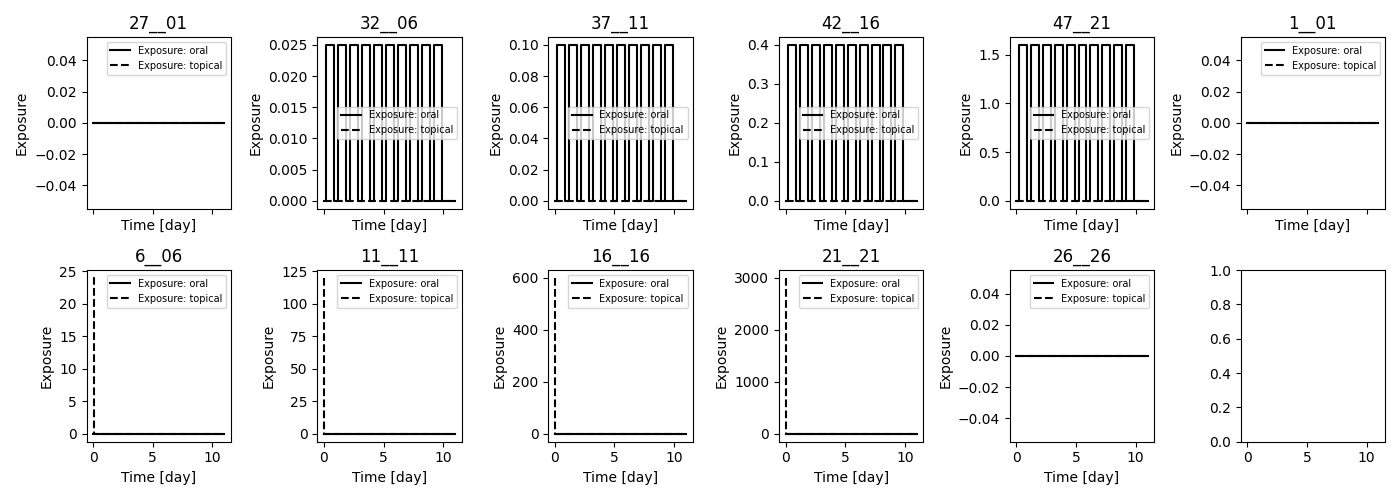}}
\caption{Exposure model
fits.\label{exposure--bufferguts--osmia-sulfoxaflor-fluctuating-exposure-16-8}}
\end{figure}

Report `model\_input' was successfully generated and saved in
`../bufferguts/results/osmia\_sulfoxaflor\_fluctuating\_exposure\_16\_8/exposure\_multipanel.png'

\subsection{Report: Model fits}\label{report-model-fits}

\subsubsection{Survival model fits}\label{survival-model-fits}

Survival observations on the unit scale with model fits. The solid line
is the average of individual survival probability predictions from
multiple draws from the posterior parameter distribution. In case a
point estimator was used the solid line indicates the best fit. Grey
uncertainty intervals indicate the uncertainty in survival
probabilities. Note that the survival probabilities indicate the
probability for a given individual or population to be alive when
observed at time t.

\begin{figure}
\centering
\pandocbounded{\includegraphics[keepaspectratio,alt={Surival model fits.}]{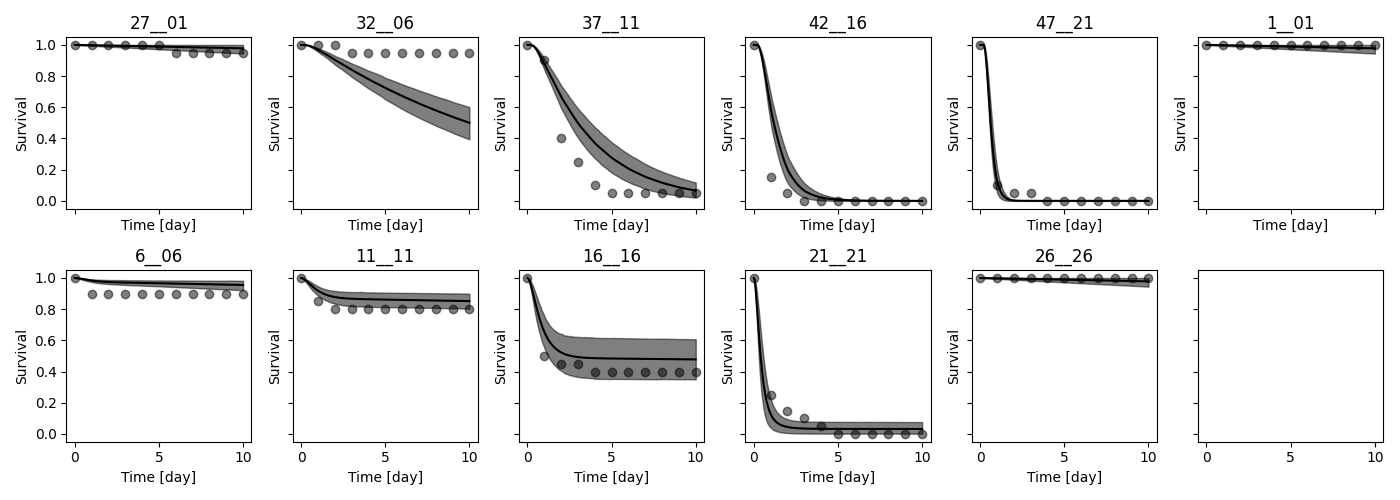}}
\caption{Surival model
fits.\label{survival_fits--bufferguts--osmia-sulfoxaflor-fluctuating-exposure-16-8}}
\end{figure}

Report `model\_fits' was successfully generated and saved in\\
`../bufferguts/results/osmia\_sulfoxaflor\_fluctuating\_exposure\_16\_8/survival\_multipanel.png'

\subsection{Report: Lcx estimates}\label{report-lcx-estimates}

\begin{footnotesize}
\begin{longtable}[]{@{}
  >{\raggedleft\arraybackslash}p{(\linewidth - 20\tabcolsep) * \real{0.0310}}
  >{\raggedleft\arraybackslash}p{(\linewidth - 20\tabcolsep) * \real{0.0698}}
  >{\raggedleft\arraybackslash}p{(\linewidth - 20\tabcolsep) * \real{0.0465}}
  >{\raggedleft\arraybackslash}p{(\linewidth - 20\tabcolsep) * \real{0.0620}}
  >{\raggedright\arraybackslash}p{(\linewidth - 20\tabcolsep) * \real{0.1628}}
  >{\raggedleft\arraybackslash}p{(\linewidth - 20\tabcolsep) * \real{0.0853}}
  >{\raggedleft\arraybackslash}p{(\linewidth - 20\tabcolsep) * \real{0.0775}}
  >{\raggedleft\arraybackslash}p{(\linewidth - 20\tabcolsep) * \real{0.0853}}
  >{\raggedleft\arraybackslash}p{(\linewidth - 20\tabcolsep) * \real{0.0775}}
  >{\raggedleft\arraybackslash}p{(\linewidth - 20\tabcolsep) * \real{0.0465}}
  >{\raggedright\arraybackslash}p{(\linewidth - 20\tabcolsep) * \real{0.2558}}@{}}
\toprule\noalign{}
\begin{minipage}[b]{\linewidth}\raggedleft
\end{minipage} & \begin{minipage}[b]{\linewidth}\raggedleft
index
\end{minipage} & \begin{minipage}[b]{\linewidth}\raggedleft
x
\end{minipage} & \begin{minipage}[b]{\linewidth}\raggedleft
time
\end{minipage} & \begin{minipage}[b]{\linewidth}\raggedright
scenario
\end{minipage} & \begin{minipage}[b]{\linewidth}\raggedleft
mean
\end{minipage} & \begin{minipage}[b]{\linewidth}\raggedleft
q05
\end{minipage} & \begin{minipage}[b]{\linewidth}\raggedleft
q95
\end{minipage} & \begin{minipage}[b]{\linewidth}\raggedleft
std
\end{minipage} & \begin{minipage}[b]{\linewidth}\raggedleft
cv
\end{minipage} & \begin{minipage}[b]{\linewidth}\raggedright
msg
\end{minipage} \\
\midrule\noalign{}
\endhead
\bottomrule\noalign{}
\endlastfoot
0 & 0 & 0.05 & 1 & acute 1hour oral & 0.06 & 0.0452 & 0.0777 & 0.0107
& 0.18 & Estimation success rate: 100.0\% \\
1 & 1 & 0.05 & 1 & acute 4hour oral & 0.0445 & 0.0342 & 0.0566 &
0.0072 & 0.16 & Estimation success rate: 100.0\% \\
2 & 2 & 0.05 & 1 & chronic oral & 0.0293 & 0.0223 & 0.0383 & 0.0051 &
0.17 & Estimation success rate: 100.0\% \\
3 & 3 & 0.05 & 1 & acute 1hour topical & 89.3906 & 65.1267 & 123.75 &
18.0145 & 0.2 & Estimation success rate: 99.2\% \\
4 & 4 & 0.05 & 1 & acute 4hour topical & 66.8085 & 47.7026 & 93.4906 &
14.5916 & 0.22 & Estimation success rate: 100.0\% \\
5 & 5 & 0.05 & 1 & chronic topical & 44.5532 & 27.6765 & 66.2923 &
11.885 & 0.27 & Estimation success rate: 100.0\% \\
6 & 6 & 0.05 & 2 & acute 1hour oral & 0.0413 & 0.0272 & 0.0615 &
0.0115 & 0.28 & Estimation success rate: 100.0\% \\
7 & 7 & 0.05 & 2 & acute 4hour oral & 0.0301 & 0.021 & 0.0427 & 0.0071
& 0.24 & Estimation success rate: 100.0\% \\
\end{longtable}
\end{footnotesize}
Report `LCx\_estimates' was successfully generated and saved in
`../bufferguts/results/osmia\_sulfoxaflor\_fluctuating\_exposure\_16\_8/report\_table\_lc\_x\_estimates.csv'

\subsection{Report: Exposure profiles}\label{report-exposure-profiles}

\subsubsection{Oral}\label{oral}

\begin{footnotesize}
\begin{longtable}[]{@{}
  >{\raggedleft\arraybackslash}p{(\linewidth - 24\tabcolsep) * \real{0.0400}}
  >{\raggedleft\arraybackslash}p{(\linewidth - 24\tabcolsep) * \real{0.0960}}
  >{\raggedleft\arraybackslash}p{(\linewidth - 24\tabcolsep) * \real{0.0800}}
  >{\raggedleft\arraybackslash}p{(\linewidth - 24\tabcolsep) * \real{0.0800}}
  >{\raggedleft\arraybackslash}p{(\linewidth - 24\tabcolsep) * \real{0.0800}}
  >{\raggedleft\arraybackslash}p{(\linewidth - 24\tabcolsep) * \real{0.0800}}
  >{\raggedleft\arraybackslash}p{(\linewidth - 24\tabcolsep) * \real{0.0800}}
  >{\raggedleft\arraybackslash}p{(\linewidth - 24\tabcolsep) * \real{0.0720}}
  >{\raggedleft\arraybackslash}p{(\linewidth - 24\tabcolsep) * \real{0.0720}}
  >{\raggedleft\arraybackslash}p{(\linewidth - 24\tabcolsep) * \real{0.0800}}
  >{\raggedleft\arraybackslash}p{(\linewidth - 24\tabcolsep) * \real{0.0800}}
  >{\raggedleft\arraybackslash}p{(\linewidth - 24\tabcolsep) * \real{0.0800}}
  >{\raggedleft\arraybackslash}p{(\linewidth - 24\tabcolsep) * \real{0.0800}}@{}}
\toprule\noalign{}
\begin{minipage}[b]{\linewidth}\raggedleft
\end{minipage} & \begin{minipage}[b]{\linewidth}\raggedleft
time
\end{minipage} & \begin{minipage}[b]{\linewidth}\raggedleft
27\_\_01
\end{minipage} & \begin{minipage}[b]{\linewidth}\raggedleft
32\_\_06
\end{minipage} & \begin{minipage}[b]{\linewidth}\raggedleft
37\_\_11
\end{minipage} & \begin{minipage}[b]{\linewidth}\raggedleft
42\_\_16
\end{minipage} & \begin{minipage}[b]{\linewidth}\raggedleft
47\_\_21
\end{minipage} & \begin{minipage}[b]{\linewidth}\raggedleft
1\_\_01
\end{minipage} & \begin{minipage}[b]{\linewidth}\raggedleft
6\_\_06
\end{minipage} & \begin{minipage}[b]{\linewidth}\raggedleft
11\_\_11
\end{minipage} & \begin{minipage}[b]{\linewidth}\raggedleft
16\_\_16
\end{minipage} & \begin{minipage}[b]{\linewidth}\raggedleft
21\_\_21
\end{minipage} & \begin{minipage}[b]{\linewidth}\raggedleft
26\_\_26
\end{minipage} \\
\midrule\noalign{}
\endhead
\bottomrule\noalign{}
\endlastfoot
0 & 0 & 0 & 0 & 0 & 0 & 0 & 0 & 0 & 0 & 0 & 0 & 0 \\
1 & 0.0416667 & 0 & 0 & 0 & 0 & 0 & 0 & 0 & 0 & 0 & 0 & 0 \\
2 & 0.0416667 & 0 & 0 & 0 & 0 & 0 & 0 & 0 & 0 & 0 & 0 & 0 \\
3 & 0.0833333 & 0 & 0 & 0 & 0 & 0 & 0 & 0 & 0 & 0 & 0 & 0 \\
4 & 0.0833333 & 0 & 0 & 0 & 0 & 0 & 0 & 0 & 0 & 0 & 0 & 0 \\
5 & 0.125 & 0 & 0 & 0 & 0 & 0 & 0 & 0 & 0 & 0 & 0 & 0 \\
6 & 0.125 & 0 & 0 & 0 & 0 & 0 & 0 & 0 & 0 & 0 & 0 & 0 \\
7 & 0.166667 & 0 & 0 & 0 & 0 & 0 & 0 & 0 & 0 & 0 & 0 & 0 \\
\end{longtable}
\end{footnotesize}

\subsubsection{Topical}\label{topical}

\begin{footnotesize}
\begin{longtable}[]{@{}
  >{\raggedleft\arraybackslash}p{(\linewidth - 24\tabcolsep) * \real{0.0400}}
  >{\raggedleft\arraybackslash}p{(\linewidth - 24\tabcolsep) * \real{0.0960}}
  >{\raggedleft\arraybackslash}p{(\linewidth - 24\tabcolsep) * \real{0.0800}}
  >{\raggedleft\arraybackslash}p{(\linewidth - 24\tabcolsep) * \real{0.0800}}
  >{\raggedleft\arraybackslash}p{(\linewidth - 24\tabcolsep) * \real{0.0800}}
  >{\raggedleft\arraybackslash}p{(\linewidth - 24\tabcolsep) * \real{0.0800}}
  >{\raggedleft\arraybackslash}p{(\linewidth - 24\tabcolsep) * \real{0.0800}}
  >{\raggedleft\arraybackslash}p{(\linewidth - 24\tabcolsep) * \real{0.0720}}
  >{\raggedleft\arraybackslash}p{(\linewidth - 24\tabcolsep) * \real{0.0720}}
  >{\raggedleft\arraybackslash}p{(\linewidth - 24\tabcolsep) * \real{0.0800}}
  >{\raggedleft\arraybackslash}p{(\linewidth - 24\tabcolsep) * \real{0.0800}}
  >{\raggedleft\arraybackslash}p{(\linewidth - 24\tabcolsep) * \real{0.0800}}
  >{\raggedleft\arraybackslash}p{(\linewidth - 24\tabcolsep) * \real{0.0800}}@{}}
\toprule\noalign{}
\begin{minipage}[b]{\linewidth}\raggedleft
\end{minipage} & \begin{minipage}[b]{\linewidth}\raggedleft
time
\end{minipage} & \begin{minipage}[b]{\linewidth}\raggedleft
27\_\_01
\end{minipage} & \begin{minipage}[b]{\linewidth}\raggedleft
32\_\_06
\end{minipage} & \begin{minipage}[b]{\linewidth}\raggedleft
37\_\_11
\end{minipage} & \begin{minipage}[b]{\linewidth}\raggedleft
42\_\_16
\end{minipage} & \begin{minipage}[b]{\linewidth}\raggedleft
47\_\_21
\end{minipage} & \begin{minipage}[b]{\linewidth}\raggedleft
1\_\_01
\end{minipage} & \begin{minipage}[b]{\linewidth}\raggedleft
6\_\_06
\end{minipage} & \begin{minipage}[b]{\linewidth}\raggedleft
11\_\_11
\end{minipage} & \begin{minipage}[b]{\linewidth}\raggedleft
16\_\_16
\end{minipage} & \begin{minipage}[b]{\linewidth}\raggedleft
21\_\_21
\end{minipage} & \begin{minipage}[b]{\linewidth}\raggedleft
26\_\_26
\end{minipage} \\
\midrule\noalign{}
\endhead
\bottomrule\noalign{}
\endlastfoot
0 & 0 & 0 & 0 & 0 & 0 & 0 & 0 & 24 & 120 & 600 & 3000 & 0 \\
1 & 0.0416667 & 0 & 0 & 0 & 0 & 0 & 0 & 24 & 120 & 600 & 3000 & 0 \\
2 & 0.0416667 & 0 & 0 & 0 & 0 & 0 & 0 & 0 & 0 & 0 & 0 & 0 \\
3 & 0.0833333 & 0 & 0 & 0 & 0 & 0 & 0 & 0 & 0 & 0 & 0 & 0 \\
4 & 0.0833333 & 0 & 0 & 0 & 0 & 0 & 0 & 0 & 0 & 0 & 0 & 0 \\
5 & 0.125 & 0 & 0 & 0 & 0 & 0 & 0 & 0 & 0 & 0 & 0 & 0 \\
6 & 0.125 & 0 & 0 & 0 & 0 & 0 & 0 & 0 & 0 & 0 & 0 & 0 \\
7 & 0.166667 & 0 & 0 & 0 & 0 & 0 & 0 & 0 & 0 & 0 & 0 & 0 \\
\end{longtable}
\end{footnotesize}

Report `exposure\_profiles' was successfully generated and saved in
`{[}'../bufferguts/results/osmia\_sulfoxaflor\_fluctuating\_exposure\_16\_8/report\_table\_exposure\_profile\_oral--bufferguts--osmia-sulfoxaflor-fluctuating-exposure-16-8.csv',
\\'../bufferguts/results/osmia\_sulfoxaflor\_fluctuating\_exposure\_16\_8/report\_table\_exposure\_profile\_topical--bufferguts--osmia-sulfoxaflor-fluctuating-exposure-16-8.csv'{]}'

\subsection{Report: Almass parameters}\label{report-almass-parameters}

\subsubsection{ALMaSS Unit conversion}\label{almass-unit-conversion}

To use the GUTS parameters in predictions with other quantities . These
parameters can be used in ALMaSSrunning with the following exposure
units:

\begin{longtable}[]{@{}
  >{\raggedright\arraybackslash}p{(\linewidth - 6\tabcolsep) * \real{0.2152}}
  >{\raggedright\arraybackslash}p{(\linewidth - 6\tabcolsep) * \real{0.2658}}
  >{\raggedright\arraybackslash}p{(\linewidth - 6\tabcolsep) * \real{0.2532}}
  >{\raggedleft\arraybackslash}p{(\linewidth - 6\tabcolsep) * \real{0.2658}}@{}}
\toprule\noalign{}
\begin{minipage}[b]{\linewidth}\raggedright
exposure\_path
\end{minipage} & \begin{minipage}[b]{\linewidth}\raggedright
Calibration Units
\end{minipage} & \begin{minipage}[b]{\linewidth}\raggedright
Prediction Units
\end{minipage} & \begin{minipage}[b]{\linewidth}\raggedleft
Conversion Factor
\end{minipage} \\
\midrule\noalign{}
\endhead
\bottomrule\noalign{}
\endlastfoot
oral & ng/$\mu$L & mg/mg & 1.12e+06 \\
topical & ng & mg & 1e+06 \\
\end{longtable}

To account for other input (exposure) units, only the \(w\) parameters
need linear rescaling, according to the parameter units relationships in
the following MAP-table.The weight for topical exposure is considered
fixed. This means, also the unitis dimensionless, so that the unit of
the threshold parameter \(z\) can takethe unit of topical exposure. If
for predictions of topical exposure a differentunit is used than during
calibration, the weight needs to be modified with aconversion parameter,
which converts from the new input unit to a the old exposure unit, which
resolves any unit conflicts.

Example for modifying the topical exposure weight if the calibration was
donein \(\mu g\) and the prediction is done in \(mg\):
\[w_{topical}~\left[\frac{\mu g}{mg}\right] = w_{topical}~[-] \cdot \frac{1000~\mu g}{mg}\]The
new unit is still dimensionless (mass/mass), but in it rescales the
unitto the correct scale, which satisfies that the unit of the
multiplication\[B_{topical} [mg] \cdot w_{topical} [\mu g / mg] \rightarrow \mu g\]

Example for modifying the oral exposure weight if the calibration was
done in \(\mu g/\mu l\) (volumetric concentration) and the prediction is
done in \(mg/mg\) (mass concentration): Let's break this down into two
steps. First scale the
nominator\[w_{oral}~\left[\mu l~\text{food} \cdot \frac{\mu g}{mg}\right] = w_{oral}~[\mu l~\text{food}] \cdot \frac{1000~\mu g}{mg}\]Second,
we will convert the volumetric concentration to a mass
concentration,assuming a density of 1.12 g/ml = 1.12 mg/$\mu$l of a 30\%
sucrose
solution\[w_{oral}~\left[mg~\text{food} \cdot \frac{\mu g}{mg}\right] = w_{oral}~[\mu l~\text{food}] \cdot \frac{1000~\mu g}{mg} \cdot \frac{1.12~mg~\text{food}}{\mu l~\text{food}}\]
Again, this satisfies the units for the
equation\[B_{oral} \left[\frac{mg}{mg~\text{food}}\right] \cdot w_{oral} \left[mg~\text{food} \cdot \frac{\mu g}{mg}\right] \rightarrow \mu g\]

\subsubsection{Maximum a posteriori (MAP) parameter
values}\label{maximum-a-posteriori-map-parameter-values}

Reporting the sample from the posterior with the highest
likelihood(\(\approx\) maximum a posteriori, MAP)

\begin{longtable}[]{@{}
  >{\raggedleft\arraybackslash}p{(\linewidth - 12\tabcolsep) * \real{0.0426}}
  >{\raggedright\arraybackslash}p{(\linewidth - 12\tabcolsep) * \real{0.1277}}
  >{\raggedleft\arraybackslash}p{(\linewidth - 12\tabcolsep) * \real{0.1702}}
  >{\raggedright\arraybackslash}p{(\linewidth - 12\tabcolsep) * \real{0.1277}}
  >{\raggedleft\arraybackslash}p{(\linewidth - 12\tabcolsep) * \real{0.2234}}
  >{\raggedleft\arraybackslash}p{(\linewidth - 12\tabcolsep) * \real{0.1809}}
  >{\raggedright\arraybackslash}p{(\linewidth - 12\tabcolsep) * \real{0.1277}}@{}}
\toprule\noalign{}
\begin{minipage}[b]{\linewidth}\raggedleft
\end{minipage} & \begin{minipage}[b]{\linewidth}\raggedright
index
\end{minipage} & \begin{minipage}[b]{\linewidth}\raggedleft
MAP original
\end{minipage} & \begin{minipage}[b]{\linewidth}\raggedright
old unit
\end{minipage} & \begin{minipage}[b]{\linewidth}\raggedleft
conversion factor
\end{minipage} & \begin{minipage}[b]{\linewidth}\raggedleft
MAP converted
\end{minipage} & \begin{minipage}[b]{\linewidth}\raggedright
new unit
\end{minipage} \\
\midrule\noalign{}
\endhead
\bottomrule\noalign{}
\endlastfoot
0 & b & 0.00213564 & 1/d/ng & 1 & 0.00213564 & 1/d/ng \\
1 & h\_b & 0.00124531 & 1/d & 1 & 0.00124531 & 1/d \\
2 & k\_d & 1.91086 & 1/d & 1 & 1.91086 & 1/d \\
3 & z & 0.0906224 & ng & 1 & 0.0906224 & ng \\
4 & w\hyperref[oral]{oral} & 1562.6 & $\mu$l & 1.12e+06 & 1.75011e+09 &
ng \\
5 & w\hyperref[topical]{topical} & 1 & & 1e+06 & 1e+06 & ng/mg \\
\end{longtable}

\subsubsection{Posterior Samples}\label{posterior-samples}

Reporting samples from the posterior scaled to ALMaSS units. These
parametersamples have the same units as the MAP parameter values.

\begin{footnotesize}
\begin{longtable}[]{@{}
  >{\raggedleft\arraybackslash}p{(\linewidth - 16\tabcolsep) * \real{0.0412}}
  >{\raggedleft\arraybackslash}p{(\linewidth - 16\tabcolsep) * \real{0.0928}}
  >{\raggedleft\arraybackslash}p{(\linewidth - 16\tabcolsep) * \real{0.0825}}
  >{\raggedleft\arraybackslash}p{(\linewidth - 16\tabcolsep) * \real{0.1237}}
  >{\raggedleft\arraybackslash}p{(\linewidth - 16\tabcolsep) * \real{0.1340}}
  >{\raggedleft\arraybackslash}p{(\linewidth - 16\tabcolsep) * \real{0.0928}}
  >{\raggedleft\arraybackslash}p{(\linewidth - 16\tabcolsep) * \real{0.1134}}
  >{\raggedleft\arraybackslash}p{(\linewidth - 16\tabcolsep) * \real{0.1340}}
  >{\raggedleft\arraybackslash}p{(\linewidth - 16\tabcolsep) * \real{0.1856}}@{}}
\toprule\noalign{}
\begin{minipage}[b]{\linewidth}\raggedleft
\end{minipage} & \begin{minipage}[b]{\linewidth}\raggedleft
chain
\end{minipage} & \begin{minipage}[b]{\linewidth}\raggedleft
draw
\end{minipage} & \begin{minipage}[b]{\linewidth}\raggedleft
b
\end{minipage} & \begin{minipage}[b]{\linewidth}\raggedleft
h\_b
\end{minipage} & \begin{minipage}[b]{\linewidth}\raggedleft
k\_d
\end{minipage} & \begin{minipage}[b]{\linewidth}\raggedleft
z
\end{minipage} & \begin{minipage}[b]{\linewidth}\raggedleft
w\hyperref[oral]{oral}
\end{minipage} & \begin{minipage}[b]{\linewidth}\raggedleft
w\hyperref[topical]{topical}
\end{minipage} \\
\midrule\noalign{}
\endhead
\bottomrule\noalign{}
\endlastfoot
0 & 0 & 0 & 0.00215915 & 0.00107392 & 1.5291 & 1.04498 & 1.84613e+09 &
1.00001e+06 \\
1 & 0 & 1 & 0.00211176 & 0.0026166 & 2.16054 & 0.0206521 & 1.90897e+09 &
999988 \\
2 & 0 & 2 & 0.00199637 & 0.00215552 & 1.71689 & 4.13085 & 2.07411e+09 &
1.00001e+06 \\
3 & 0 & 3 & 0.00178811 & 0.011158 & 1.40833 & 7.86911 & 1.84143e+09 &
999994 \\
4 & 0 & 4 & 0.00175703 & 0.00159007 & 1.49219 & 0.614089 & 1.82224e+09 &
1.00002e+06 \\
5 & 0 & 5 & 0.00219964 & 0.00341549 & 1.58512 & 0.514841 & 1.55228e+09 &
1.00002e+06 \\
6 & 0 & 6 & 0.00257267 & 0.000771103 & 2.268 & 0.195835 & 1.3254e+09 &
999994 \\
7 & 0 & 7 & 0.00211959 & 0.00270542 & 1.78554 & 0.807085 & 1.66276e+09 &
1.00001e+06 \\
8 & 0 & 8 & 0.0018401 & 0.000664292 & 1.59174 & 0.0900573 & 2.26895e+09
& 999993 \\
9 & 0 & 9 & 0.00200638 & 0.00414157 & 2.33167 & 0.138918 & 1.65811e+09 &
1.00001e+06 \\
10 & 0 & 10 & 0.00172053 & 0.000393532 & 1.38449 & 0.728299 &
2.79334e+09 & 999990 \\
11 & 0 & 11 & 0.00161698 & 0.0025904 & 2.30088 & 0.348249 & 2.59796e+09
& 1.00001e+06 \\
12 & 0 & 12 & 0.00258736 & 0.000867956 & 1.52789 & 0.85384 & 1.04398e+09
& 999988 \\
13 & 0 & 13 & 0.00232552 & 0.00174187 & 1.92628 & 2.15121 & 1.39506e+09
& 999996 \\
14 & 0 & 14 & 0.00228096 & 0.0020595 & 1.6375 & 2.86885 & 1.45465e+09 &
999995 \\
15 & 0 & 15 & 0.00224862 & 0.00240523 & 1.89863 & 3.0669 & 1.38422e+09 &
999999 \\
16 & 0 & 16 & 0.00186013 & 0.00267178 & 1.6955 & 0.170419 & 2.29782e+09
& 999998 \\
17 & 0 & 17 & 0.00209788 & 0.000261169 & 1.77183 & 0.0650263 &
1.65603e+09 & 999999 \\
18 & 0 & 18 & 0.00222837 & 0.00242614 & 2.01299 & 0.118818 & 1.56801e+09
& 1e+06 \\
19 & 0 & 19 & 0.00273045 & 0.00539858 & 2.47891 & 0.16663 & 1.05398e+09
& 999997 \\
20 & 0 & 20 & 0.00220192 & 0.00279349 & 2.52361 & 0.799739 & 1.66577e+09
& 999988 \\
21 & 0 & 21 & 0.00191914 & 0.00352607 & 2.04787 & 0.0564138 &
1.77392e+09 & 999990 \\
22 & 0 & 22 & 0.0018302 & 0.00221885 & 1.89787 & 1.14251 & 1.635e+09 &
999996 \\
23 & 0 & 23 & 0.00242806 & 0.000840493 & 1.93038 & 0.0585593 &
1.59939e+09 & 1e+06 \\
24 & 0 & 24 & 0.00214931 & 0.00334651 & 2.1131 & 2.36113 & 1.60804e+09 &
999997 \\
25 & 0 & 25 & 0.00212744 & 0.00311526 & 2.61472 & 0.572342 & 1.56537e+09
& 1.00001e+06 \\
26 & 0 & 26 & 0.00248947 & 0.000459459 & 1.83902 & 0.0292086 &
1.82052e+09 & 1e+06 \\
27 & 0 & 27 & 0.00236385 & 0.00104668 & 2.20194 & 0.317328 & 1.89917e+09
& 999984 \\
28 & 0 & 28 & 0.00236897 & 0.00108218 & 2.15592 & 0.343203 & 1.64543e+09
& 999982 \\
29 & 0 & 29 & 0.00272557 & 0.000685009 & 2.24685 & 0.0955135 &
1.37199e+09 & 1.00002e+06 \\
30 & 0 & 30 & 0.00204391 & 0.000345498 & 1.48033 & 0.284863 &
1.67842e+09 & 1e+06 \\
31 & 0 & 31 & 0.00251132 & 0.00139078 & 2.05025 & 0.680157 & 1.25981e+09
& 999996 \\
32 & 0 & 32 & 0.00172126 & 0.000843544 & 1.85178 & 0.108004 & 2.5416e+09
& 1.00001e+06 \\
33 & 0 & 33 & 0.00233889 & 0.000232086 & 1.85852 & 0.140285 &
2.03027e+09 & 999994 \\
34 & 0 & 34 & 0.00205755 & 0.00183302 & 1.95416 & 2.07027 & 1.24625e+09
& 1e+06 \\
35 & 0 & 35 & 0.00204607 & 0.000299679 & 2.06271 & 0.0373003 &
2.03037e+09 & 999990 \\
36 & 0 & 36 & 0.00231942 & 0.00570742 & 1.89662 & 2.44806 & 1.32213e+09
& 1.00001e+06 \\
37 & 0 & 37 & 0.00236489 & 0.000539948 & 1.96071 & 1.20882 & 2.14869e+09
& 1.00002e+06 \\
38 & 0 & 38 & 0.00207512 & 0.000695518 & 1.60025 & 0.300008 &
2.52259e+09 & 999996 \\
39 & 0 & 39 & 0.00201344 & 0.00197759 & 2.12834 & 0.178976 & 1.37491e+09
& 999996 \\
40 & 0 & 40 & 0.00227226 & 0.0010919 & 1.8021 & 2.53608 & 2.30498e+09 &
999982 \\
41 & 0 & 41 & 0.00210868 & 0.00198125 & 1.49655 & 3.90227 & 2.06215e+09
& 1.00001e+06 \\
42 & 0 & 42 & 0.00215527 & 0.0015777 & 2.38702 & 0.0301932 & 1.58825e+09
& 1e+06 \\
43 & 0 & 43 & 0.00177075 & 0.00402002 & 1.54167 & 2.00613 & 2.28924e+09
& 1e+06 \\
44 & 0 & 44 & 0.0021965 & 0.000445876 & 1.70337 & 0.123667 & 1.81147e+09
& 999997 \\
45 & 0 & 45 & 0.00233933 & 0.00315624 & 2.06228 & 1.99113 & 1.4129e+09 &
1.00001e+06 \\
46 & 0 & 46 & 0.00200262 & 0.00367238 & 3.02578 & 0.165813 & 1.69728e+09
& 999997 \\
47 & 0 & 47 & 0.00278514 & 0.00433969 & 3.97109 & 1.31668 & 1.06344e+09
& 1e+06 \\
48 & 0 & 48 & 0.00296946 & 0.00525322 & 3.70388 & 0.219367 & 8.92491e+08
& 1.00001e+06 \\
49 & 0 & 49 & 0.00361505 & 0.00623117 & 4.03587 & 0.0449436 &
8.47668e+08 & 1e+06 \\
\end{longtable}
\end{footnotesize}

\section{Osmia bicornis female population sizes}

\begin{figure}[htb]
    \centering
    \begin{subfigure}{0.9\textwidth}
         \centering
         \includegraphics[width=\linewidth]{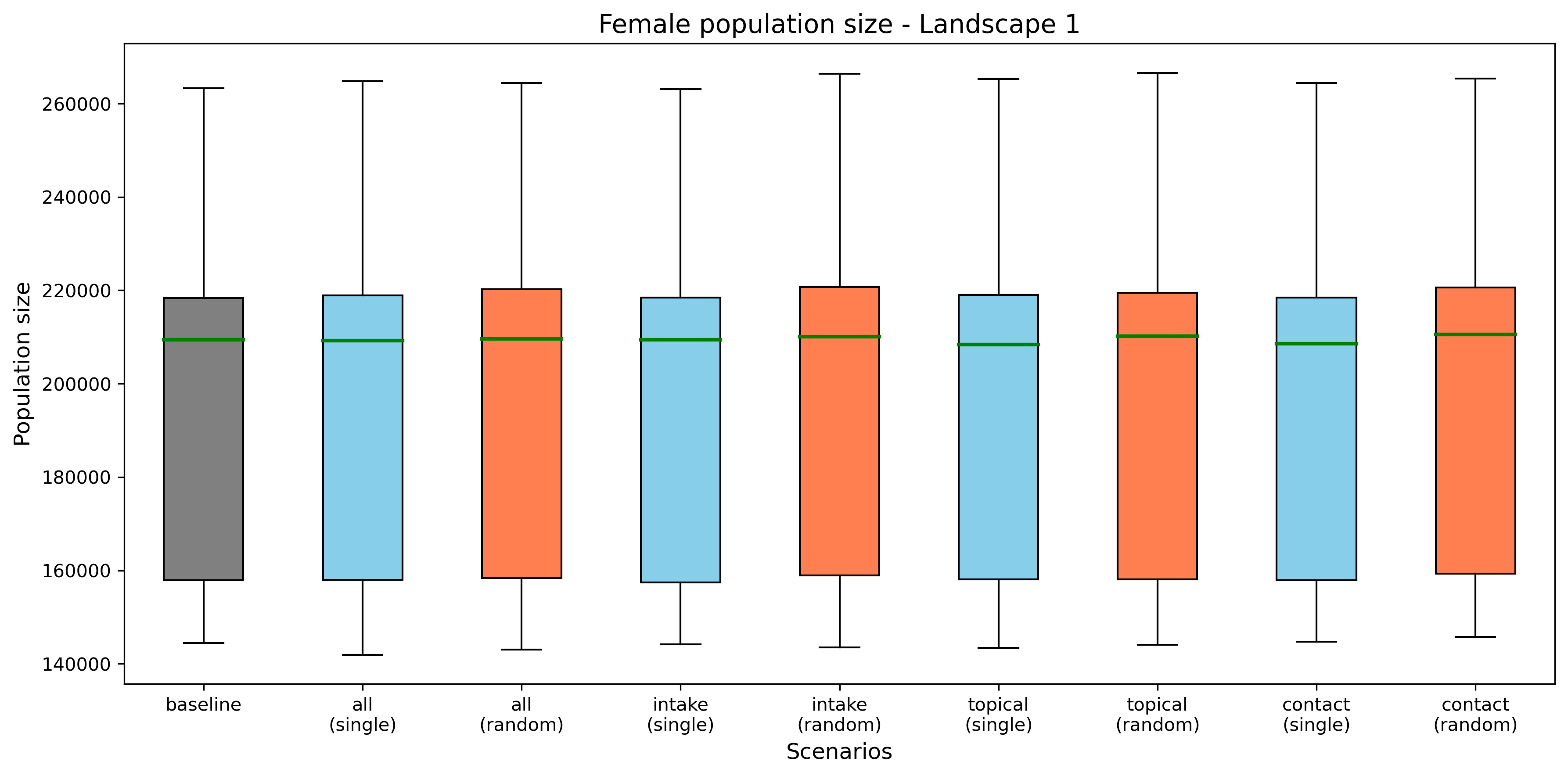}
    \end{subfigure}
    \begin{subfigure}{0.9\textwidth}
         \centering
         \includegraphics[width=\linewidth]{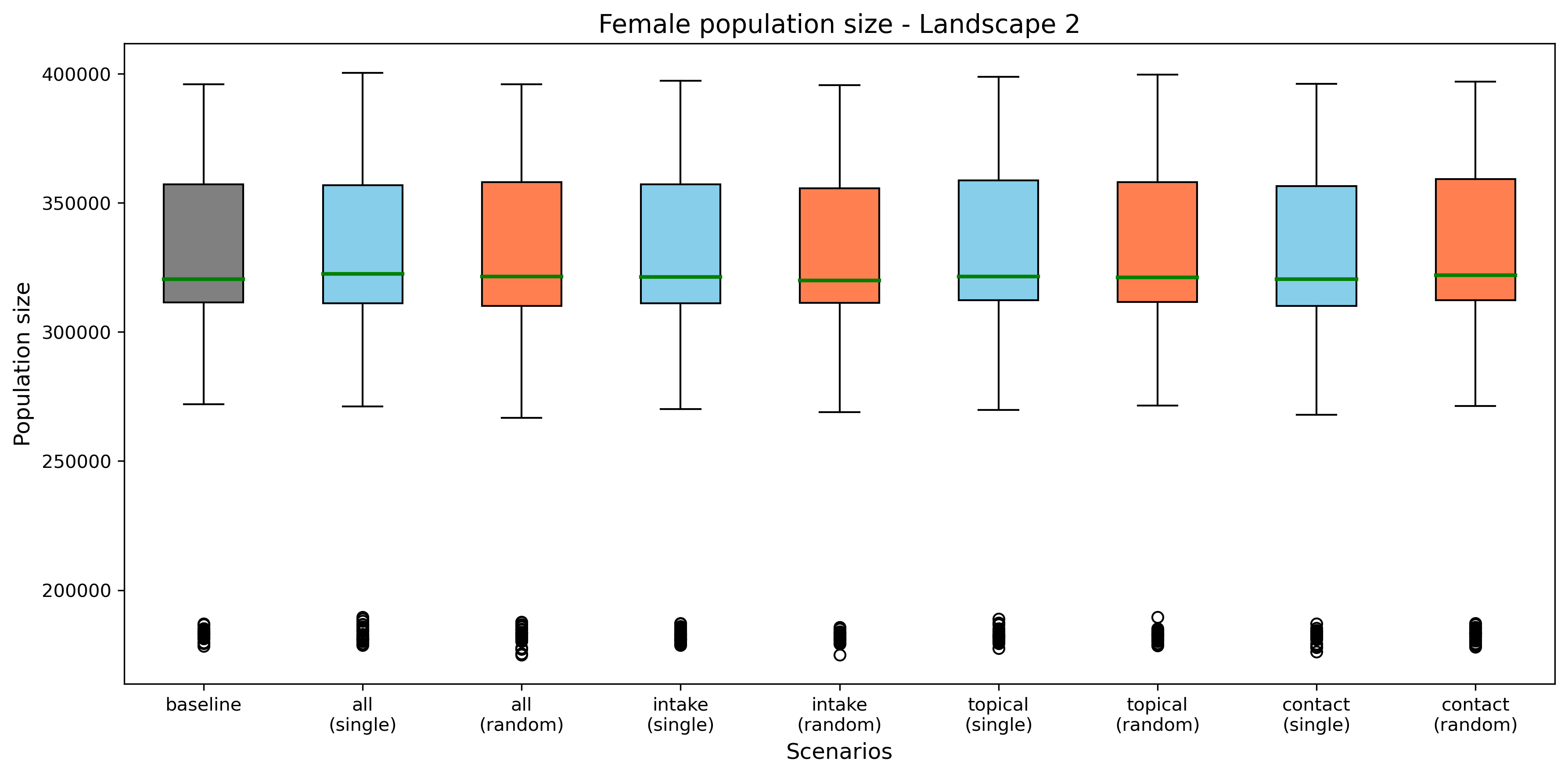}
    \end{subfigure}
    \caption{early female population size on Landscape 1 and 2 for the five scenarios. Box plots show the median (green line), interquartile range (box), and range (whiskers) of female population across 30 replications over the last 10 years of simulation.}
    \label{fig:accumulated_females}
\end{figure}

\end{document}